\documentclass{IEEEojcsys}

\usepackage[colorlinks,urlcolor=blue,linkcolor=blue,citecolor=blue]{hyperref}

\usepackage{subcaption}
\usepackage[font={small,bf,sf}]{caption}
\usepackage[utf8]{inputenc}
\usepackage{amsmath}
\usepackage{color,array}
\usepackage{romannum}
\usepackage{textcomp}
\usepackage{balance}
\usepackage{array}
\usepackage{xcolor}
\usepackage{textcomp}
\usepackage{stfloats}
\usepackage[most]{tcolorbox}
\usepackage{cleveref}
\newtcolorbox{mymathbox}[1][]{colback=white, sharp corners, #1}
\usepackage{mathtools, nccmath}
\usepackage{graphicx}
\usepackage{amsfonts}
\usepackage{amssymb}
\usepackage{algorithm}
\usepackage{algorithmic}
\usepackage{hyperref}
\usepackage{tikz}
\usepackage{arydshln}
\usepackage{multirow}
\usepackage{bm}
\usepackage{epstopdf}
\usepackage{cite}
\usepackage{siunitx}
\usepackage{lipsum}
\usepackage{wrapfig}
\usepackage{amsthm}
\Crefname{figure}{\text{Fig.}}{\text{Figs.}}
\Crefname{equation}{}{}
\usepackage{titlesec}

\titleformat{\subsubsection}[runin] 
  {\itshape\normalsize}          
  {\thesubsubsection)}              
  {0.3em}                             
  {}                                

\jvol{00}
\jnum{XX}
\paper{1234567}
\pubyear{2021}
\receiveddate{XX September 2021}
\accepteddate{XX October 2021}
\publisheddate{XX November 2021}
\currentdate{XX November 2021}
\doiinfo{OJCSYS.2021.Doi Number}

\newtheorem{theorem}{Theorem}

\newtheorem{remark}{Remark}
\newtheorem{definition}{Definition}
\newtheorem{problem}{Problem}

\begin{document}

\sptitle{Article Category}

\title{Learning-Enabled 
 Iterative Convex Optimization for Safety-Critical \\ Model Predictive Control} 

\editor{This paper was recommended by Associate Editor F. A. Author.}

\author{SHUO LIU\affilmark{1} (Student Member, IEEE)}

\author{ZHE HUANG\affilmark{1}}

\author{JUN ZENG\affilmark{2} (Member, IEEE)}

\author{KOUSHIL SREENATH \affilmark{2} (Senior Member, IEEE)}

\author{AND CALIN A. BELTA \affilmark{3} (Fellow, IEEE)}

\affil{Department of Mechanical Engineering, Boston
University, Boston, MA 02215 USA} 
\affil{Department of Mechanical Engineering, University of California, Berkeley, Berkeley, CA 94720 USA} 
\affil{Departments of Electrical and Computer Engineering and Computer Science, University of Maryland, College Park, MD 20742 USA} 

\corresp{CORRESPONDING AUTHOR: SHUO LIU (e-mail: \href{mailto:liushuo@bu.edu}{liushuo@bu.edu})}
\authornote{This work was supported in part by the NSF under grants IIS-2024606 and CMMI-1944722.}

\markboth{LEARNING-ENABLED ITERATIVE CONVEX OPTIMIZATION FOR SAFETY-CRITICAL MODEL PREDICTIVE CONTROL}{LIU {\itshape ET AL}.}

\begin{abstract}
Safety remains a central challenge in control of dynamical systems, particularly when the boundaries of unsafe sets are complex (e.g., nonconvex, nonsmooth)
or unknown.
This paper proposes a learning-enabled framework for safety-critical Model Predictive Control (MPC) that integrates Discrete-Time High-Order Control Barrier Functions (DHOCBFs) with iterative convex optimization. Unlike existing methods that primarily address CBFs of relative degree one with fully known unsafe set boundaries, our approach generalizes to arbitrary relative degrees and addresses scenarios where only samples are available for the unsafe set boundaries.
We extract pixels from unsafe set boundaries and train a neural network to approximate local linearizations. The learned models are incorporated into the linearized DHOCBF constraints at each time step within the MPC framework. An iterative convex optimization procedure is developed to accelerate computation while maintaining formal safety guarantees.
The benefits of computational performance and safe avoidance of obstacles with diverse shapes are examined and confirmed through numerical results.
By bridging model-based control with learning-based environment modeling, this framework advances safe autonomy for discrete-time systems operating in complex and partially known settings.
\end{abstract}


\maketitle

\section{INTRODUCTION}

\subsection{MOTIVATION}

Safety-critical optimal control is a central problem in  autonomous systems.
For example, reaching a goal while avoiding obstacles and minimizing energy can be formulated as a constrained optimal control problem by using continuous-time Control Barrier Functions (CBFs)~\cite{ames2014control, ames2016control}. By dividing the timeline into small intervals, the problem is reduced to a (possibly large) number of  quadratic programs, which can be solved at real-time speeds. However, this approach may be overly aggressive because it does not anticipate future events.

In \cite{zeng2021safety}, the authors use discrete-time CBFs as constraints inside a discrete-time Model Predictive Control (MPC) problem, which provides a safer control policy as it incorporates information about future states over a receding horizon to establish greater safety margins.
However, the computational time is large and grows rapidly with a larger horizon, since the optimization is usually nonlinear and non-convex. An additional problem with this nonlinear MPC formulation is the feasibility of the optimization. Moreover, it is difficult to obtain the CBFs from complex environments, i.e., the boundaries of unsafe sets in the environment are difficult to describe with known equations.

In this article, we address the above challenges by proposing a convex MPC with linearized, discrete-time CBFs, under an iterative approach. In contrast with the real-time iteration (RTI) approach introduced in \cite{diehl2005real}, which solves the problem through iterative Newton steps, and the learning MPC (LMPC) framework \cite{rosolia2018learning}, which builds invariant safe sets from iteratively stored trajectories and requires repeated executions of the same task with an initial feasible trajectory for convergence, our approach iteratively solves a convex MPC at each time step by constructing linear safety constraints from unsafe sets, enabling safety from the first execution with no prior data or trial runs needed. While classical constrained MPC~\cite{borrelli2017predictive} addresses linear dynamics with polyhedral constraints, our formulation embeds high-order CBF constraints, a learned boundary detector, and bounded slack variables, enabling safety constraint enforcement for iteratively linearized systems and unsafe sets. If the boundaries of the unsafe sets are simple and known, we directly linearize them; if complex and unknown, a Deep Neural Network (DNN), referred to as a Safety Boundary Detector (SBD), predicts local linear approximations, which are then transformed into linearized, discrete-time CBFs to ensure safety. 

\subsection{RELATED WORK}

\subsubsection{Model Predictive Control (MPC):}\hspace{0.1em} This paper can be seen in the context of MPC with safety constraints. MPC is widely used in control systems, such as controller design in robotic manipulation and locomotion~\cite{paolo2017mpc, scianca2020mpc}, to obtain a control strategy as a solution to an optimization problem. Stability was achieved in \cite{grandia2020nonlinear} by incorporating Discrete-time Control Lyapunov Functions (DCLFs) into a general MPC-based optimization problem to realize real-time control on a robotic system with limited computational resources.  The authors of \cite{son2019safety} emphasize safety in robot design and deployment. When safety means obstacle avoidance, some works achieve it through the introduction of repelling functions~\cite{ames2014control, eklund2011switched}, while others  regard obstacle avoidance as one concrete example of safety criteria for robots \cite{frasch2013auto, liniger2015optimization, zhang2020optimization}. In these works,  safety is formulated as constraints in optimization problems.

\subsubsection{Control Barrier Functions (CBFs):}\hspace{0.1em}
CBFs are used to enforce safety, i.e.,
rendering a set forward invariant for a dynamical system. Recent studies have demonstrated that to ensure the stabilization of an affine control system, while also adhering to safety constraints and control limitations, CBFs can be integrated with Control Lyapunov Functions (CLFs). This integration facilitates the creation of a sequence of single-step optimization programs \cite{ames2014control, ames2016control, galloway2015torque, ames2019control}.
If the cost is quadratic, the optimizations are quadratic programs (QP), and the solutions can be deployed in real time \cite{ames2014control, nguyen20163d}. For safety constraints expressed using functions with high relative degree with respect to the dynamics of the system, exponential CBFs~\cite{nguyen2016exponential} and high-order CBFs (HOCBFs)~\cite{xiao2021high, tan2021high, usevitch2022adversarial} were proposed. Adaptive, robust, stochastic and feasibility-guaranteed versions of safety-critical control with CBFs were introduced in~\cite{xiao2021adaptive, liu2023auxiliary, Nguyen2015_RobustCLF, jankovic2018robust, clark2021control, liu2023feasibility}. Discrete-time CBFs (DCBFs) were introduced in \cite{agrawal2017discrete} as a means to enable safety-critical control for discrete-time systems. They were used in a nonlinear MPC (NMPC) framework to create NMPC-DCBF \cite{zeng2021safety}, wherein the DCBF constraint was enforced through a predictive horizon.
Generalized discrete-time CBFs (GCBFs) and discrete-time high-order CBFs (DHOCBFs) were proposed in \cite{ma2021feasibility} and \cite{xiong2022discrete} respectively, where the DCBF constraint only acted on the first time-step, i.e., a single-step constraint.
MPC with DCBF has been used in various fields, such as autonomous driving \cite{he2022autonomous} and legged robotics \cite{li2022bridging}. To improve the optimization feasibility, a decay-rate relaxing technique \cite{zeng2021decay} was introduced for NMPC with DCBF \cite{zeng2021enhancing} for every time-step to simultaneously boost safety and feasibility. However, the overall optimization remains nonlinear and non-convex, resulting in slow processing times for systems with large horizons and nonlinear dynamics (see \cite{zeng2021enhancing}, which is limited to relative-degree one).
In this paper, we extend the relaxation techniques introduced in \cite{zeng2021enhancing} to accommodate high relative-degree CBFs, and largely improve the computational time compared to all existing work.

\subsubsection{Learning-Based CBFs:}\hspace{0.1em}
Many CBF-based methods require explicit equations for the boundaries of the safety sets. When these are not available, they can be learned. In \cite{srinivasan2020synthesis}, the authors utilized a Support Vector Machine (SVM) to parameterize a CBF. They employed a supervised learning approach to classify regions of the state-space as either safe or unsafe. In \cite{saveriano2019learning}, the authors proposed a method that incrementally learns a linear CBF by grouping expert demonstrations into linear subspaces and creating low-dimensional representations through fitting. The authors of \cite{peng2023safe} proposed to construct polynomial CBFs to represent complex obstacles or unreachable regions using logistic regression. However, the above works do not provide theoretical guarantees of correctness of the learned CBFs. The work in \cite{robey2020learning} proposed and analyzed an optimization-based method for learning a CBF that provides provable safety guarantees given some mild assumptions. Nevertheless, it still requires that the equation of the boundary of the safe set is known. Recently, the authors of \cite{lavanakul2024safety} defined a learning-based discriminating hyperplane trained on trajectory data from a black-box system to eliminate the dependence on any specific equations for the safe sets. However, the trained hyperplane can only work for that system. We use a DNN to predict the nearest boundary point; the outward normal is then computed from this point, defining a local linear (tangent) boundary of the safe set that we embed as a DHOCBF. The run-time constraint is linear, but mapping raw geometry to the nearest point is nonlinear, which motivates using a DNN.

\subsubsection{Beyond CBF-Based Methods:}\hspace{0.1em}
Many works tackle complex or unknown obstacles—reachability-based safety filters \cite{fisac2019tac}, planning on reconstructed or occupancy maps \cite{wang2024risk}, and certified MPC \cite{wabersich2021tac}. They typically scale poorly beyond low dimensions or fixed maps, lack per-step safety certificates under the true dynamics and input bounds, or require costly recomputation or large polyhedral encodings. Although safe reinforcement learning \cite{achiam2017constrained}, temporal logic based planning \cite{belta2017formal}, and multi-agent safety methods \cite{qinlearning} have considered high-dimensional problems, these approaches generally rely on set-based abstractions or task-level specifications rather than explicit handling of arbitrary and irregular obstacle geometries. Hence, they are not directly applicable to environments with complex-shaped obstacles. In contrast, we achieve fast, convex per-step safety constraints via DHOCBFs in convex MPC with learned linear boundaries from complex-shaped obstacles—without global discretization of the environment.  

\subsection{CONTRIBUTIONS}
This paper introduces a novel approach to safety-critical MPC that is significantly faster than existing approaches and enhances safety within complex environments where unsafe set boundaries are irregular and poorly defined. The core of our methodology is a learning-based technique for predicting and linearizing the boundaries of unsafe sets. 
The key contributions of our work are as follows: 

\begin{itemize}
\item A control strategy that ensures safety by enforcing constraints derived from linearized DHOCBFs. We relax these constraints using bounded slack variables to improve feasibility while preserving safety. (Sec. \textnormal{IV-}\ref{subsec:linearization-dhocbf})
  
\item A model predictive control framework that integrates linearized system dynamics and DHOCBFs as constraints within a convex optimization problem. This problem is solved iteratively, achieving fast computation speeds ideal for real-time applications. (Sec. \ref{sec:iterative-opt})
    
\item A machine learning-based method to accurately predict the linearized boundaries of unsafe sets. The complement of these boundaries defines linearized approximations of safe sets, which are then incorporated into DHOCBF constraints. This is particularly effective in scenarios where unsafe set boundaries are complex, irregular, or otherwise difficult to model with analytical equations (e.g., unknown boundaries). (Sec. \textnormal{IV-C-}\ref{subsec: SBD unknown})
\end{itemize}

This work significantly extends our previous conference paper~\cite{liu2023iterative}, where we first introduced the idea of linearizing DHOCBFs under the assumption of known and simple unsafe set boundaries (e.g., circles). In this manuscript, we go beyond that by developing a general framework to learn linearized approximations of complex, unknown boundaries and integrating this framework into a convex finite-time optimal control scheme. Specifically, the learned linearized boundaries are embedded as DHOCBFs to ensure safety requirements. Furthermore, we provide the details of the technique based on slack variables within linearized DHOCBFs for improving both feasibility and safety — these details were omitted in~\cite{liu2023iterative}. Compared with \cite{liu2023iterative}, which considered only a unicycle model, this work additionally includes a higher-dimensional vehicle model to examine the scalability of the proposed framework, and validates it in simulations where both robots navigate narrow passages with convex and nonconvex obstacles, achieving markedly better performance in computational efficiency, safety, and feasibility than existing methods.

\begin{table}[]
\centering
\caption{List of acronyms.}
\begin{tabular}{|c|c|}
\hline
Acronyms & Meaning \\ \hline
 CBF &  Control Barrier Function\\ 
 DCBF & Discrete-Time Control Barrier Function \\
 DHOCBF & Discrete-Time High-Order Control Barrier Function \\
 MPC &  Model Predictive Control\\
 NMPC & Nonlinear Model Predictive Control \\
 iMPC & Iterative Model Predictive Control \\
 DNN &  Deep Neural Network\\
 SBD &  Safety Boundary Detector \\
CFTOC & Convex Finite-Time Optimal Control  \\
 MSE &  Mean Squared Error\\
\hline
\end{tabular}
\label{tab:acronyms}
\end{table}
\section{PRELIMINARIES}
\label{sec: Preliminaries}


In this work, safety is defined as forward invariance of a set $\mathcal C$, i.e., a system is said to be {\em safe} if it stays in $\mathcal C$ for all times, given that it is initialized in $\mathcal C$. We consider the set  $\mathcal C$ as the superlevel set of a function $h: \mathbb{R}^{n}\to \mathbb{R}$:
\begin{equation}
\label{eq:safe-set}
\mathcal{C} \coloneqq \{\mathbf{x} \in \mathbb{R}^{n}: h(\mathbf{x} )\geq0 \}.
\end{equation}
We consider a discrete-time control system in the form  
\begin{equation}
\label{eq:discrete-dynamics}
\mathbf{x}_{t+1}=f(\mathbf{x}_{t},\mathbf{u}_{t}),
\end{equation}
where $\mathbf{x}_{t} \in \mathcal X \subset \mathbb{R}^{n}$ represents the state of the system at time step $t\in\mathbb{N}, \mathbf{u}_{t}\in \mathcal U \subset \mathbb{R}^{q}$ is the control input, and the function $f: \mathbb{R}^{n}\times \mathbb{R}^{q}\to \mathbb{R}^{n}$ is locally Lipschitz.

\begin{definition}[Relative degree~\cite{sun2003initial}]
\label{def:relative-degree}
The output $y_{t}=h(\mathbf{x}_{t})$ of system \eqref{eq:discrete-dynamics} is said to have relative degree $m$ if
\begin{equation}
\begin{split}
&y_{t+i}=h(\bar{f}_{i-1}(f(\mathbf{x}_{t},\mathbf{u}_{t}))), \ i \in \{1,2,\dots,m\},\\
 \text{s.t.} & \ \frac{\partial y_{t+m}}{\partial \mathbf{u}_{t}} \ne \textbf{0}_{q}, \frac{\partial y_{t+i}}{\partial \mathbf{u}_{t}}= \textbf{0}_{q},  \ i \in \{1,2,\dots,m-1\},
\end{split}
\end{equation}
i.e., $m$ is the number of steps (delay) in the output $y_{t}$ in order for any component of the control input $\mathbf{u}_{t}$ to explicitly appear ($\textbf{0}_{q}$ is the zero vector of dimension $q$). 
\end{definition}

In the above definition, we use $\bar{f}(\mathbf{x})$ to denote the uncontrolled state dynamics $f(\mathbf{x}, 0)$. The subscript $i$ of function $\bar{f}(\cdot)$ denotes the $i$-times recursive compositions of $\bar{f}(\cdot)$, i.e.,  $\bar{f}_{i}(\mathbf{x})=\underset{i\text{-times}~~~~~~~~~~~~~}{\underbrace{\bar{f}(\bar{f}(\dots,\bar{f}}(\bar{f}_{0}(\mathbf{x}))))}$ with $\bar{f}_{0}(\mathbf{x})=\mathbf{x}$. Note that in discrete time the relative degree $m$ only means that the control input influences the safety function after $m$ steps; this look-ahead is not the same as a physical actuator delay.

We assume that $h(\mathbf{x})$ has
relative degree $m$ with respect to system (\ref{eq:discrete-dynamics}) based on Def. \ref{def:relative-degree}.
Starting with $\psi_{0}(\mathbf{x}_{t})\coloneqq h(\mathbf{x}_{t})$, we define a sequence of discrete-time functions $\psi_{i}:  \mathbb{R}^{n}\to\mathbb{R}$, $i=1,\dots,m$ as:
\begin{equation}
\label{eq:high-order-discrete-CBFs}
\psi_{i}(\mathbf{x}_{t})\coloneqq \bigtriangleup \psi_{i-1}(\mathbf{x}_{t})+\alpha_{i}(\psi_{i-1}(\mathbf{x}_{t})), 
\end{equation}
where $\bigtriangleup \psi_{i-1}(\mathbf{x}_{t})\coloneqq \psi_{i-1}(\mathbf{x}_{t+1})-\psi_{i-1}(\mathbf{x}_{t})$, and $\alpha_{i}(\cdot)$ denotes the $i^{th}$ class $\kappa$ function which satisfies $\alpha_{i}(\psi_{i-1}(\mathbf{x}_{t}))\le \psi_{i-1}(\mathbf{x}_{t})$ for $i=1,\ldots, m$.
A sequence of sets $\mathcal {C}_{i}$ is defined based on \eqref{eq:high-order-discrete-CBFs} as
\begin{equation}
\label{eq:high-order-safety-sets}
\mathcal {C}_{i}\coloneqq \{\mathbf{x}\in \mathbb{R}^{n}:\psi_{i}(\mathbf{x})\ge 0\}, \ i \in\{0,\ldots,m-1\}.
\end{equation}
Imposing $\psi_i(\mathbf x_t)\ge0$ at time $t$ enforces safety $i$ steps ahead, since Eq.~\eqref{eq:high-order-discrete-CBFs} sequentially ties $\psi_i$ to $\mathbf x_{t},\dots,\mathbf x_{t+i}$.

\begin{definition}[DHOCBF~\cite{xiong2022discrete}]
\label{def:high-order-discrete-CBFs}
Let $\psi_{i}(\mathbf{x}), \ i\in \{1,\dots,m\}$ be defined by \eqref{eq:high-order-discrete-CBFs} and $\mathcal {C}_{i},\ i\in \{0,\dots,m-1\}$ be defined by \eqref{eq:high-order-safety-sets}. A function $h:\mathbb{R}^{n}\to\mathbb{R}$ is a Discrete-time High-Order Control Barrier Function (DHOCBF) with relative degree $m$ for system \eqref{eq:discrete-dynamics} if there exist $\psi_{m}(\mathbf{x})$ and $\mathcal {C}_{i}$ such that
\begin{equation}
\label{eq:highest-order-CBF}
\psi_{m}(\mathbf{x}_{t})\ge 0, \ \forall x_{t}\in \mathcal{C}_{0}\cap \dots \cap \mathcal {C}_{m-1}, t\in\mathbb{N}.
\end{equation}
\end{definition}

\begin{theorem}[Safety Guarantee \cite{xiong2022discrete}]
\label{thm:forward-invariance}
Given a DHOCBF $h(\mathbf{x})$ from Def. \ref{def:high-order-discrete-CBFs} with corresponding sets $\mathcal{C}_{0}, \dots,\mathcal {C}_{m-1}$ defined by \eqref{eq:high-order-safety-sets}, if $\mathbf{x}_{0} \in \mathcal {C}_{0}\cap \dots \cap \mathcal {C}_{m-1},$ then any Lipschitz controller $\mathbf{u}_{t}$ that satisfies the constraint in \eqref{eq:highest-order-CBF}, $\forall t\ge 0$ renders $\mathcal {C}_{0}\cap \dots \cap \mathcal {C}_{m-1}$ forward invariant for system \eqref{eq:discrete-dynamics}, $i.e., \mathbf{x}_{t} \in \mathcal {C}_{0}\cap \dots \cap \mathcal {C}_{m-1}, \forall t\ge 0.$
\end{theorem}

\begin{remark}
\label{rem:sufficient-condition}
The function $\psi_{i}(\mathbf{x})$ in \eqref{eq:high-order-discrete-CBFs} is called an $i^{th}$ order Discrete-time CBF (DCBF). If the constraints in an optimal control problem include only DCBF constraints ($\psi_{i}(\mathbf{x}_{t})\ge0$), we must formulate the DCBF constraints up to the $m^{th}$ order to ensure that the control input $\mathbf{u}_{t}$ is explicitly represented based on Def. \ref{def:relative-degree}. However, if the constraints also include the system constraint \eqref{eq:discrete-dynamics}, where the control input $\mathbf{u}_{t}$ is already explicitly represented, we can flexibly select a suitable order for the DCBF constraints to minimize computational demands. In other words, the highest order for DCBF could be $m_{\text{cbf}}$ with $m_{\text{cbf}}\le m$. We can simply define an $i^{th}$ order DCBF $\psi_{i}(\mathbf{x})$ in \eqref{eq:high-order-discrete-CBFs} as
\begin{equation}
\label{eq:simple-high-order-discrete-CBFs}
\psi_{i}(\mathbf{x}_{t})\coloneqq \bigtriangleup \psi_{i-1}(\mathbf{x}_{t})+\gamma_{i}\psi_{i-1}(\mathbf{x}_{t}),
\end{equation}
where $\alpha(\cdot)$ is linear and $0<\gamma_{i}\le 1, i\in \{1,\dots,m_{\text{cbf}}\}$.
\end{remark}

The expression presented in \eqref{eq:simple-high-order-discrete-CBFs} adheres to the structure of the first-order DCBF introduced in \cite{agrawal2017discrete} and can be applied to define a DHOCBF with any relative degree for the classical constrained, safety-critical optimal control problem: 
\begin{subequations}
\label{eq:classical ocp}
\begin{align}
J(\mathbf{u}_{t},\mathbf{x}_{k})&=\min_{\mathbf{u}_{t}} \mathbf{u}_{t}^{T}\mathbf{u}_{t}+\sum_{k=t}^{t+1}(\mathbf{x}_{k}-\mathbf{x}_{e})^{T}(\mathbf{x}_{k}-\mathbf{x}_{e})\label{subeq:obj}\\
s.t. \ \ &\mathbf{x}_{t+1}=f(\mathbf{x}_{t},\mathbf{u}_{t}),\label{subeq:system1}\\
&\psi_{i-1}(\mathbf{x}_{t+1}) \ge (1-\gamma_{i})\psi_{i-1}(\mathbf{x}_{t}), ~0<\gamma_{i}\le1,\label{subeq:dcbf1}\\
&\mathbf{u}_{t}\in \mathcal U \subset \mathbb{R}^{q},~\mathbf{x}_{k} \in \mathcal X \subset \mathbb{R}^{n}\label{subeq:control and state limits},
\end{align}
\end{subequations}
where \eqref{subeq:obj} defines the objective
function of the optimization problem, $\mathbf{x}_{e}$ denotes a reference state, $0\le t\le T, t \in \mathbb{N}$ and the function $\psi_{i-1},i\in \{1,\dots,m_{\text{cbf}}\}$ in \eqref{subeq:dcbf1} ensures that the state $\mathbf{x}_{k}$ of system \eqref{subeq:system1} stays within a safe set $\mathcal{C}$ according to \eqref{eq:safe-set}. The constraint \eqref{subeq:control and state limits} bounds the control input and state, which may conflict with constraints \eqref{subeq:system1} and \eqref{subeq:dcbf1}, leading to infeasibility. Note that solving the problem described above can only yield the current optimal control input, resulting in a greedy control policy that considers short-term (one-step) safety and may overlook safer solutions. As discussed in \cite{zeng2021safety}, there exists a fundamental trade-off between safety and feasibility in control design, and enhancing both simultaneously is challenging. In \cite{zeng2021safety}, the authors show that MPC with DCBFs can give a safer control policy, as it takes future state and control information into account. A version of nonlinear MPC that incorporates DCBFs (called NMPC-DCBF) with a relaxation technique that simultaneously enhances feasibility and safety was developed in \cite{zeng2021enhancing}. This approach, although it can flexibly incorporate the system's physical constraints and safety requirements as soft or hard constraints into the control strategy, often leads to non-convex optimizations, resulting in inherently high computational complexity. In this paper, we show how to linearize the system and DCBF constraints to obtain a convex finite-time optimal control (CFTOC) framework within an MPC setting, which significantly improves computational efficiency.

\section{PROBLEM FORMULATION AND APPROACH}
\label{sec:Problem Formulation and Approach}
Our objective is to find a closed-loop control strategy for system \eqref{eq:discrete-dynamics} over a time interval $[0,T]$ that minimizes the deviation from a reference state and satisfies safety requirements and constraints on states and control inputs. 

\textbf{Safety Requirement:} System \eqref{eq:discrete-dynamics} should always satisfy a safety requirement of the form: 
\begin{equation}
\label{eq:Safety constraint}
h(\mathbf{x}_{t})\ge 0, ~\mathbf{x}_{t} \in \mathbb{R}^{n}, ~0\le t \le T,
\end{equation}
where $h:\mathbb{R}^{n}\to\mathbb{R}$. 

\textbf{Control and State Limitation Requirements:} The controller $\mathbf{u}_{t}$ and state $\mathbf{x}_{t}$ must satisfy \eqref{subeq:control and state limits} for $0\le t \le T.$

\textbf{Objective:} We consider the following cost:  
\begin{equation}
\label{eq:cost-function-1}
\begin{split}
J(\mathbf{u}_{t,k},\mathbf{x}_{t,k})=\sum_{k=0}^{N-1}\mathbf{u}_{t,k}^{T}\mathbf{u}_{t,k}+\sum_{k=0}^{N}(\mathbf{x}_{t,k}-\mathbf{x}_{e})^{T}(\mathbf{x}_{t,k}-\mathbf{x}_{e}),
\end{split}
\end{equation}
over a receding horizon $N<T$, where $\mathbf{x}_{t, k}$, $\mathbf{u}_{t, k}$ are the state and input predictions (according to the system's dynamics) at time $t+k$ made at the current time $t$, $0\le t \le T$. We denote $\mathbf{x}_{t, 0}=\mathbf{x}_{t}$, $\mathbf{u}_{t, 0}=\mathbf{u}_{t}$ and the reference state as $\mathbf{x}_{e}$.
A control policy is {\em feasible} if all the constraints guaranteeing the aforementioned requirements are mutually non-conflicting for all $0\le t \le T$. In this paper, we consider the following problem:

\begin{problem}
\label{prob:Path-prob}
Find a feasible control policy for system \eqref{eq:discrete-dynamics} such that the safety requirement, control and state limitations are satisfied, and the cost \eqref{eq:cost-function-1} is minimized. The safety requirement is defined with respect to unsafe sets whose boundaries may be either explicitly \textbf{known} (e.g., analytic functions) or \textbf{unknown} and only available through samples. 
\end{problem}

To satisfy the safety requirement, the authors of \cite{zeng2021enhancing} incorporated the DCBF constraint
\begin{equation}
\label{subeq:mpc-cbf-cons2}
\begin{split}
h(\mathbf{x}_{t, k+1}) \ge \omega_{t, k}(1-\gamma)h(\mathbf{x}_{t, k}),~ 0<\gamma\le1
\end{split}
\end{equation}
into MPC with a relaxation variable $\omega_{t, k}\in \mathbb{R}$, which simultaneously enhances feasibility and safety. They also formulated \eqref{eq:discrete-dynamics} and \eqref{subeq:control and state limits} as constraints in the MPC to meet additional requirements, which results in the entire optimization problem being nonconvex, as \eqref{subeq:mpc-cbf-cons2} and \eqref{eq:discrete-dynamics} are often nonlinear. This method, referred to as NMPC-DCBF, can lead to slow processing times for systems with large horizons. Another issue is that, when the boundary of the unsafe set is difficult to express with explicit equations, it becomes challenging to obtain a DCBF candidate $h(\mathbf{x})$.
Some approaches \cite{zeng2021safety,thirugnanam2022duality} exist for the case when the boundaries of the unsafe sets are circular or polytopic. In many real-world situations, this assumption is restrictive.          Meanwhile, using a conservative known surrogate (e.g., polygon or circle) to replace the ground-truth obstacle set yields a smaller safe set (see e.g., \cite{thirugnanam2022duality}, where the authors used NMPC to ensure safe path following for a unicycle robot). However, due to the presence of numerous polytopic obstacles in the map, defining the safe region using linear equations corresponding to the edges of the polytopes results in a very small safe region. This limited the NMPC to handling only nearby obstacles, making distant ones difficult to address.

\textbf{Approach:} In all tested scenarios, the proposed iMPC-DHOCBF achieved a remarkable online computation speedup of tens to hundreds of times over the NMPC baseline, while consistently preserving safety and maintaining high feasibility rates under tight input bounds. These results highlight that our approach not only accelerates online optimization but also ensures consistent safety performance. The key to these gains lies in transforming the original nonconvex safety constraints into efficiently solvable convex forms via a Safety Boundary Detector (SBD). This SBD obtains linearized unsafe set boundaries either directly, when they are simple and \textbf{known}, or via a trained deep neural network, when they are complex or \textbf{unknown}. Each linearized boundary equation serves as a DHOCBF candidate 
$h(\mathbf{x})$  for constructing its corresponding constraint. Since each unsafe set corresponds to a single DHOCBF constraint—unlike in \cite{thirugnanam2022duality}, where multiple DCBF constraints are used per unsafe set—the resulting safe set, defined as the intersection of all DHOCBF constraints, achieves broader coverage with fewer constraints. We integrate these constraints into an MPC framework that also incorporates state and input bounds, relaxing each DHOCBF constraint with a bounded slack variable to improve feasibility while still guaranteeing safety. The system dynamics constraint is additionally linearized, yielding a fast-computing convex optimization problem designed to be solved iteratively, with decision variable errors reduced after a certain number of iterations.

\section{ITERATIVE CONVEX MPC WITH DHOCBF}
\label{sec:iterative-opt}
In this section, we present an iterative convex MPC for general DHOCBFs defined in Sec. \ref{sec: Preliminaries}.
\begin{figure*}
    \centering
    \includegraphics[scale=0.36]{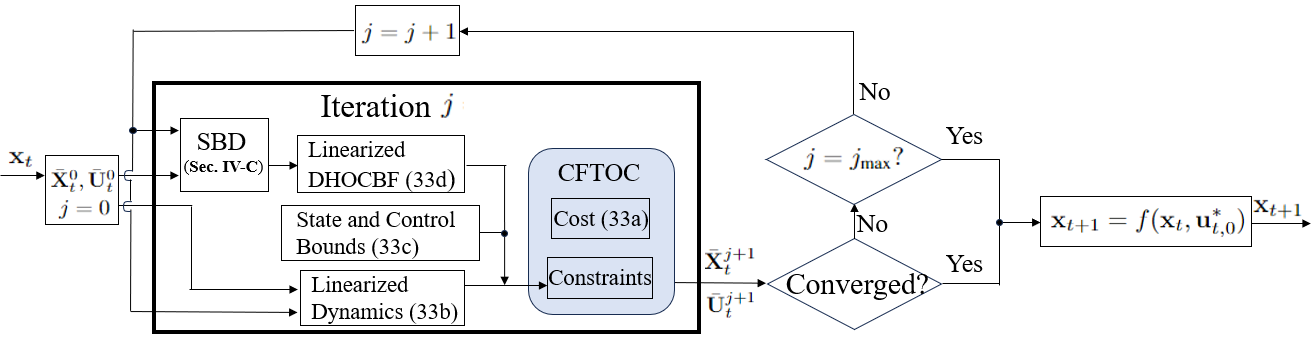}
    \caption{Schematic of the iterative process for solving convex MPC at time \( t \). The safety boundary detector (SBD) generates linearized boundaries, which are converted into DHOCBF constraints. At each iteration, a convex finite-time optimal control (CFTOC) problem with linearized dynamics and quadratic cost is solved. The problem remains convex due to linear constraints and a positive semidefinite weight matrix. A convergence check determines whether to iterate or update the state \( \mathbf{x}_{t+1} \).}
    \label{fig:iteration-module}
\end{figure*}

\subsection{ITERATIVE CONVEX MPC}
\label{subsec:iterative-convex-optimization}

Our proposed approach, which is depicted in Fig. \ref{fig:iteration-module}, involves iterative optimizations at each time step $t$, which is denoted as \emph{iterative} MPC-DHOCBF (iMPC-DHOCBF).  
Our iterative optimization problem contains three parts for each iteration $j$: (1) solving a CFTOC problem with linearized dynamics and DHOCBF to get optimal values of states and inputs $\mathbf{X}_{t}^{*, j}=[\mathbf{x}_{t, 0}^{*, j},\dots, \mathbf{x}_{t, N}^{*, j}], \mathbf{U}_{t}^{*, j}=[\mathbf{u}_{t,0}^{*, j},\dots, \mathbf{u}_{t,N-1}^{*, j}],$ (2) checking convergence, (3) updating the state and input vectors for next iteration, i.e., $\bar{\mathbf{X}}_{t}^{j+1} = \mathbf{X}_{t}^{*, j}, \bar{\mathbf{U}}_{t}^{j+1} = \mathbf{U}_{t}^{*, j}.$
Notice that the open-loop trajectory with updated states $\bar{\mathbf{X}}_{t}^{j} = [\bar{\mathbf{x}}_{t, 0}^{j},\dots, \bar{\mathbf{x}}_{t, N-1}^{j}]$ and inputs $\bar{\mathbf{U}}_{t}^{j} = [\bar{\mathbf{u}}_{t, 0}^{j},\dots, \bar{\mathbf{u}}_{t, N-1}^{j}]$ is passed between iterations, which allows iterative linearization for both system dynamics and DHOCBF locally.

The iteration is finished when the convergence error function $e(\mathbf{X}_{t}^{*, j}, \mathbf{U}_{t}^{*, j}, \bar{\mathbf{X}}_{t}^{j}, \bar{\mathbf{U}}_{t}^{j})$ is within a user-defined normalized convergence criterion.
To restrict the number of iterations, we limit $j < j_{\text{max}}$, where $j_{\text{max}}$ denotes the maximum numbers of iterations.
Therefore, the iterative optimization process halts upon reaching a local minimum for the cost function, meeting the convergence criterion, or when the iteration count equals $j_{\text{max}}$.
The optimized states $\mathbf{X}_{t}^{*}$ and inputs $\mathbf{U}_{t}^{*}$ from the last iteration are passed to the iMPC-DHOCBF for the next time instant as $\bar{\mathbf{U}}_{t+1}^{0},\bar{\mathbf{X}}_{t+1}^{0}$. 
At each time step, we record the updated states $\mathbf{x}_{t+1}$ propagated by the system dynamics $\mathbf{x}_{t+1}=f(\mathbf{x}_{t},\mathbf{u}_{t,0}^{*})$, enabling us to extract a closed-loop trajectory. Note that if we have the boundaries of unsafe sets, the SBD can directly linearize safe boundaries by finding the point nearest to the robot's current location and drawing a tangent surface at that point. If we do not have the boundaries of unsafe sets, SBD relies on a DNN that is trained before the start of the iMPC. The advantage of this approach is that we have pre-trained a DNN which can learn in advance information about all unsafe sets in the environment, especially when the boundaries of the unsafe sets cannot be accurately identified and described. Consequently, the time required to perform inference with this DNN for obtaining the linearized DHOCBF is essentially fixed in each iteration and can gradually decrease as adjustments are made to the DNN, ensuring the efficiency of solving each CFTOC.

\subsection{LINEARIZATION OF DYNAMICS}
\label{subsec:linearization-dynamics}
At iteration $j$, a control vector $\mathbf{u}_{t,k}^{j}$ is obtained by linearizing the system around $\bar{\mathbf{x}}_{t,k}^{j}, \bar{\mathbf{u}}_{t,k}^{j}$:
\begin{equation}
\begin{split}
\label{eq:linearized-dynamics}
\mathbf{x}_{t,k+1}^{j}{-}\bar{\mathbf{x}}_{t,k+1}^{j}{=}A_{t,k}^{j}(\mathbf{x}_{t,k}^{j}{-}\bar{\mathbf{x}}_{t,k}^{j})+B_{t,k}^{j}(\mathbf{u}_{t,k}^{j}{-}\bar{\mathbf{u}}_{t,k}^{j})+c_{t,k}^{j},
\end{split}
\end{equation}
where $c_{t,k}^{j}=f(\bar{\mathbf{x}}_{t,k}^{j}, \bar{\mathbf{u}}_{t,k}^{j})-\bar{\mathbf{x}}_{t,k+1}^{j}$, $0 \leq j < j_{\text{max}}$; $k$ and $j$ represent open-loop time step and iteration indices, respectively. We also have
\begin{equation}
A_{t,k}^{j}=D_{\mathbf{x}}f(\bar{\mathbf{x}}_{t,k}^{j}, \bar{\mathbf{u}}_{t,k}^{j}), \ B_{t,k}^{j}=D_{\mathbf{u}}f(\bar{\mathbf{x}}_{t,k}^{j}, \bar{\mathbf{u}}_{t,k}^{j}),
\end{equation}
where $D_{\mathbf{x}}$ and $D_{\mathbf{u}}$ denote the Jacobian of the system dynamics $f(\mathbf{x}, \mathbf{u})$ with respect to the state $\mathbf{x}$ and the input $\mathbf{u}$.
This approach allows us to linearize the system at $(\bar{\mathbf{x}}_{t,k}^{j}, \bar{\mathbf{u}}_{t,k}^{j})$ locally between iterations. The convex system dynamics constraints are provided in \eqref{eq:linearized-dynamics} since all the nominal vectors $(\bar{\mathbf{x}}_{t,k}^{j}, \bar{\mathbf{u}}_{t,k}^{j})$ at the current iteration are constant and constructed from the previous iteration $j-1$.
 \begin{remark}[\textit{Warm Start}]
 \label{rem: warm start}
We linearize the safe set constraint by projecting the nominal prediction to the nearest boundary and enforcing the outward-normal half-space, which yields an approximation of the safe set. This approximation holds only if the nominal $N$-step rollout remains outside obstacles. In this case, we use a cheap safe warm start: for $t>0$ we shift the previous optimal input sequence and pad the tail; for $t=0$ we select from sampled admissible inputs whose nominal rollout ($\bar{\mathbf{X}}_{0}^{0}$) remains outside obstacles over the $N$-step horizon. This improves feasibility, while the DHOCBF constraints enforce closed-loop safety at each step.
\end{remark}

\subsection{LINEARIZED SBD-GENERATED DCBF AND DHOCBF}
\label{subsec:linearization-dhocbf}
\subsubsection{Safety Boundary Detector for \textbf{Known} Unsafe Boundaries:}
\begin{figure}
    \centering
\includegraphics[width=\linewidth]{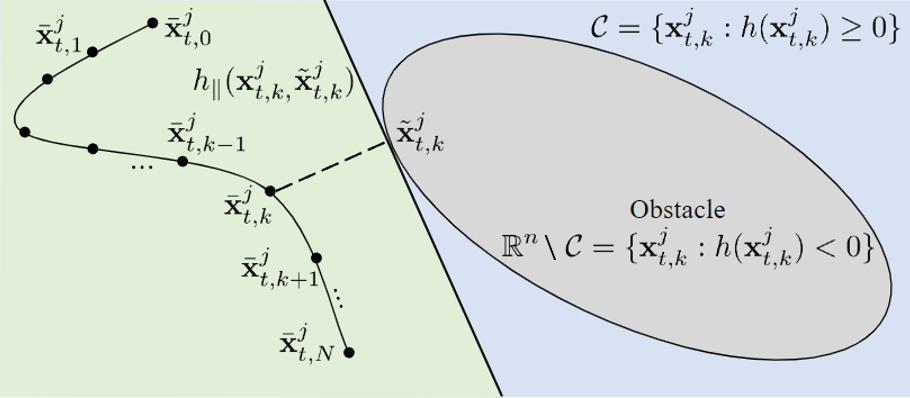}
    \caption{Linearization of DHOCBF: $h_{\parallel}(\mathbf{x}_{t, k}^{j},\tilde{\mathbf{x}}_{t, k}^{j})\ge 0$ represents the linearized safe set locally and is colored in green. Note that $h_{\parallel}(\mathbf{x}_{t, k}^{j},\tilde{\mathbf{x}}_{t, k}^{j})\ge 0$ guarantees $h(\mathbf{x}_{t,k}^{j}) \ge 0$ (colored in blue plus green), which ensures collision avoidance (outside the grey region).}
    \label{fig:linearization-dhocbf}
\end{figure}

\hspace{0.1em}
\label{subsec: SBD known}
In this section, we show how to linearize the DCBF up to the highest order with known unsafe boundary $h(\mathbf{x})=0.$ At iteration $j$ and time step $t+k,$ as shown in Fig. \ref{fig:linearization-dhocbf}, in order to linearize $h(\mathbf{x}_{t,k}^{j})$, an explicit dashed line is projected in the state space to the nearest point $\tilde{\mathbf{x}}_{t, k}^{j}$ on the boundary of the unsafe set from each state $\bar{\mathbf{x}}_{t, k}^{j}$. Note that $\bar{\mathbf{x}}_{t, k}^{j}$ is the nominal state vector from iteration $j-1$ for the linearization at iteration $j$, which means $\bar{\mathbf{x}}_{t, k}^{j}=\mathbf{x}_{t,k}^{j-1}$. The hyperplane (solid line in the 2D illustration) perpendicular to the dashed line and passing through the nearest point $\tilde{\mathbf{x}}^j_{t,k}$ is denoted as $h_{\parallel}(\mathbf{x}^j_{t,k}, \tilde{\mathbf{x}}^j_{t,k})$.  
This allows us to incorporate the above process into the SBD to define a linearized safe set by $h_{\parallel}(\mathbf{x}_{t, k}^{j}, \tilde{\mathbf{x}}_{t, k}^{j})\ge 0$, $\forall t \in \mathbb{N}$ by the green region. Note that $\tilde{\mathbf{x}}_{t, k}^{j}$ generally represents the optimized value of the minimum distance problem with distance function $h(\cdot)$ between $\bar{\mathbf{x}}_{t, k}^{j}$ and safe set $\mathcal{C}$. Since $h(\cdot)$ is known and continuous, for common shapes of unsafe sets, the expression of $\tilde{\mathbf{x}}_{t, k}^{j}$ as a function of $\bar{\mathbf{x}}_{t, k}^{j}$ is explicit.
For example, when $h(\cdot)$ describes a $l_2$-norm function with the unsafe set being a circular shape, $\tilde{\mathbf{x}}_{t, k}^{j}$ is exactly the intersection point of the line determined by $\bar{\mathbf{x}}_{t, k}^{j}$ and the center of the circular region with the circumference.
The relative degree of $h_{\parallel}(\mathbf{x}_{t,k}^{j},\tilde{\mathbf{x}}_{t, k}^{j})$ with respect to system \eqref{eq:discrete-dynamics} is still $m$ when the relative degree of $h(\mathbf{x}_{t,k}^{j})$ is $m$.
Thus, in order to guarantee safety with forward invariance based on Thm.~\ref{thm:forward-invariance} and Rem.~\ref{rem:sufficient-condition}, two sufficient conditions need to be satisfied: (1) the sequence of linearized DHOCBF $\tilde{\psi}_{0}(\cdot),\dots, \tilde{\psi}_{m_{\text{cbf}}-1}(\cdot)$  is larger or equal to zero at the initial condition $\mathbf{x}_{t}$, and (2) the highest-order DCBF constraint $\tilde{\psi}_{m_{\text{cbf}}}(\mathbf{x}) \ge 0$ is always satisfied, where $\tilde{\psi}_{i}(\cdot)$ is defined as:
\begin{equation}
\label{eq:linearized-CBFs}
\begin{split}
 \tilde{\psi}_{0}(\mathbf{x}_{t,k}^{j}) \coloneqq & h_{\parallel}(\mathbf{x}_{t,k}^{j},\tilde{\mathbf{x}}_{t, k}^{j}) \\
 \tilde{\psi}_{i}(\mathbf{x}_{t,k}^{j}) \coloneqq & \tilde{\psi}_{i-1}(\mathbf{x}_{t,k+1}^{j}){-}\tilde{\psi}_{i-1}(\mathbf{x}_{t,k}^{j}){+}\gamma_{i}\tilde{\psi}_{i-1}(\mathbf{x}_{t,k}^{j}). 
 \end{split}
\end{equation}
Here, we have $0 <\gamma_{i} \le 1$, $i\in\{1,\dots,m_{\text{cbf}}\}$, and $m_{\text{cbf}} \le m$ (as in \eqref{eq:simple-high-order-discrete-CBFs}). From Rem.~\ref{rem:sufficient-condition}, it follows that that $m_{\text{cbf}}$ is not necessarily equal to $m$. A detailed discussion on this can be found in~\cite{ma2021feasibility, zeng2021enhancing}.

An important issue is {\em feasibility}.
It is possible that $\psi_{i}(\mathbf{x}_{t,k}^{0}) \ge 0$, $1\le i \le m_{\text{cbf}}-1$, with $k\in\{0,\dots,N\}$ is not satisifed since the linearized DHOCBF functions $\tilde{\psi}_0(\cdot),\dots, \tilde{\psi}_{m_{\text{cbf}}-1}(\cdot)$ are more conservative than the original forms $\psi_0(\cdot),\dots, \psi_{m_{\text{cbf}}-1}(\cdot)$.
This problem can occur when the horizon is too large, or the linearization is too conservative. 
In order to handle this issue, we introduce a slack variable $\omega_{t,k,i}^{j}$ with a corresponding decay rate $(1-\gamma_{i})$:
\begin{equation}
\label{eq:nonconvex-hocbf-constraint}
  \tilde{\psi}_{i-1}(\mathbf{x}_{t,k+1}^{j})\ge \omega_{t,k,i}^{j}(1-\gamma_{i}) \tilde{\psi}_{i-1}(\mathbf{x}_{t,k}^{j}),\ \omega_{t,k,i}^{j} \ge 0,
\end{equation}
where $i\in\{1,\dots,m_{\text{cbf}}\}$.
The slack variable $\omega_{t,k,i}^{j}$ is selected by minimizing a cost function term to satisfy DCBF constraints at initial condition at any time step \cite{zeng2021enhancing}.

Another challenge stemming from the linearization of the DCBF is that the constraints in \eqref{eq:nonconvex-hocbf-constraint} might become non-convex, e.g., if $i=k=1,$ equation \eqref{eq:nonconvex-hocbf-constraint} becomes a nonconvex inequality: 
\begin{equation}
\label{eq:nonconvex-hocbf-constraint-eg1}
  \tilde{\psi}_{0}(\mathbf{x}_{t,2}^{j})\ge \omega_{t,1,1}^{j}(1-\gamma_{1}) \tilde{\psi}_{0}(\mathbf{x}_{t,1}^{j}),\ \omega_{t,1,1}^{j} \ge 0,
\end{equation}
where both $\omega_{t,1,1}^{j}$ and $\mathbf{x}_{t,1}^{j}$ are optimization variables. It is important to note that $\tilde{\psi}_{0}(\mathbf{x}_{t,0}^{j})$ is always constant; therefore, we can position $\omega_{t,k,1}^{j}$ in front of $\tilde{\psi}_{0}(\mathbf{x}_{t,0}^{j})$ and relocate the other optimization variables to the opposite side of the inequalities as 
\begin{equation}
\label{eq:nonconvex-hocbf-constraint-eg2}
  \tilde{\psi}_{0}(\mathbf{x}_{t,2}^{j})\ge \omega_{t,1,1}^{j}(1-\gamma_{1})^{2} \tilde{\psi}_{0}(\mathbf{x}_{t,0}^{j}),\ \omega_{t,1,1}^{j} \ge 0,
\end{equation}
which is linear and convex. This motivates the convex formulation \eqref{eq:convex-hocbf-constraint-form}, which replaces \eqref{eq:nonconvex-hocbf-constraint} with convex constraints. The following constraints of $\tilde{\psi}_{i}(\mathbf{x}_{t,k+1}^{j})$ are stacked by order $i$ and, within each order, by prediction step $k$: first all $i{=}0$ constraints for $k=0,\ldots,k{-}1$, then $i{=}1$ for $k=0,\ldots,k{-}1$, and so on up to $i{=}m_{\text{cbf}}$.

\begin{equation}
\label{eq:convex-hocbf-constraint-form}
\begin{split}
  \tilde{\psi}_{0}(\mathbf{x}_{t,1}^{j})\ge \omega_{t,0,1}^{j}(1-\gamma_{1}) \tilde{\psi}_{0}&(\mathbf{x}_{t,0}^{j}),\ \omega_{t,0,1}^{j} \ge 0, \\
\tilde{\psi}_{0}(\mathbf{x}_{t,2}^{j})\ge \omega_{t,1,1}^{j} (1-\gamma_{1})^{2}& \tilde{\psi}_{0}(\mathbf{x}_{t,0}^{j}),\ \omega_{t,1,1}^{j} \ge 0,\\
&\vdots\\
\tilde{\psi}_{0}(\mathbf{x}_{t,k}^{j})\ge \omega_{t,k-1,1}^{j} (1-\gamma_{1})^{k} \tilde{\psi}_{0}&(\mathbf{x}_{t,0}^{j}),\ \omega_{t,k-1,1}^{j} \ge 0 ,\\
 \tilde{\psi}_{1}(\mathbf{x}_{t,1}^{j})+(\gamma_{2}-1)\tilde{\psi}_{0}(\mathbf{x}_{t,1}^{j})\ge &
 \omega_{t,0,2}^{j}(1-\gamma_{2})(\gamma_{1}-1)\\
 \tilde{\psi}_{0}(\mathbf{x}_{t,0}^{j}), \ \omega_{t,0,2}^{j} \le &\frac{(\gamma_{1}-1+\gamma_{2}-1)\tilde{\psi}_{0}(\mathbf{x}_{t,1}^{j})}{(1-\gamma_{2})(\gamma_{1}-1)\tilde{\psi}_{0}(\mathbf{x}_{t,0}^{j})},\\
 \tilde{\psi}_{1}(\mathbf{x}_{t,2}^{j})+(\gamma_{2}-1)\tilde{\psi}_{0}(\mathbf{x}_{t,2}^{j}) \ge &
 \omega_{t,1,2}^{j}(\gamma_{2}-1)(1-\gamma_{1})^{2} \\
 \tilde{\psi}_{0}(\mathbf{x}_{t,0}^{j}),\ \omega_{t,1,2}^{j} \le &\frac{(\gamma_{1}-1+\gamma_{2}-1)\tilde{\psi}_{0}(\mathbf{x}_{t,2}^{j})}{(\gamma_{2}-1)(1-\gamma_{1})^{2}\tilde{\psi}_{0}(\mathbf{x}_{t,0}^{j})},\\
 &\vdots\\
  \tilde{\psi}_{m_{\text{cbf}}-1}(\mathbf{x}_{t,k}^{j}) + \sum_{\nu =1}^{m_{\text{cbf}}}(\gamma_{m_{\text{cbf}}}-1)&Z_{\nu,m_{\text{cbf}}}(1-\gamma_{1})^{k-1}\tilde{\psi}_{0}(\mathbf{x}_{t,\nu}^{j}) \\
 \ge \quad \omega_{t,k,m_{\text{cbf}}}^{j}(1-\gamma_{m_{\text{cbf}}}) Z_{0,m_{\text{cbf}}}&(1-\gamma_{1})^{k-1}\tilde{\psi}_{0}(\mathbf{x}_{t,0}^{j}), \\
\end{split}
\end{equation}

where $\omega_{t,k,m_{\text{cbf}}}^{j}, Z_{\nu,m_{\text{cbf}}},Z_{0,m_{\text{cbf}}}$ can be found in \eqref{eq:convex-hocbf-constraint-2}, \eqref{eq:z-value}.
The above constraints can be summarized as 
\begin{equation}
\label{eq:convex-hocbf-constraint}
\begin{split}
 \tilde{\psi}_{i-1}(\mathbf{x}_{t,k}^{j})& + \sum_{\nu =1}^{i}(\gamma_{i}-1)Z_{\nu,i}(1-\gamma_{1})^{k-1}\tilde{\psi}_{0}(\mathbf{x}_{t,\nu}^{j})\ge \\
\omega_{t,k,i}^{j}&(1-\gamma_{i}) Z_{0,i}(1-\gamma_{1})^{k-1}\tilde{\psi}_{0}(\mathbf{x}_{t,0}^{j}), \\
\end{split}
\end{equation}
where 
\begin{small}
\begin{equation}
\label{eq:convex-hocbf-constraint-2}
\begin{split}
j \le j_{\text{max}} \in \mathbb{N}^{+}, \ i&\in \{1,\dots,m_{\text{cbf}}\},\ 1\le k\le N, \ \omega_{t,k,1}^{j}\ge 0,\\
\omega_{t,k,i}^{j}(1-\gamma_{i}) Z_{0,i}&(1-\gamma_{1})^{k-1}\tilde{\psi}_{0}(\mathbf{x}_{t,0}^{j}) \ge \sum_{\nu =0}^{i-2}Z_{\nu,i}(1-\gamma_{1})^{k-1}\\
\tilde{\psi}_{0}(\mathbf{x}_{t,\nu+1}^{j})+ 
\sum_{\nu =1}^{i}&(\gamma_{i}-1)Z_{\nu,i}(1-\gamma_{1})^{k-1}\tilde{\psi}_{0}(\mathbf{x}_{t,\nu}^{j})\ \text{if}\ i\ge 2.
\end{split}
\end{equation}
\end{small}

$Z_{\nu,i}$ is a constant that can be obtained recursively by reformulating $\tilde{\psi}_{i-1}(\cdot)$ back to $\tilde{\psi}_{0}(\cdot)$ given $\nu \in\{0,..,i\}$.
We define $Z_{\nu,i}$ as follows. When $2\le i, \nu \le i-2$, we have
\begin{equation}
\label{eq:z-value}
\begin{split}
Z_{\nu,i}=\sum_{l=1}^{l_{\text{max}}}[(\gamma_{\zeta_{1}}-1)(\gamma_{\zeta_{2}}-1)\cdots(\gamma_{\zeta_{i-\nu-1}}-1)]_{l},\\
\zeta_{1}<\zeta_{2}<\dots<\zeta_{i-\nu-1},\zeta_{s}\in\{1,2,\dots,i-1\}, 
\end{split}
\end{equation}
where $l_{\text{max}}=\binom{i-1}{i-\nu-1},$ $\zeta_{s}$ denotes $\zeta_{1},\zeta_{2},\dots,\zeta_{i-\nu-1}.$ $[\cdot]_{l}$ denotes the $l^{th}$ combination of the product of the elements in parenthesis, e.g., if $i=4, \nu=1, Z_{1,4}=(\gamma_{1}-1)(\gamma_{2}-1)+(\gamma_{1}-1)(\gamma_{3}-1)+(\gamma_{2}-1)(\gamma_{3}-1).$ 
For the case $\nu=i-1$, we define $Z_{\nu,i}=1$. Besides that, we define $Z_{\nu,i}=0$ for the case $\nu=i$ (\eqref{eq:nonconvex-hocbf-constraint-eg2} shows the case where $Z_{0,1}=1, Z_{1,1}=0$). 
With the linearized DHOCBF in place, we now connect the construction to closed-loop safety. Assuming that at each step there exists a control input satisfying \eqref{eq:convex-hocbf-constraint}
under \eqref{eq:convex-hocbf-constraint-2}, the theorem below establishes forward invariance of the (linearized) safe set.

\begin{theorem}
\label{thm:safety-feasibility}
Given a linearized DHOCBF $h_{\parallel}(\mathbf{x})$ for system \eqref{eq:discrete-dynamics} with corresponding functions $\tilde{\psi}_{0}(\mathbf{x}), \tilde{\psi}_{1}(\mathbf{x}) \dots,\tilde{\psi}_{m_{\text{cbf}}}(\mathbf{x})$ defined by \eqref{eq:linearized-CBFs}, if $\tilde{\psi}_{0}(\mathbf{x}_{0})\ge 0$, then any Lipschitz controller $\mathbf{u}_{t}$ that satisfies the constraints in \eqref{eq:convex-hocbf-constraint} given conditions in \eqref{eq:convex-hocbf-constraint-2}, $\forall t\ge 0$ renders $\tilde{\mathcal{C}}_{0}\coloneqq \{\mathbf{x}\in \mathbb{R}^{n}:\tilde{\psi}_{0}(\mathbf{x})\ge 0\}$ forward invariant for system \eqref{eq:discrete-dynamics}, $i.e., \mathbf{x}_{t} \in  \tilde{\mathcal{C}}_{0}, \forall t\ge 0$.
\end{theorem}
\begin{proof}
By rewriting \eqref{eq:linearized-CBFs}, if $i\ge 2, k\ge 1$, we obtain 
\begin{equation}
\label{eq:convex-hocbf-constraint-proof1}
\begin{split}
 \tilde{\psi}_{i}(\mathbf{x}_{t,k}^{j})=
\sum_{\nu =0}^{i+1}Z_{\nu,i+1}(1-\gamma_{1})^{k-1}\tilde{\psi}_{0}(\mathbf{x}_{t,\nu}^{j}).\\
\end{split}
\end{equation}
Then, we have 
\begin{equation}
\label{eq:convex-hocbf-constraint-proof2}
\begin{split}
 \tilde{\psi}_{i}(\mathbf{x}_{t,k}^{j})=\tilde{\psi}_{i-1}(\mathbf{x}_{t,k+1}^{j})+(\gamma_{i}-1)\tilde{\psi}_{i-1}(\mathbf{x}_{t,k}^{j})\ge 0\\
 \Longrightarrow \sum_{\nu =0}^{i}Z_{\nu,i}(1-\gamma_{1})^{k-1}\tilde{\psi}_{0}(\mathbf{x}_{t,\nu+1}^{j})+\\
 \sum_{\nu =0}^{i}(\gamma_{i}-1)Z_{\nu,i}(1-\gamma_{1})^{k-1}\tilde{\psi}_{0}(\mathbf{x}_{t,\nu}^{j})\ge 0 \\
 \Longrightarrow Z_{i-1,i}(1-\gamma_{1})^{k-1}\tilde{\psi}_{0}(\mathbf{x}_{t,i}^{j})+\sum_{\nu =0}^{i-2}Z_{\nu,i}(1-\gamma_{1})^{k-1}\\
 \tilde{\psi}_{0}(\mathbf{x}_{t,\nu+1}^{j})+\sum_{\nu =1}^{i}(\gamma_{i}-1)Z_{\nu,i}(1-\gamma_{1})^{k-1}\tilde{\psi}_{0}(\mathbf{x}_{t,\nu}^{j})\ge \\
(1-\gamma_{i}) Z_{0,i}(1-\gamma_{1})^{k-1}\tilde{\psi}_{0}(\mathbf{x}_{t,0}^{j}).
\end{split}
\end{equation}
Since $\tilde{\psi}_{0}(\mathbf{x}_{t,0}^{j})$ is a constant, we position the slack variable $\omega_{t,k,i}^{j}$ in front of it and equation \eqref{eq:convex-hocbf-constraint-proof2} becomes
\begin{equation}
\label{eq:convex-hocbf-constraint-proof3}
\begin{split}
 Z_{i-1,i}(1-\gamma_{1})^{k-1}\tilde{\psi}_{0}(\mathbf{x}_{t,i}^{j})\ge \omega_{t,k,i}^{j}(1-\gamma_{i}) Z_{0,i}(1-\gamma_{1})^{k-1}\\
 \tilde{\psi}_{0}(\mathbf{x}_{t,0}^{j})-
 \sum_{\nu =0}^{i-2}Z_{\nu,i}(1-\gamma_{1})^{k-1}
 \tilde{\psi}_{0}(\mathbf{x}_{t,\nu+1}^{j})-\\
 \sum_{\nu =1}^{i}(\gamma_{i}-1)Z_{\nu,i}(1-\gamma_{1})^{k-1}\tilde{\psi}_{0}(\mathbf{x}_{t,\nu}^{j}).\\
\end{split}
\end{equation}
Note that $\tilde{\psi}_{0}(\mathbf{x}_{t,i+k-1}^{j})\ge (1-\gamma_{1})^{k-1}\tilde{\psi}_{0}(\mathbf{x}_{t,i}^{j})$. If $i=1$, we need to ensure
\begin{equation}
\label{eq:convex-hocbf-constraint-proof4}
\begin{split}
\tilde{\psi}_{0}(\mathbf{x}_{t,k}^{j})\ge \omega_{t,k-1,1}^{j}& (1-\gamma_{1})^{k} \tilde{\psi}_{0}(\mathbf{x}_{t,0}^{j}).
\end{split}
\end{equation}
Equations \eqref{eq:convex-hocbf-constraint-proof3}, \eqref{eq:convex-hocbf-constraint-proof4} are also equivalent to \eqref{eq:convex-hocbf-constraint}. Since $Z_{i-1,i}=1$, in order to make $\tilde{\psi}_{0}(\mathbf{x}_{t,i}^{j})\ge 0$, we  need
\begin{equation}
\label{eq:convex-hocbf-constraint-proof5}
\begin{split}
\omega_{\scriptscriptstyle t,k,i}^{j} (1 - \gamma_i) Z_{\scriptscriptstyle 0,i} &(1 - \gamma_1)^{\scriptscriptstyle k-1}
\tilde{\psi}_0(\mathbf{x}_{\scriptscriptstyle t,0}^{j}) \ge
\sum_{\nu=0}^{i-2} Z_{\scriptscriptstyle \nu,i} (1 - \gamma_1)^{\scriptscriptstyle k-1}\\
 \tilde{\psi}_{0}(\mathbf{x}_{t,\nu+1}^{j})+
 \sum_{\nu =1}^{i}&(\gamma_{i}-1)Z_{\nu,i}(1-\gamma_{1})^{k-1}\tilde{\psi}_{0}(\mathbf{x}_{t,\nu}^{j})\ \text{if}\ i\ge 2,\\
 &\omega_{t,k-1,1}^{j} \ge 0\ \text{if}\ i= 1,
\end{split}
\end{equation}
which is the same as \eqref{eq:convex-hocbf-constraint-2}. This means that by satisfying constraints in \eqref{eq:convex-hocbf-constraint}, given the conditions in \eqref{eq:convex-hocbf-constraint-2}, $\forall t\ge 0$, we have $\tilde{\psi}_{0}(\mathbf{x}_{t,i}^{j})\ge 0, j \le j_{\text{max}} \in \mathbb{N}^{+}, \ i\in \{1,\dots,m_{\text{cbf}}\}, \forall t\ge 0$, thus $\tilde{\psi}_{0}(\mathbf{x}_{t})\ge 0$ is guaranteed.
\end{proof}

\begin{remark}
\label{rem: different-relax-techniques}
Note that if we position $\omega_{t,k,i}^{j}$ in front of $\sum_{\nu =1}^{i}(\gamma_{i}-1)Z_{\nu,i}(1-\gamma_{1})^{k-1}\tilde{\psi}_{0}(\mathbf{x}_{t,\nu}^{j})$ in \eqref{eq:convex-hocbf-constraint}, equation \eqref{eq:nonconvex-hocbf-constraint} is the same as \eqref{eq:convex-hocbf-constraint}. This illustrates that the decay rate in \eqref{eq:convex-hocbf-constraint} used by the iMPC-DHOCBF is partially relaxed compared to the one in \eqref{eq:nonconvex-hocbf-constraint} due to the requirement of the linearization. This can affect the feasibility of the optimization. Allowing $\omega_{t,k,i}^{j}\in\mathbb{R}$ (unbounded) can further enlarge feasibility, but it does not enforce $\tilde{\psi}_{0}(\mathbf{x})\ge 0$ and thus may violate safety. By constraining $\omega_{t,k,i}^{j}$ via (20), we guarantee $\tilde{\psi}_{0}(\mathbf{x})\ge 0$ (safety) while still enlarging feasibility relative to the no‑slack case; hence we adopt (20). Note, however, that a bounded slack only improves \textbf{per-step} feasibility and does not ensure \textbf{recursive feasibility}: without a control-invariant terminal set/tube, feasibility at time $t$ does not imply feasibility at $t{+}1$, since under bounded inputs and a finite horizon the next-step linearized constraints may conflict with the dynamics/bounds or demand a slack beyond its admissible range.
\end{remark}

\subsubsection{Safety Boundary Detector for \textbf{Unknown} Unsafe Boundaries:}\hspace{0.1em}
\label{subsec: SBD unknown}
Notice that $\tilde{\mathbf{x}}_{t, k}^{j}$ in Sec. \textnormal{IV-C-}\ref{subsec: SBD known} could be implicit if $h(\cdot)$ is unknown or discontinuous, or for unsafe sets of irregular shapes (like general ellipse),  but it could still be numerically approximated as the values of $\bar{\mathbf{x}}_{t, k}^{j}$ known at iteration $j$ before the linearization, i.e., if we could predict the nearest point $\tilde{\mathbf{x}}_{t, k}^{j}$ for the $i^{th}$ unsafe set. For a map, the linearized boundary of the unsafe set can be expressed as
\begin{equation}
\label{eq:linearized safe set}
\begin{split}
h_{\parallel}(\mathbf{x}_{t,k}^{j},\tilde{\mathbf{x}}_{t, k}^{j})=(\tilde{\mathbf{x}}_{t, k}^{j}-\bar{\mathbf{x}}_{t, k}^{j})^{T}(\mathbf{x}_{t,k}^{j}-\tilde{\mathbf{x}}_{t, k}^{j}),
\end{split}
\end{equation}
which can also be used as a linearized DHOCBF in Sec. \textnormal{IV-C-}\ref{subsec: SBD known}. Specifically, it involves finding the nearest point on the boundary of an unsafe set or an inaccessible region relative to the current position, and then constructing a tangent line or plane through that point to serve as the linearized unsafe boundary. Next we will show how to design a Deep Neural Network (DNN) to get the approximate nearest point $\tilde{\mathbf{x}}_{t, k}^{j}.$

The designed DNN learns a specific map to identify and return the nearest point on each unsafe set relative to the system. A line perpendicular to the vector connecting the robot's current location and this point defines the linearized boundary. During training, nearest points are collected as ground truth for every safe point, enabling the DNN to learn via supervised learning. To efficiently store this information, we focus on architectures with strong associative memory, as conventional DNNs excel at pattern recognition but lack memory capacity. Models such as Hopfield Networks, Neural Turing Machines, Autoencoders \cite{kramer1992autoassociative}, and Transformers are well-suited for this task.

\textbf{Model Selection:}
We adopt an Autoencoder framework for its ability to encode complex patterns, but use only the encoder to streamline the model and accelerate inference. Instead of reconstructing inputs, our focus is on efficient nearest-point retrieval. During training, the encoder compresses the map into a low-dimensional space by minimizing the loss between predicted and ground-truth nearest points. Once trained, it acts as a memory register, recalling nearest points efficiently. Omitting the decoder reduces computational overhead, and the encoder delivers consistent inference times regardless of map complexity (e.g., the number of unsafe sets).

\begin{figure}
    \centering
    \includegraphics[scale=0.23]{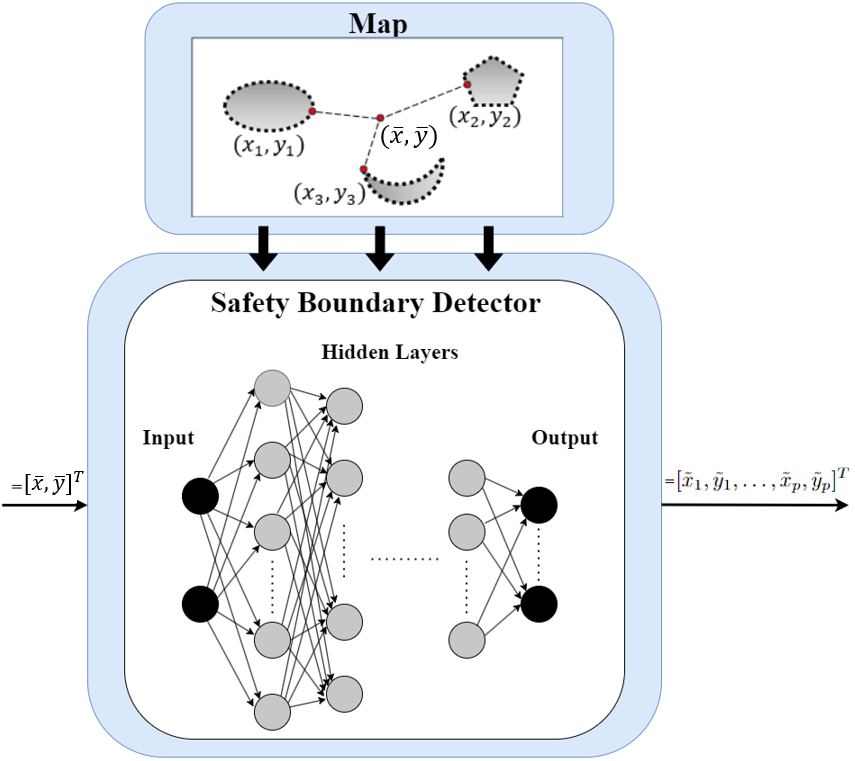}
    \caption{System identification for the Safety Boundary Detector (SBD), a DNN \( f_\theta(\bar{\mathbf{x}}) \) trained on map data. Given the robot’s location $(\bar{x}, \bar{y})$, the SBD outputs the nearest point ($\tilde x_i, \tilde y_i$) on each of the $p$ unsafe set boundaries, one per set.}
    \label{fig:neural-network}
\end{figure}

\textbf{Map Processing:}
We begin by identifying the boundaries of unsafe sets on a pixel-based canvas that simulates convex and non-convex shapes. Using image processing, we detect boundary pixels without requiring precise mathematical descriptions. To improve efficiency, we select a subset of boundary pixels $\partial \mathcal{X}^{i}_{\text{unsafe}}$ to form discrete contours representing $i^{\text{th}}$ unsafe set ($i\le p$). This method generalizes well across different maps and shapes. The processed boundaries are shown in Fig.~\ref{fig:neural-network}, where \(\bar{\mathbf{x}}= [\bar{x}, \bar{y}]^T \) denotes the system's current position, and \( [x_1, y_1, \dots, x_p, y_p]^T \) represent the nearest boundary points $\mathbf{x}_{1},\dots,\mathbf{x}_{p}$, computed via pixel-wise comparison: 
\begin{equation}
\label{eq:nearest points}
\mathbf{x}_i=\Pi_{\partial \mathcal{X}^{i}_{\text{unsafe}}}( \bar{\mathbf{x}}):= \arg\min_{z_i \in \partial \mathcal{X}^{i}_{\text{unsafe}}} \| \bar{\mathbf{x}} - z_i \|,
\end{equation}
where $z_i$ is a point on the boundary of the $i^{\text{th}}$ unsafe set, and $\mathbf{x}_i = \Pi_{\partial \mathcal{X}^{i}_{\text{unsafe}}}(\bar{\mathbf{x}})$ denotes the nearest boundary point to $\bar{\mathbf{x}}$. Both $\bar{\mathbf{x}}$ and $\mathbf{x}_i$ are shown as red dots.

\textbf{Network Design and Training:}
As described above, we use the encoder of an Autoencoder neural network. Figure \ref{fig:neural-network} shows its structure, where the number of nodes decreases layer by layer to extract essential map information. Each layer is fully connected to ensure comprehensive processing. The network takes the system's current location \(\bar{\mathbf{x}}= [\bar{x}, \bar{y}]^T \) as input and outputs predicted nearest points \( [\tilde{x}_1, \tilde{y}_1, \dots, \tilde{x}_p, \tilde{y}_p]^T \) on nearby unsafe sets. Since the number of unsafe sets varies across maps, the output size may change, requiring network re-training for adaptation. The first hidden layer is denoted as $h_{0}$. The subsequent layers’ dimensions are defined by:
\begin{equation}
\label{eq:node number}
h_{i}=h_{0}-i*floor((h_{0}-l_{out})/L),
\end{equation}
in which $h_{i}$ stands for the number of nodes in the $i^{th}$ of hidden layers, $l_{out}$ is the output layer size, and $L$ denotes the total number of hidden layers. The floor function rounds down a real number to the nearest integer less than or equal to that number. This model architecture, which uniformly reduces layer by layer, can adaptively adjust to the size of the output and ensure smoothness in dimension reduction.

We train a neural network \( f_\theta(\bar{\mathbf{x}}) \) with parameters \( \theta \) to approximate the nearest points by minimizing the mean squared error loss:
\begin{equation}
\label{eq:learning_loss}
\mathcal{L}(\theta)=\min_{\theta} \frac{1}{p} \sum_{i=1}^p \| f_\theta(\bar{\mathbf{x}}^{\text{train}}) - \mathbf{x}_i^{\text{train}} \|^2,
\end{equation}
where the system locations \(\bar{\mathbf{x}}^{\text{train}}\) are sampled across the map and their nearest boundary points $\mathbf{x}_{1}^{\text{train}},\dots,\mathbf{x}_{p}^{\text{train}}$ are identified by \eqref{eq:nearest points} as ground truth. If multiple nearest points exist (e.g., with nonconvex boundaries), one is selected arbitrarily. More generally, the learning objective can be viewed as minimizing the expected loss over the data distribution \( \mathcal{D} \):
\begin{equation}
\mathcal{L}(\theta)=\min_{\theta} \mathbb{E}_{\bar{\mathbf{x}}^{\text{train}} \sim \mathcal{D}} \left[ \| f_\theta(\bar{\mathbf{x}}^{\text{train}}) - \Pi_{\partial \mathcal{X}_{\text{unsafe}}^{i}}(\bar{\mathbf{x}}^{\text{train}}) \|^2 \right].
\end{equation}
To quantify generalization, we evaluate the performance of the trained model on unseen test data using:
\begin{equation}
\label{eq:MSE}
\text{MSE}_{\text{test}} = \frac{1}{p} \sum_{j=1}^{p} \| f_\theta(\bar{\mathbf{x}}^{\text{test}}) - \mathbf{x}_j^{\text{test}} \|^2.
\end{equation}
Note that training samples $\bar{\mathbf{x}}^{\text{train}}$ include system locations both inside and outside the unsafe sets, which improves the network’s generalization. In contrast, testing samples $\bar{\mathbf{x}}^{\text{test}}$ contain only locations outside the unsafe sets, to evaluate prediction accuracy.
Since the network's output $\tilde{\mathbf{x}}_{i}=[\tilde{x}_{i},\tilde{y}_{i}]^{T}=f_\theta(\bar{\mathbf{x}}),  i\in \{1,\dots,p\}$ could be used as the approximate nearest points, we can incorporate the above process into the SBD to define linearized safe sets from \eqref{eq:linearized safe set} by $h_{\parallel}(\mathbf{x}_{t, k}^{j},\tilde{\mathbf{x}}_{t, k}^{j})\ge 0$, $\forall t \in \mathbb{N}$ similar to Fig. \ref{fig:linearization-dhocbf}. The linearization of the DHOCBF $h_{\parallel}(\mathbf{x}_{t, k}^{j},\tilde{\mathbf{x}}_{t, k}^{j})$ up to the highest order will be the same as shown in Sec. \textnormal{IV-C-}\ref{subsec: SBD known}.

\begin{remark}[\textit{Continuity of $h_{\parallel}(\cdot)$}]
\label{rem: continuity}
Whether the equation of a linearized safe boundary \eqref{eq:linearized safe set} to a continuous boundary is continuous with respect to point $\mathbf{x}_{t,k}^{j}$ depends on the properties and smoothness of the curve. For smooth boundaries, such as circles or ellipses, the nearest point $\tilde{\mathbf{x}}_{t, k}^{j}$ to an external point $\mathbf{x}_{t,k}^{j}$ can typically be calculated, and this point changes continuously as $\mathbf{x}_{t,k}^{j}$ moves. In these cases, \eqref{eq:linearized safe set} is usually continuous as well. However, for shapes with sharp corners or discontinuities, like stars, the location of the nearest point may jump abruptly with minor movements of $\mathbf{x}_{t,k}^{j}$ (jumping from one vertex to another or from an edge to a vertex), leading to discontinuities in \eqref{eq:linearized safe set} at these points. The pixel-based boundary approximation introduced in Sec. \textnormal{IV-C-}\ref{subsec: SBD unknown} will also bring discontinuity. Discontinuity affects the optimality of the obtained controller. However, pixel-based discontinuity could be ameliorated by increasing the number of pixels (minimizing the distance between adjacent pixels). 
\end{remark}
\begin{remark}[\textit{Motivation for Using Neural Networks as Nearest-Point Detectors}]
\label{rem: motivation of DNN}
While traditional methods such as Euclidean distance functions (EDF \cite{oleynikova2017voxblox}) and spatial tree structures (e.g., k-d trees \cite{muja2009fast}) are widely used for nearest-boundary queries, our motivation for adopting a DNN-based approach is three-fold. First, it is memory-efficient, avoiding the need to store dense distance maps or indexing structures, whose memory usage scales with map size and resolution. In contrast, the trained DNN serves as a compact function approximator with fixed memory. Second, when boundary pixels are sparse, EDF and spatial-tree methods rely on limited samples, potentially leading to suboptimal nearest-point queries. Our DNN can interpolate from sparse data to predict locally optimal points, mitigating discretization issues. Third, under noisy observations, traditional methods may fail due to reliance on disturbed pixels, while the DNN can learn to filter outliers and predict smooth, reliable points by leveraging spatial patterns in training. Its accuracy can further improve through model and hyperparameter tuning.
\end{remark}
\begin{remark}[\textit{The Convexity of Unsafe Sets}]
\label{rem: convexity}
Let $O\subset\mathbb{R}^n$ be the unsafe set and $C=\mathbb{R}^n\!\setminus\!O$ the safe set.
At the current location $\bar{\mathbf{x}}$, the SBD returns the nearest boundary point $\tilde {\mathbf{x}}\in\partial O$. We form the local linear constraint
$H(\bar{\mathbf{x}}):=\{\mathbf{x}:\,h_{\parallel}\ge 0\}
$ where $h_{\parallel}$ is defined by \eqref{eq:linearized safe set}.
If $O$ is convex, the hyperplane at $\tilde x$ is supporting and $H(\bar{\mathbf{x}})\subset C$, so enforcing $h_{\parallel}\!\ge\!0$ implies the original safety constraint along the horizon (as shown in Fig. \ref{fig:linearization-dhocbf}).
If $O$ is nonconvex, $H(\bar{\mathbf{x}})$ is a \textbf{local} approximation of $C$ near the boundary component containing $\tilde{\mathbf{x}}$. In either case, the closed‑loop safety can be strengthened by re‑linearizing at every iMPC iteration, using a small discretization step $\Delta t$ to limit inter‑step drift, and (optionally) intersecting a few tangents, $\bigcap_{m=1}^{l} H^{(m)}(\bar{\mathbf{x}})$, i.e., having the SBD return the $l$ nearest boundary points $\tilde{\mathbf{x}}^{\,j,(m)}_{t,k}$ to $\bar{\mathbf{x}}$ for each obstacle and forming multiple tangents to obtain a tighter local approximation of $C$, or adding a margin $h_{\parallel}\ge \varepsilon>0$.
Because the SBD is agnostic to convexity and supplies nearest boundary points for arbitrary shapes, the procedure applies to complex unsafe sets.
\end{remark}
\begin{remark}[\textit{DNN Prediction Errors and Certified Bounds}]
Note that our DNN-based Safety Boundary Detector does not provide a strict a priori bound on the prediction error. Since our setting involves low-dimensional inputs (e.g., 2D location states), existing techniques such as Lipschitz-based certificates \cite{tsuzuku2018lipschitz}, neural network verification tools \cite{katz2017reluplex}, or robust training with certified bounds \cite{wong2018provable} could, in principle, be applied to derive worst-case guarantees. Incorporating such certified error bounds into our framework will be pursued as future work, which can further strengthen the theoretical safety guarantees. At the same time, extending these verification techniques to high-dimensional sensory inputs (e.g., images or point clouds) and real-time MPC remains an open challenge for the broader community.
\end{remark}
\subsection{CFTOC PROBLEM}
\label{subsec:convex-mpc-dhocbf}

In Secs.~\textnormal{IV-}\ref{subsec:linearization-dynamics} and~\textnormal{IV-}\ref{subsec:linearization-dhocbf}, we have demonstrated the linearization of system dynamics and the safety constraints using DHOCBF. This enables us to incorporate them as constraints in a convex MPC formulation at each iteration, which we refer to as Convex Finite-Time Constrained Optimization Control (CFTOC). This is solved at iteration $j$ with optimization variables $\mathbf{U}_{t}^{j} = [ \mathbf{u}_{t,0}^{j}, \dots, \mathbf{u}_{t,N-1}^{j}]$ and $\Omega_{t, i}^{j} = [\omega_{t,0,i}^{j},\dots, \omega_{t,N,i}^{j}]$, where $i\in \{1,\dots, m_{\text{cbf}}\}$.

\noindent\rule{\columnwidth}{0.4pt}
  \textbf{CFTOC of iMPC-DHOCBF at iteration $j$:}  
\begin{subequations}
{\small
\label{eq:impc-dcbf}
\begin{align}
\label{eq:impc-dcbf-cost}
  & \min_{\mathbf{U}_{t}^{j},\Omega_{t,1}^{j},\dots, \Omega_{t,m_{\text{cbf}}}^{j}} p(\mathbf{x}_{t,N}^{j})+\sum_{k=0}^{N-1} q(\mathbf{x}_{t,k}^{j},\mathbf{u}_{t,k}^{j},\omega_{t,k,i}^{j}) \\
   \text{s.t.} \ & \mathbf{x}_{t,k+1}^{j}{-} \bar{\mathbf{x}}_{t,k+1}^{j}{=}A_{t,k}^{j}(\mathbf{x}_{t,k}^{j}-\bar{\mathbf{x}}_{t,k}^{j}){+}B_{t,k}^{j}(\mathbf{u}_{t,k}^{j}-\bar{\mathbf{u}}_{t,k}^{j})+c_{t,k}^{j}, \label{subeq:impc-dcbf-linearized-dynamics}\\
    & \mathbf{u}_{t,k}^{j} \in \mathcal U, \  \mathbf{x}_{t,k}^{j} \in \mathcal X, \ \omega_{t,k,i}^{j} \ \text{s.t.} \ \eqref{eq:convex-hocbf-constraint-2},\label{subeq:impc-dcbf-variables-bounds}\\
    & \tilde{\psi}_{i-1}(\mathbf{x}_{t,k}^{j})+ \sum_{\nu=1}^{i}(\gamma_{i}-1)Z_{\nu,i}  (1-\gamma_{1})^{k-1}\tilde{\psi}_{0}(\mathbf{x}_{t,\nu}^{j})  \ge \nonumber \\ 
    & \omega_{t,k,i}^{j}(1-\gamma_{i})Z_{0,i}(1-\gamma_{1})^{k-1}\tilde{\psi}_{0}(\mathbf{x}_{t,0}^{j}) \label{subeq:impc-dcbf-linearized-hocbf}.
\end{align}
}
\end{subequations}
\noindent\rule{\columnwidth}{0.4pt}
In the CFTOC, the linearized dynamics constraints in~\eqref{eq:linearized-dynamics} and the linearized DHOCBF constraints in~\eqref{eq:convex-hocbf-constraint} are enforced with constraints~\eqref{subeq:impc-dcbf-linearized-dynamics} and ~\eqref{subeq:impc-dcbf-linearized-hocbf} at each open-loop time step $k\in\{0,\dots, N-1\}$.
The state and input constraints are considered in~\eqref{subeq:impc-dcbf-variables-bounds}. 
The slack variables are subject to constraints in \eqref {eq:convex-hocbf-constraint-2} to enhance feasibility while guaranteeing safety, as discussed in Thm. \ref{thm:safety-feasibility}.
Note that, for ensuring the safety guarantee established by the DHOCBF, the constraints \eqref{subeq:impc-dcbf-linearized-hocbf} are enforced with $i\in\{0,\dots, m_{\text{cbf}}\}$, where $Z_{\nu,i}\in \mathbb{R}$ is as defined in \eqref{eq:z-value} with $\nu \in\{0,..,i\}$.
The optimal decision variables of \eqref{eq:impc-dcbf} at iteration $j$ is a list of control input vectors as $\mathbf{U}_{t}^{*,j}=[\mathbf{u}_{t,0}^{*,j},\dots,\mathbf{u}_{t,N-1}^{*,j}]$ and a list of slack variable vectors as $\Omega_{t,i}^{*,j}=[\omega_{t,0,i}^{*,j},\dots,\omega_{t,N-1,i}^{*,j}]$. For avoidance of multiple unsafe sets, the DHOCBF constraint \eqref{subeq:impc-dcbf-linearized-hocbf} in the CFTOC framework \eqref{eq:impc-dcbf} originally considers only a single set. To handle \( p \) unsafe sets, the slack variable vectors \( \Omega_{t,i}^{j} \) (used in both the cost \eqref{eq:impc-dcbf-cost} and \eqref{subeq:impc-dcbf-linearized-hocbf}) and the constraint itself must be expanded accordingly. The CFTOC is solved iteratively in the proposed iMPC-DHOCBF framework, and a solution is extracted once the convergence criteria or maximum iteration number \( j_{\text{max}} \) is reached, as shown in Fig.~\ref{fig:iteration-module}.

\begin{figure*}[!t]
    \vspace*{-0.4cm}
    \centering
    \begin{subfigure}[t]{0.24\linewidth}
        \centering
        \includegraphics[width=1\linewidth]{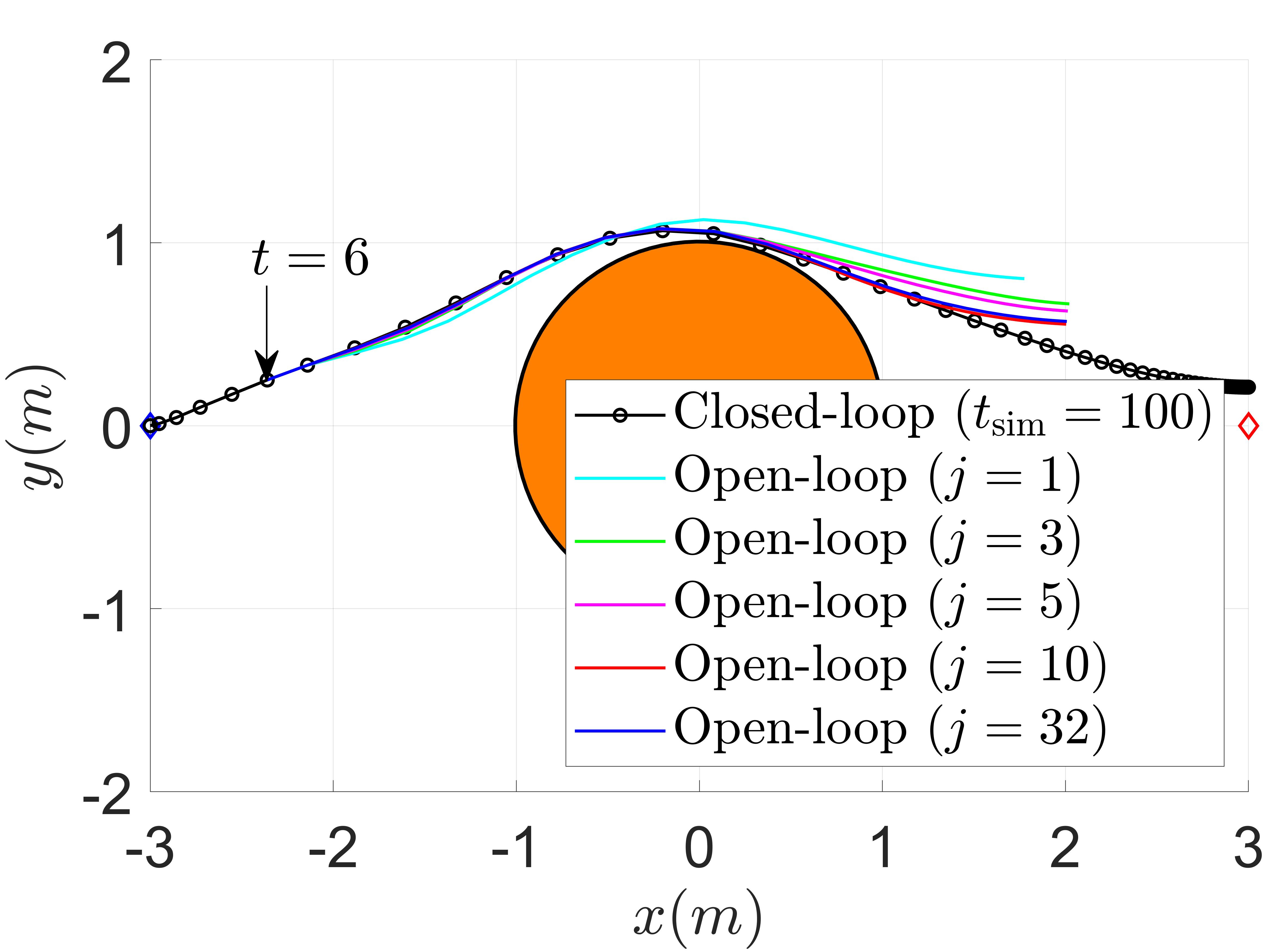}
        \caption{iMPC-DHOCBF when $N=24$, $\gamma_1=\gamma_2 = 0.4$.}
        \label{fig:openloop-snapshots}
    \end{subfigure}
    \begin{subfigure}[t]{0.24\linewidth}
        \centering
        \includegraphics[width=1\linewidth]{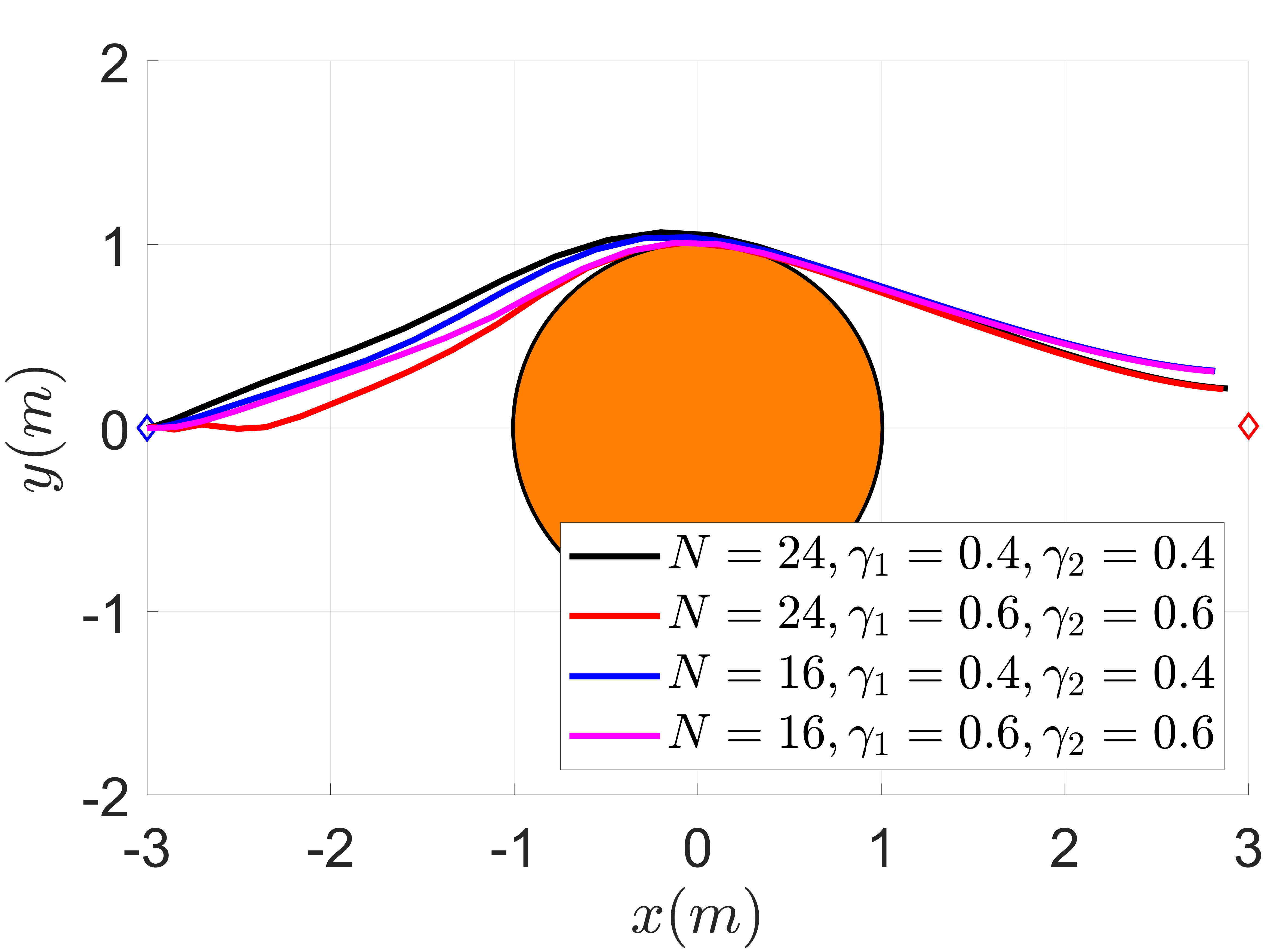}
        \caption{iMPC-DHOCBF with $m_{\text{cbf}} = 2$.}
        \label{fig:closedloop-snapshots1}
    \end{subfigure}  
    \begin{subfigure}[t]{0.24\linewidth}
        \centering
        \includegraphics[width=1\linewidth]{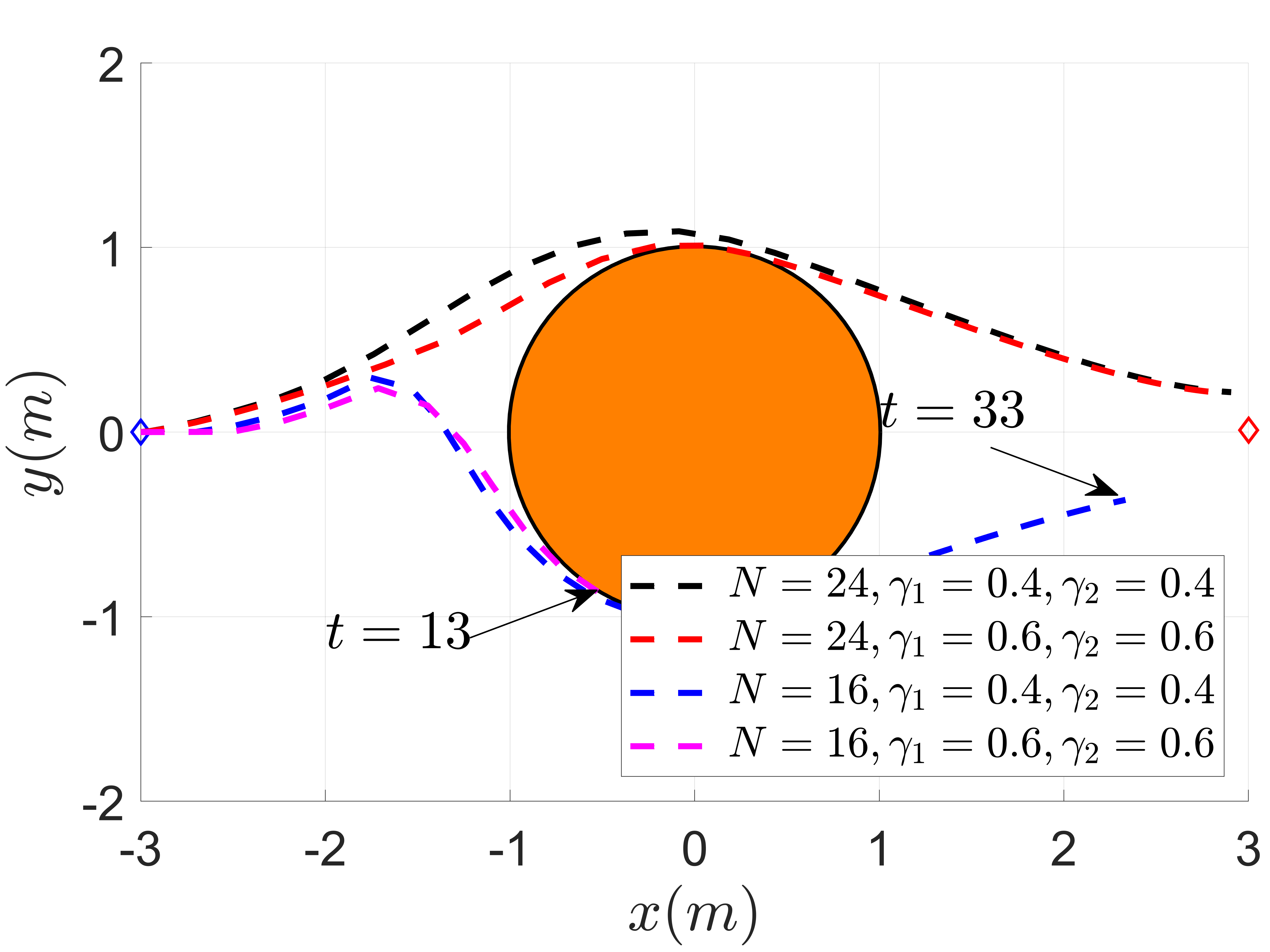}
        \caption{NMPC-DHOCBF with $m_{\text{cbf}} = 2$.}
        \label{fig:closedloop-snapshots2}
    \end{subfigure}
    \begin{subfigure}[t]{0.24\linewidth}
        \centering
        \includegraphics[width=1\linewidth]{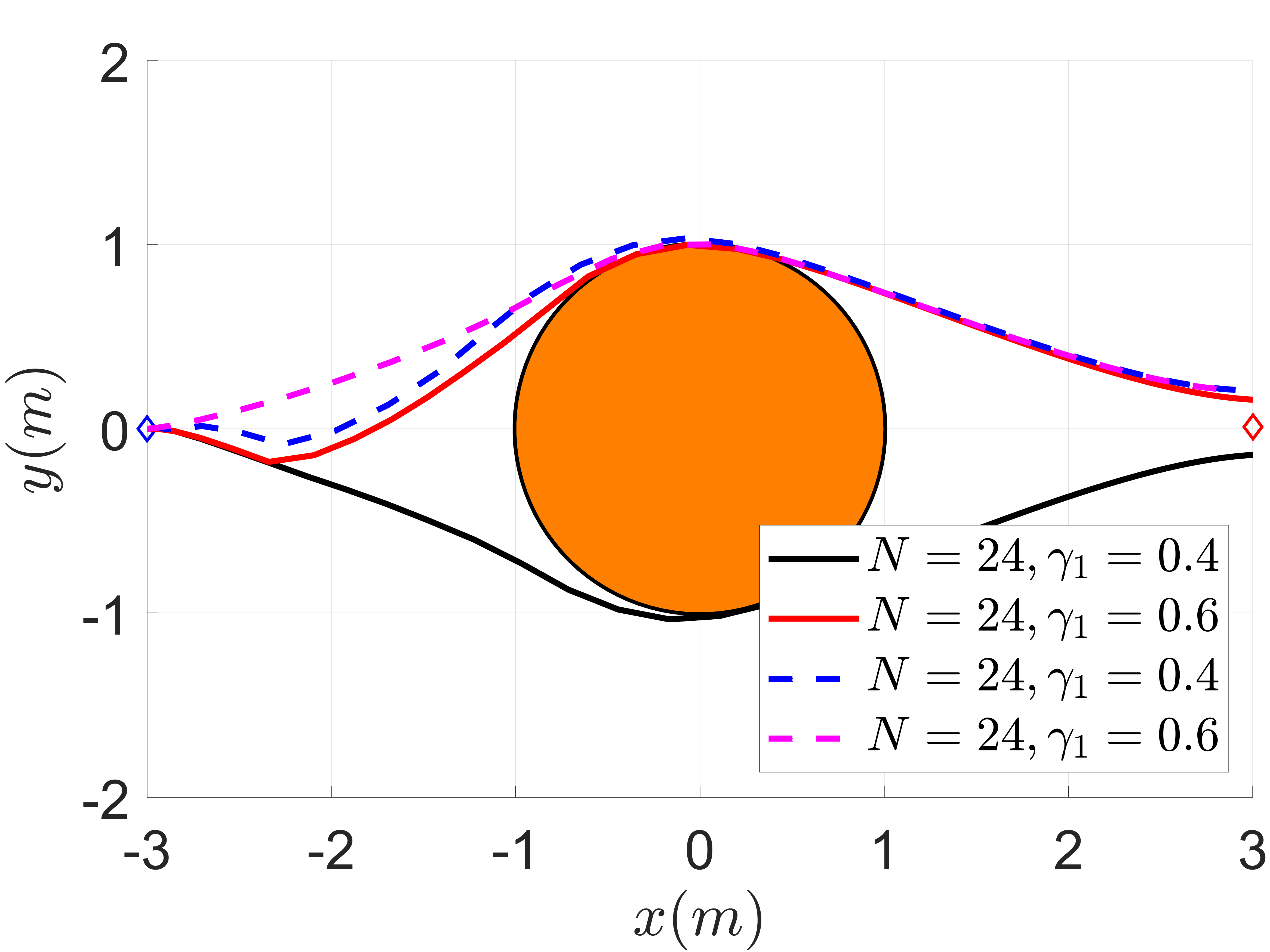}
        \caption{iMPC-DHOCBF and NMPC-DHOCBF with $m_{\text{cbf}}=1$.}
        \label{fig:closedloop-snapshots3}
    \end{subfigure}
    \caption{Open-loop and closed-loop trajectories with controllers iMPC-DHOCBF (solid lines) and NMPC-DHOCBF (dashed lines): (a) several open-loop trajectories at different iterations predicted at $t=6$ and one closed-loop trajectory with controller iMPC-DHOCBF; (b) closed-loop trajectories with controller iMPC-DHOCBF with different choices of $N$ and $\gamma$; (c) closed-loop trajectories with controller NMPC-DHOCBF with different choices of $N$ and $\gamma$. Note that two trajectories stop at $t=13$ and $t=33$ because of infeasibility; (d) closed-loop trajectories with controllers iMPC-DHOCBF and NMPC-DHOCBF with $m_{\text{cbf}}=1$. Both methods work well for safety-critical navigation. This figure demonstrates that at a specific time step, iMPC-DHOCBF can iteratively drive the open-loop trajectory to converge to a local minimum while ensuring the safety of the closed-loop trajectory over $t_{\text{sim}}$.
    } 
    \label{fig:open-closed-loop}
\end{figure*}

\begin{figure*}[t]
    \centering
    \begin{subfigure}[t]{0.24\linewidth}
        \centering
        \includegraphics[width=1.0\linewidth]{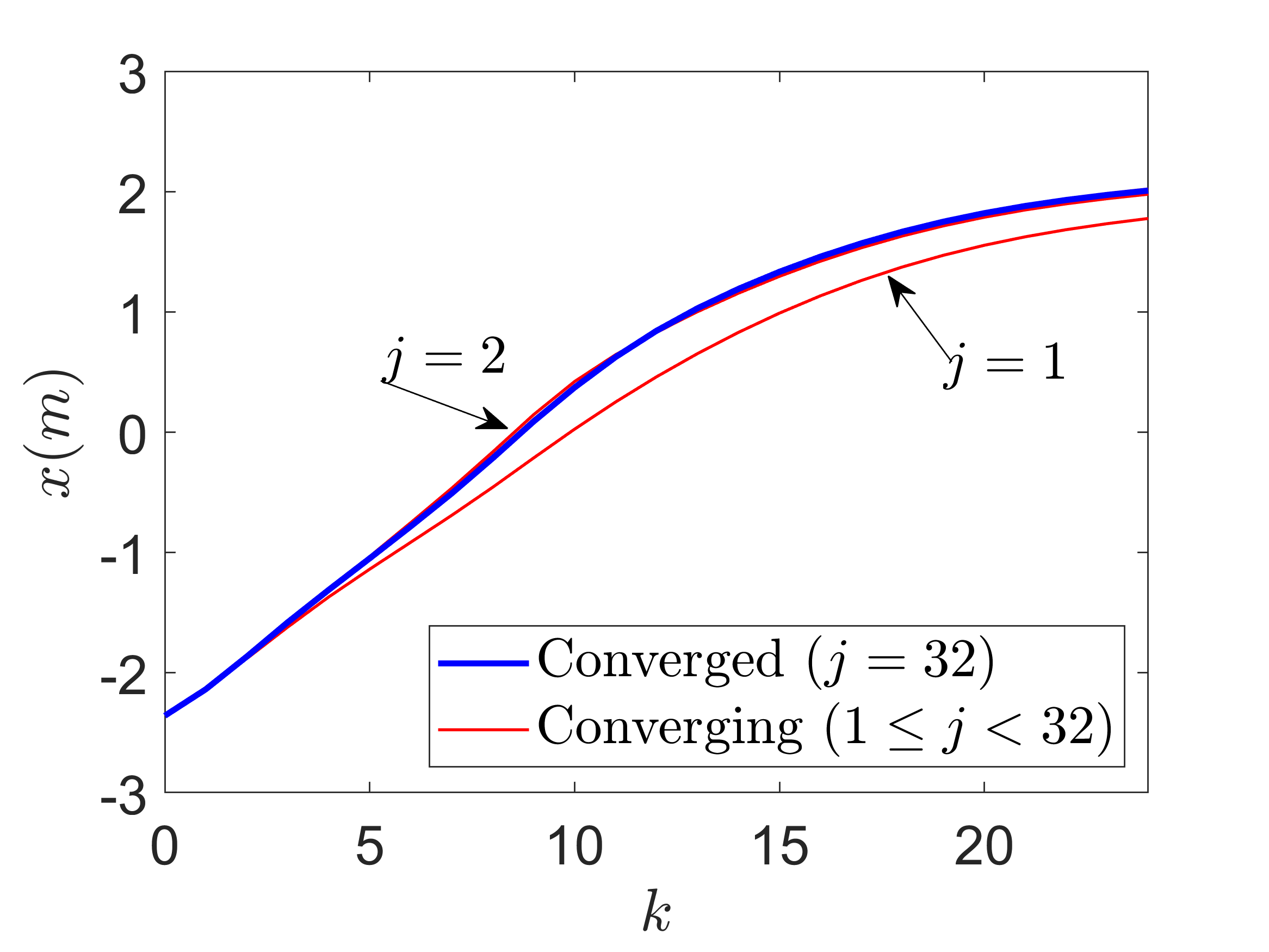}
        \caption{Location $x$}
        \label{subfig:convergence-state-1}
    \end{subfigure}
    \begin{subfigure}[t]{0.24\linewidth}
        \centering
        \includegraphics[width=1.0\linewidth]{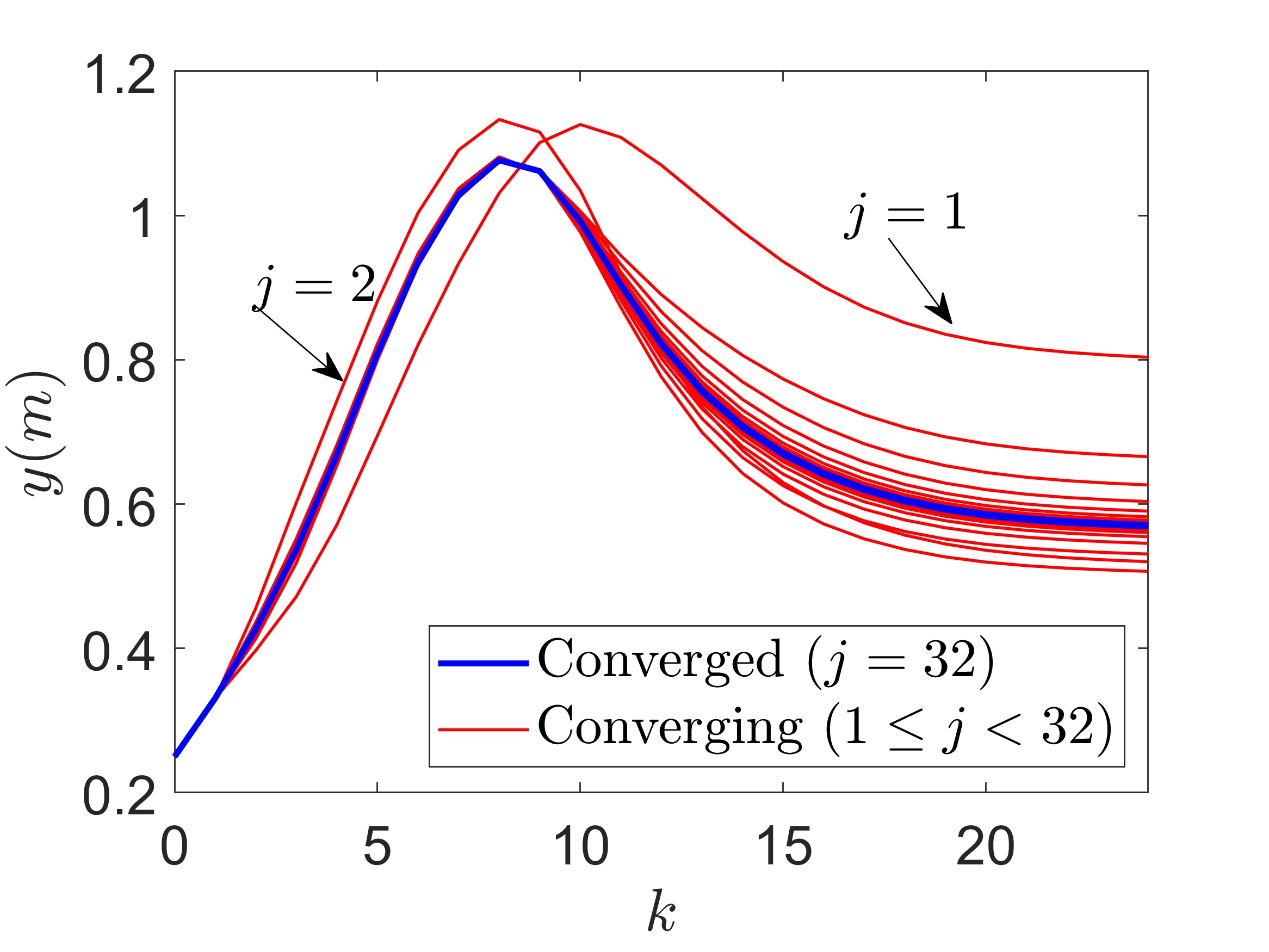}
        \caption{Location $y$}
        \label{subfig:convergence-state-2}
    \end{subfigure}  
    \begin{subfigure}[t]{0.24\linewidth}
        \centering
        \includegraphics[width=1.0\linewidth]{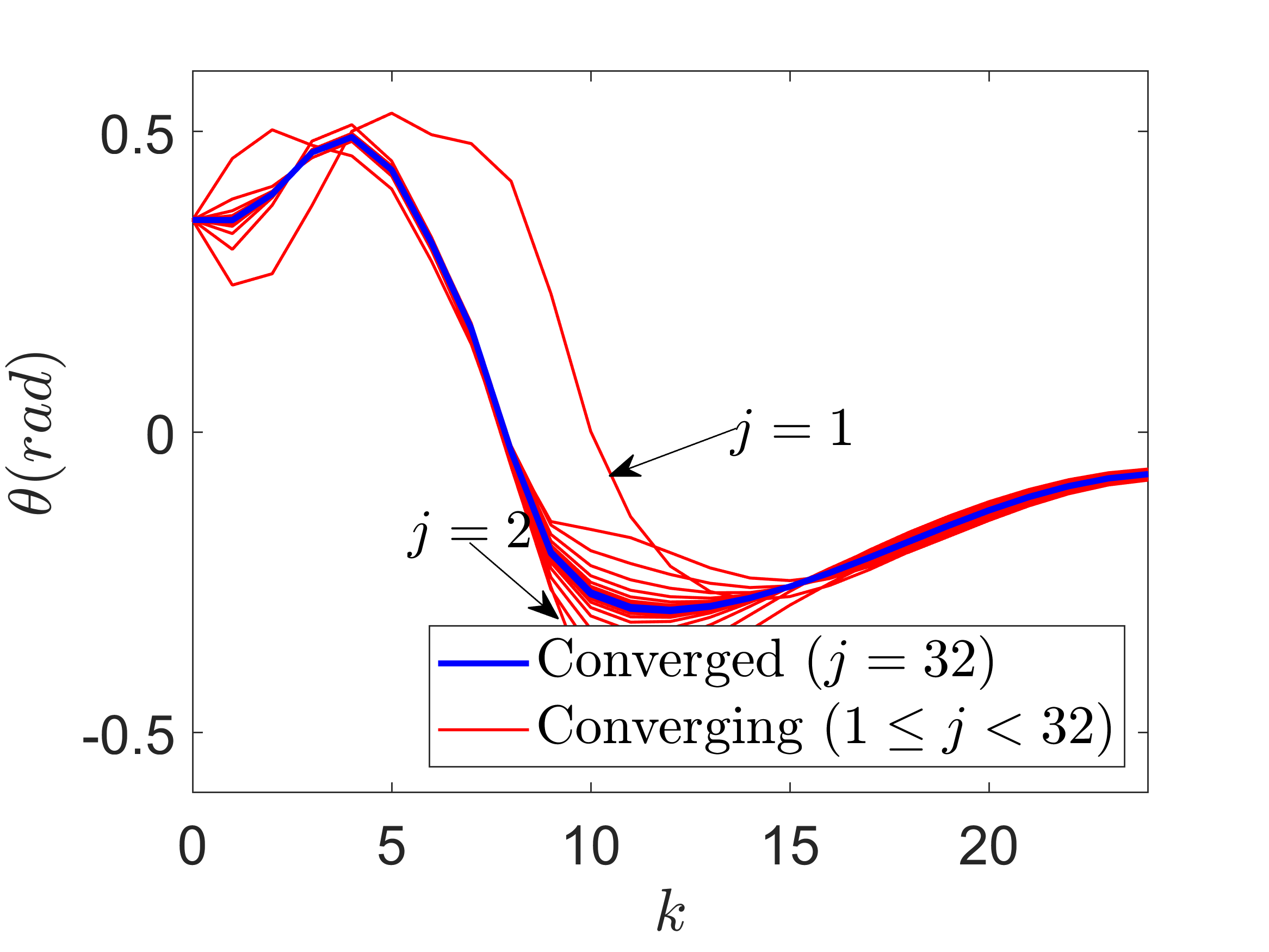}
        \caption{Orientation $\theta$}
        \label{subfig:convergence-state-3}
    \end{subfigure}
    \begin{subfigure}[t]{0.24\linewidth}
        \centering
        \includegraphics[width=1.0\linewidth]{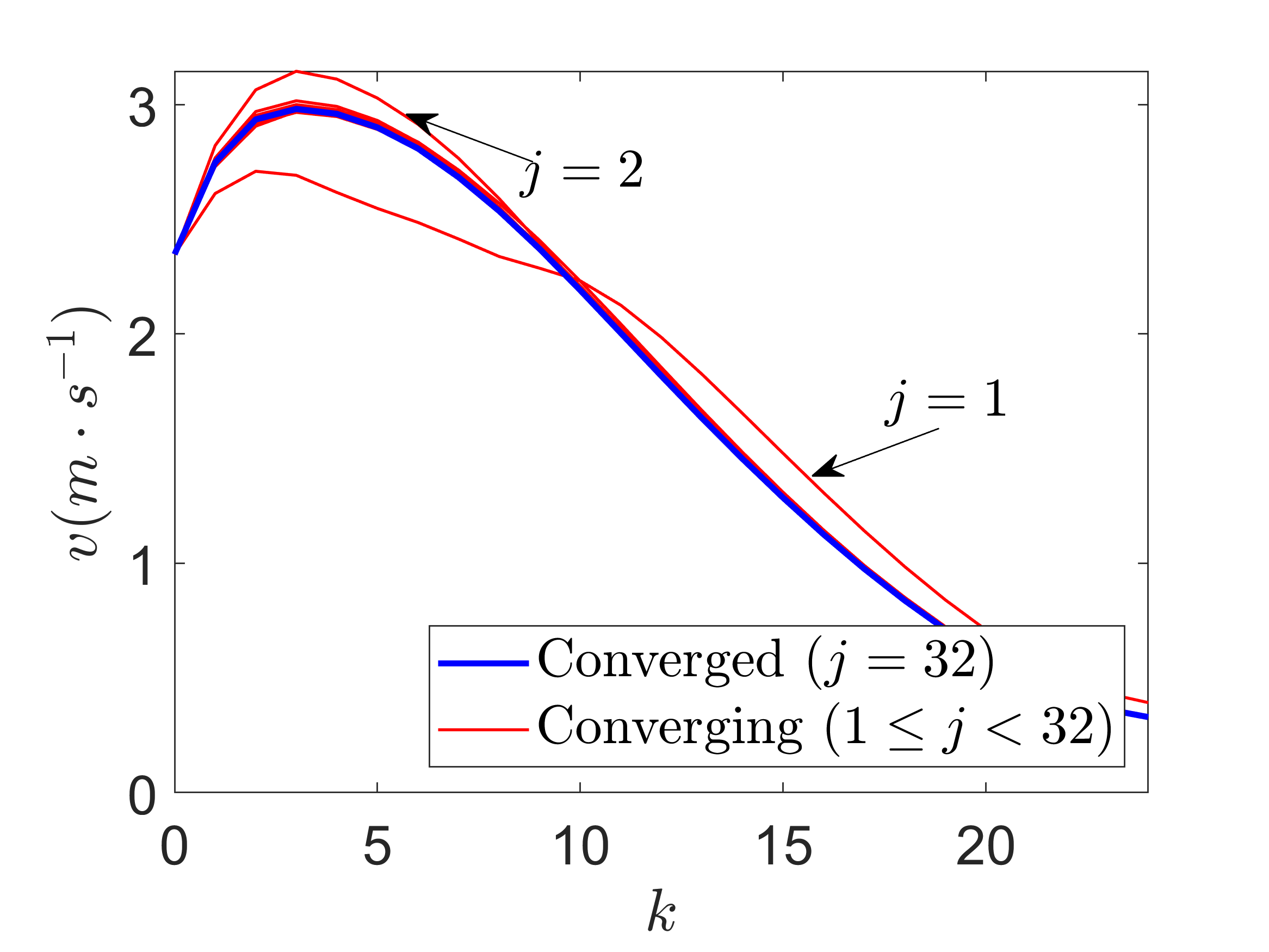}
        \caption{Speed $v$}
        \label{subfig:convergence-state-4}
    \end{subfigure}
    \caption{Iterative convergence of all states at converged iteration $j_{6,\text{conv}}=32$ with $N=24, m_{\text{cbf}}=2$, $\gamma_{1}=\gamma_{2}=0.4$. iMPC does help to optimize the cost function to reach local optimal minimum.}
    \label{fig:convergence-state}
\end{figure*}
\subsection{COMPLEXITY OF iMPC-DHOCBF}

\subsubsection{Size of Map and Network:}\hspace{0.1em}
In Sec. \textnormal{IV-C-}\ref{subsec: SBD unknown}, before we train the network, we need to gather as many sampled system's locations as possible across the map, while also identifying the nearest points on the  boundaries of unsafe sets relative to the system. This means that the size of the map will influence the number of sampled locations and ground truth data. The larger the map, the greater the number of sampled locations and ground truth needed, which in turn extends the time required to train the network. Additionally, the more complex the network structure (such as more hidden layers or more nodes), the longer it will take to train the network.

\subsubsection{Number of Unsafe Sets:}\hspace{0.1em}
In \eqref{eq:impc-dcbf}, every system's state $\mathbf{x}_{t,k}^{j}$ corresponds to a nearest point $\tilde{\mathbf{x}}_{t, k}^{j}$ on each unsafe set, and each nearest point corresponds to a linearized DHOCBF \eqref{subeq:impc-dcbf-linearized-hocbf}. Therefore, the more unsafe sets there are, the more DHOCBF constraints exist, leading to increased complexity in the CFTOC. One can choose to only consider unsafe sets within a certain range based on the current location to reduce complexity. This means that when the horizon $N$ is large, the predictive capability of the MPC will be limited. Nonetheless, the complexity related to unsafe sets in our algorithm is lower than the state-of-the-art \cite{thirugnanam2022duality}, because each polytopic unsafe set corresponds to multiple DHOCBFs.
\subsubsection{Horizon Length:}\hspace{0.1em}
In \eqref{eq:impc-dcbf}, the horizon $N$ results in a linear increase in the number of all constraints. Compared to distance constraints, DHOCBF constraints incorporate additional higher-order sufficient constraints to ensure safety. As discussed in \cite{thirugnanam2022duality}, DHOCBF constraints enable effective avoidance behavior for unsafe sets with a smaller horizon length. One could consider to require the horizon of DHOCBF $N_{\text{cbf}}$ smaller than $N$ to reduce the complexity.

\subsubsection{Convergence Criterion:}\hspace{0.1em}
In Fig. \ref{fig:iteration-module}, the number of iterations at each time step is determined by the convergence criterion or $j_{\text{max}}$. One can flexibly choose which variables to include in the convergence criterion, e.g., state $\mathbf{X}_{t}^{*, j},$ input $\mathbf{U}_{t}^{*, j},$ or both state and input. Appropriately relaxing the convergence conditions (such as increasing the allowable error) or reducing 
$j_{\text{max}}$ can both decrease the complexity of the algorithm.
\subsubsection{Highest Order of DHOCBF:}\hspace{0.1em}
In \eqref{eq:impc-dcbf}, the highest order of DHOCBF $m_{\text{cbf}}$ results in a linear increase in the number of DHOCBF constraints \eqref{subeq:impc-dcbf-linearized-hocbf}. As discussed in Rem. \ref{rem:sufficient-condition}, it is not necessary to formulate DCBF constraints up to the $m^{th}$ order. In other words, the highest order for DHOCBF could be $m_{\text{cbf}}$ with $m_{\text{cbf}}\le m$. Making $m_{\text{cbf}}$ smaller than $m$ but larger than one will reduce the complexity of the algorithm without significantly compromising safety. 
\subsubsection{Complexity Scaling with State and Input Dimensions:}\hspace{0.1em}
\label{subsubsec:state input scale}
Because each prediction step carries its own copy of the state vector $x_{t,k}\!\in\!\mathbb R^{n}$ and input vector $u_{t,k}\!\in\!\mathbb R^{q}$, enlarging $n$ or $q$ inflates the stacked decision vector that the CFTOC must solve. Every extra state or input triggers additional rows for the linearized dynamics, box constraints, and any DHOCBF or slack-variable terms. Since the number of constraints already scales linearly with the horizon $N$, the overall problem size grows roughly $\mathcal O\!\bigl(N(n+q)\bigr)$ in memory. More states also mean larger Jacobians $A_{t,k}^{j}, B_{t,k}^{j}$ and extra slack variables—each adding per-iteration linearization and matrix-assembly cost. 

\section{CASE STUDIES}

We focus on a unicycle robot. Case Study \Romannum{1} compares our method with the baseline NMPC-DHOCBF using a circular obstacle as the unsafe set. The baseline is extended with a relaxed DHOCBF based on \eqref{subeq:mpc-cbf-cons2}, following \cite[Rem. 4]{zeng2021enhancing}. Case studies \Romannum{2} and \Romannum{3} demonstrate the proposed iMPC-DHOCBF on more complex maps. Case study \Romannum{3} also includes results for a vehicle model. Animation videos are available at \url{https://youtu.be/G9S7y90LRig}.%

\subsection{CASE STUDY \Romannum{1}: POINT ROBOT AVOIDANCE OF A CIRCULAR OBSTACLE WITH KNOWN BOUNDARY}
\label{subsec: case study 1}
Since the nearest point on a circle from a point outside the circle can be expressed by an accurate geometric equation, SBD is equivalent to finding tangent lines on this equation using derivatives, so we do not need to train a DNN here.
\subsubsection{System Dynamics:}\hspace{0.1em}
Consider a discrete-time unicycle model in the form
\begin{equation}
\label{eq:unicycle-model}
\begin{bmatrix} x_{t+1}{-}x_t \\ y_{t+1}{-}y_t \\ \theta_{t+1}{-}\theta_t \\ v_{t+1}{-}v_t \end{bmatrix}{=}\begin{bmatrix} v_{t} \cos(\theta_{t}) \Delta t \\ v_{t}\sin(\theta_{t}) \Delta t \\ 0 \\ 0 \end{bmatrix}{+}\begin{bmatrix} 0 & 0 \\ 0 & 0 \\ \Delta t & 0 \\ 0 & \Delta t \end{bmatrix}
\begin{bmatrix} u_{1,t} \\ u_{2,t} \end{bmatrix},
\end{equation}
where $\mathbf{x}_{t}=[x_{t},y_{t},\theta_{t},v_{t}]^{T}$ captures the 2-D location, heading angle, and linear speed; $\mathbf{u}_{t}=[u_{1,t},u_{2,t}]^{T}$ represents angular velocity ($u_{1}$) and linear acceleration ($u_{2}$), respectively.
The system is discretized with $\Delta t = 0.1$ and the total number of discretization steps $t_{\text{sim}}$ equals $T$ in Problem. \ref{prob:Path-prob}. System \eqref{eq:unicycle-model} is subject to the state and input constraints $\mathcal X$ and $\mathcal U$ which can be found in~\cite{liu2023iterative}.

\subsubsection{System Configuration:}\hspace{0.1em}
The initial state is $[-3,0,0,0]^{T}$ and the target state is $\mathbf{x}_{r}=[3,0.01,0,0]^{T},$ which are marked as blue and red diamonds in Fig.~\ref{fig:open-closed-loop}. The circular obstacle is centered at $(0,0)$ with $r=1,$ which is displayed in orange. The other reference vectors are $\mathbf{u}_{r}=[0,0]^{T}$ and $\omega_{r}=[1,1]^{T}$. 
We use the offset $y=0.01m$ in $\mathbf{x}_{r}$ to prevent singularity of the optimization problem. 

\subsubsection{DHOCBF:}\hspace{0.1em}
As a candidate DHOCBF function $\psi_{0}(\mathbf{x}_{t})$, we choose a quadratic distance function for circular obstacle avoidance $h(\mathbf{x}_{t})= (x_{t}-x_{0})^{2}+(y_{t}-y_{0})^{2}-r^{2}$, where $(x_{0},y_{0})$ and $r$ denote the obstacle center location and radius, respectively. The linearized DHOCBF $\tilde{\psi}_{0}(\mathbf{x}_{t,k}^{j})$ in \eqref{eq:linearized-CBFs} follows the formulation presented in~\cite{liu2023iterative}.

\subsubsection{MPC Design:}\hspace{0.1em}
The CFTOC cost penalizes deviations from reference vectors and includes slack variables to enhance feasibility while ensuring safety. The details of the MPC design can be found in~\cite{liu2023iterative}.

\subsubsection{Convergence Criteria:}\hspace{0.1em}
\label{subsubsec:Convergence Criteria-f}
The iterative optimization terminates when absolute or relative convergence criteria are met, as defined in Fig. \ref{fig:iteration-module}. See~\cite{liu2023iterative} for a similar formulation. The maximum iteration number is set as $j_{\text{max}} = 1000$.
To ensure a fair comparison with NMPC-DHOCBF, the hyperparameters $P, Q, R, S$ are kept consistent across all configurations.

\subsubsection{Solver and CPU Specs:}\hspace{0.1em}
For iMPC-DHOCBF, we used OSQP~\cite{stellato2020osqp} to solve the convex optimizations at all iterations.
The baseline approach NMPC-DHOCBF is open-source, and was solved using IPOPT~\cite{biegler2009large} with the modeling language Yalmip~\cite{lofberg2004yalmip}.
We used a Windows desktop with Intel Core i7-8700 (CPU 3.2 GHz) running Matlab.

\subsubsection{Iterative Convergence:}\hspace{0.1em}
\label{subsubsec:iterative-convergence}
The iterative convergence is shown in Figs. \ref{fig:openloop-snapshots} and \ref{fig:convergence-state}.
Figure \ref{fig:openloop-snapshots} shows the closed-loop trajectory (the black line) generated by solving the iMPC-DHOCBF until the converged iteration $j_{t,\text{conv}}$ from $t=0$ to $t=t_{\text{sim}}=100$ and open-loop iteratively converging trajectories (colored lines) at different iterations at $t=6$.
Figure~\ref{fig:convergence-state} presents more details on the iterative convergence of states at different iterations at $t=6$ with number of iterations $j_{t,\text{conv}} = 32$.
We note that, after around 10 iterations, the converging lines for the states (red lines) nearly overlap with the converged line (blue line) in Fig. \ref{fig:convergence-state}. This verifies the relations of the converging trajectory (red line) and the converged trajectory (blue line) in Fig. \ref{fig:openloop-snapshots}. The convergence behavior of iMPC-DHOCBF varies over time and with different $\gamma$ values, typically requiring fewer than 100 iterations except near obstacles (see~\cite{liu2023iterative} for details).

\begin{table*}[t]
\vspace*{-0.4cm}
\centering
\caption{Statistical benchmark for computation time and feasibility between NMPC-DHOCBF and iMPC-DHOCBF with randomized states.
The target position is shared among four approaches and the hyperparameters are fixed as $\gamma_{1}=\gamma_{2}=0.4$ for all random scenarios.}
\label{tab:compuation-time1}

\resizebox{0.99\textwidth}{!}{
\begin{tabular}{|cc|cccccc|}
\hline
\multicolumn{2}{|c|}{Approaches} & $N = 4$ & $N = 8$ & $N = 12$ & $N = 16$ & $N = 20$ & $N = 24$ \\ \hline
\multicolumn{1}{|c|}{\multirow{2}{*}{\begin{tabular}[c]{@{}c@{}}NMPC-DHOCBF\\ ($m_{\text{cbf}} = 2$)\end{tabular}}} & mean / std (s) & $3.687\pm6.360$ & $23.882\pm17.988$ & $27.329\pm20.115$ & $28.953\pm22.058$ & $30.970\pm23.564$ & $29.929\pm22.105$ \\
\multicolumn{1}{|c|}{} & infeas. rate & 5.8\% & 27.5\% & 21.1\%  &  16.4\% & 14.5\% & 14.4\% \\ \hline
\multicolumn{1}{|c|}{\multirow{2}{*}{\begin{tabular}[c]{@{}c@{}}NMPC-DHOCBF\\ ($m_{\text{cbf}}  = 1$)\end{tabular}}} & mean / std (s) & $2.933\pm4.678$ & $19.077\pm14.024$ & $20.418\pm15.401$ & $22.749\pm17.039$ & $24.053\pm17.811$ & $25.365\pm18.211$ \\
\multicolumn{1}{|c|}{} & infeas. rate & 6.3\% & 13.9\% & 13.0\% & 14.6\% & 13.8\% & 15.4\% \\ \hline
\multicolumn{1}{|c|}{\multirow{2}{*}{\begin{tabular}[c]{@{}c@{}}iMPC-DHOCBF\\ ($m_{\text{cbf}}  = 2$)\end{tabular}}} & mean / std (s) & $0.135\pm0.294$ & $0.104\pm0.242$ & $0.102\pm0.217$ & $0.131\pm0.301$ & $0.165\pm0.400$ & $0.135\pm0.274$ \\
\multicolumn{1}{|c|}{} & infeas. rate & 6.3\% & 8.0\% & 10.4\% & 10.9\% & 10.9\% & 10.2\% \\ \hline
\multicolumn{1}{|c|}{\multirow{2}{*}{\begin{tabular}[c]{@{}c@{}}iMPC-DHOCBF\\ ($m_{\text{cbf}}  = 1$)\end{tabular}}} & mean / std (s) & $0.131\pm0.286$ & $0.114\pm0.260$ & $0.109\pm0.237$ & $0.137\pm0.316$ & $0.173\pm0.414$ & $0.152\pm0.317$ \\
\multicolumn{1}{|c|}{} & infeas. rate & 6.3\% & 8.0\% & 10.4\% & 10.9\% & 10.9\% & 11.1\% \\ \hline
\end{tabular}
}
\end{table*}

\subsubsection{Convergence with Different Hyperparameters:}\hspace{0.1em}
Figs. \ref{fig:closedloop-snapshots1}, \ref{fig:closedloop-snapshots2}, and \ref{fig:closedloop-snapshots3} show the closed-loop trajectories generated by solving iMPC-DHOCBF (solid lines) and NMPC-DHOCBF (dashed lines) at converged iteration $j_{t,\text{conv}}$ from $t=0$ to $t=t_{\text{sim}}=45$ with different hyperparameters.
Both controllers show good performance on obstacle avoidance.
Based on black, red, blue and magenta lines with the highest order of CBF constraint $m_{\text{cbf}}=2$ in Figs. \ref{fig:closedloop-snapshots1} and \ref{fig:closedloop-snapshots2}, as $\gamma_{1}, \gamma_{2}$ become smaller, the system tends to turn further away from the obstacle when it is getting closer to obstacle, which indicates a safer control strategy.
From the lines in Fig. \ref{fig:closedloop-snapshots3} where $m_{\text{cbf}}=1$, we can see that the system can still safely navigate around the obstacle, although it turns away from the obstacle later than when having one more CBF constraint in Figs. \ref{fig:closedloop-snapshots1} and \ref{fig:closedloop-snapshots2}, indicating that having CBF constraints up to the relative degree enhances safety.
The blue and magenta dashed lines in Fig. \ref{fig:closedloop-snapshots2} stop at $t=33$ and $t=13$, respectively, due to infeasibility with $N=16$. Although the model has passed the obstacle, NMPC-DHOCBF may still become infeasible under certain hyperparameter settings due to overly conservative constraints. This illustrates its sensitivity to horizon selection. In contrast, iMPC-DHOCBF generates complete closed-loop trajectories with both $N=16$ and $N=24$, showing improved robustness to horizon choices, as seen in Fig. \ref{fig:closedloop-snapshots1}.

\subsubsection{Computation Time:}\hspace{0.1em}
To compare computational times between iMPC-DHOCBF and the baseline NMPC-DHOCBF, we generate 1000 randomized safe states within the constraint set \( \mathcal{X} \). Both methods use the same \( N \) and \( m_{\text{cbf}} \), and are evaluated on computation time and feasibility at each sampled state. Table~\ref{tab:compuation-time1} shows the distributions of computation times and infeasibility rates for one-step trajectory generation. For NMPC-DHOCBF, the mean and standard deviation of computation times grow significantly with increasing \( N \) and \( m_{\text{cbf}} \), whereas iMPC-DHOCBF remains largely unaffected. Overall, iMPC-DHOCBF achieves much faster computation—up to \( 100\sim 300 \times \) speedup over the baseline—depending on the chosen hyperparameters.

\subsubsection{Optimization Feasibility:}\hspace{0.1em}
For iMPC-DHOCBF, the infeasibility rate increases with a longer horizon \( N \) or smaller \( m_{\text{cbf}} \), while NMPC-DHOCBF shows less sensitivity to these parameters. As \( N \) increases, iMPC-DHOCBF achieves lower infeasibility rates than NMPC-DHOCBF. This difference arises from variations in convergence criteria, warm starts, and relaxation techniques, as discussed in Rems.~\ref{rem: warm start} and~\ref{rem: different-relax-techniques}. NMPC-DHOCBF, solved via IPOPT, applies stricter convergence criteria but benefits from a more refined warm start and relaxed nonlinear DCBF constraints \eqref{eq:nonconvex-hocbf-constraint}. In contrast, iMPC-DHOCBF uses relaxed linear CBF constraints \eqref{eq:convex-hocbf-constraint}, which reduce the feasible region (Fig.~\ref{fig:linearization-dhocbf}). This explains the slightly lower feasibility at small \( N \), but as \( N \) grows, the flexible convergence criteria in iMPC-DHOCBF lead to significantly improved feasibility, as confirmed in Tab.~\ref{tab:compuation-time1}.

\subsection{CASE STUDY \Romannum{2}: POINT ROBOT AVOIDANCE OF IRREGULAR OBSTACLES WITH UNKNOWN BOUNDARIES}
\label{subsec: case study 2}
Here we consider both convex and nonconvex obstacles. The nearest points to the robot on their boundaries are difficult to represent with continuous and accurate equations. We train a DNN for the SBD to predict the nearest points. 
\subsubsection{System Dynamics:}\hspace{0.1em}
We consider the same dynamics as in Sec. \textnormal{V-}\ref{subsec: case study 1} with $\Delta t = 0.05.$ System~\eqref{eq:unicycle-model} is subject to the following state and input constraints:
\begin{equation}
\begin{split}
\label{eq:state-input-constraint2}
\mathcal{X}&=\{\mathbf{x}_{t}\in \mathbb{R}^{4}: -10\cdot \mathcal{I}_{4\times1} \le \mathbf{x}_{t}\le 10\cdot \mathcal{I}_{4\times1}\},\\
\mathcal{U}&=\{\mathbf{u}_{t}\in \mathbb{R}^{2}: -10\cdot \mathcal{I}_{2\times1} \le \mathbf{u}_{t}\le 10\cdot \mathcal{I}_{2\times1}\}.
\end{split}
\end{equation}

\subsubsection{System Configuration:}\hspace{0.1em}
The five different start points are $(-2, -2)$, $(-2, 2)$, $(-0.8, -0.8)$, $(0, 0.8)$, $(1.2, 0.4)$ and the corresponding end points are $(2, 2)$, $(2, -2)$, $(0, 0.8)$, $(1.2, 0.4)$, $(0, 0),$ which are marked as solid dots in Fig. \ref{fig:five obstacles navigation}. The robot can stop when it reaches inside the target areas centered at end points with a radius of 0.1. The initial and reference linear speeds are 0 and the initial and reference heading angles are calculated by $\theta_{0}=\theta_{r}=atan2(\frac{y_{r}-y_{t}}{x_{r}-x_{t}}),$ where $(x_{t}, y_{t})$ denotes the current location and $(x_{r}, y_{r})$ denotes the end point. Five obstacles are displayed in red, with convex and nonconvex shapes. The other reference vectors are $\mathbf{u}_{r}=[0,0]^{T}$ and $\omega_{r}=\mathcal{I}_{1\times10}$. 
\subsubsection{Map Processing:}\hspace{0.1em} The scope of the map is $x\in[-2.5, 2.5]$ and $y\in[-2, 2].$ The coordinates of pixels on the boundaries of obstacles are manually chosen, and we limit the number of boundary pixels for each obstacle to 100.
\subsubsection{SBD Training and Evaluation:}\hspace{0.1em}
\label{subsubsec:SBD structure}
Our neural network has 9 hidden layers and begins with 512 nodes in the first hidden layer, denoted as $h_{0}$. The subsequent layers’ dimensions are defined by \eqref{eq:node number}. Within the range of the map, we extracted 154993 data entries (random locations across the entire map) for training and another 25741 (random locations outside of obstacles) for testing. For $p$ obstacles, we use the Mean Squared Error (MSE) equation \eqref{eq:MSE} to evaluate the accuracy of the SBD's predictions of the nearest points on obstacle boundaries relative to the robot.
\subsubsection{DHOCBF:}\hspace{0.1em}
\label{subsubsec: DHOCBF}
Since the predicted nearest points $\tilde{\mathbf{x}}_{t,k}^{j}$ can be obtained from the output of the SBD, we choose \eqref{eq:linearized safe set} as a linearized DHOCBF whose relative degree is 2. Note that five obstacles correspond to five linearized DHOCBFs. The sequence of linearized DHOCBF is defined by \eqref{eq:linearized-CBFs}. From \eqref{eq:z-value}, we have $Z_{0,2}=\gamma_{1}-1,\ Z_{1,2}=1,\ Z_{0,1}=1,\ Z_{2,2}=Z_{1,1}=0$.

\subsubsection{MPC Design:}\hspace{0.1em}
The cost function of the CFTOC problem \eqref{eq:impc-dcbf} consists of stage cost
$q(\mathbf{x}_{t,k}^j,\mathbf{u}_{t,k}^{j},\omega_{t,k,i}^{j})= \sum_{k=0}^{N-1} (||\mathbf{x}_{t,k}^{j}-\mathbf{x}_{r}||_Q^2 + ||\mathbf{u}_{t,k}^{j}-\mathbf{u}_{r}||_R^2 +||\omega_{t,k,i}^{j}-\omega_{r}||_S^2)$
and terminal cost $p(\mathbf{x}_{t,N}^{j})=||\mathbf{x}_{t,N}^{j}-\mathbf{x}_{r}||_P^2$, where $Q=P=10\cdot \mathcal{I}_{4}, R= \mathcal{I}_{2}$ and $S=1000\cdot \mathcal{I}_{10}$. In this case study, the slack variable $\omega_{t,k,i}^{j}$ is subject to the constraint in \eqref {eq:convex-hocbf-constraint-2} for iMPC-DHOCBF to enhance feasibility while guaranteeing safety, as discussed in Thm. \ref{thm:safety-feasibility}. 

\subsubsection{Convergence Criteria:}\hspace{0.1em}
\label{subsubsec: convergence criteria}
We use the same absolute or relative convergence criteria from Sec. \textnormal{V-A-}\ref{subsubsec:Convergence Criteria-f}.
The hyperparameter related to the decay rate is set to a relatively smaller value as $\gamma_{1}=0.1$ to reduce the needed maximum iteration number which is set as $j_{\text{max}} = 30$.

\subsubsection{Solver and CPU Specs:}\hspace{0.1em}
\label{subsubsec: specs}
For iMPC-DHOCBF, we used OSQP \cite{stellato2020osqp} in Python to solve the convex optimizations at all iterations.
For model training and inference, we used Pytorch. The training was conducted on a Linux desktop equipped with a Nvidia RTX 4090 graphics card. For inference, we used a Linux laptop with an AMD Ryzen 7 5800U.

\begin{figure}
    \centering
    \includegraphics[scale=0.26]{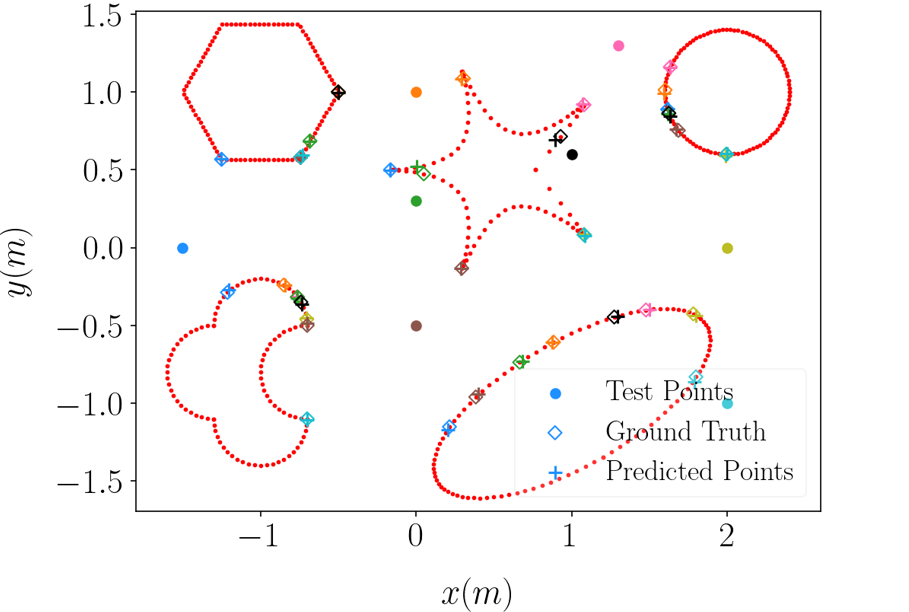}
    \caption{Demonstration of SBD prediction performance. Solid dots indicate the positions of eight robots. Diamonds show the ground-truth nearest points on obstacles, while plus signs denote the SBD predictions. Most predictions closely match the ground truth, except for the one corresponding to the green dot, which shows a larger deviation.}
    \label{fig:neural-network2}
\end{figure}

\begin{figure*}[!t]
    \vspace*{-0.4cm}
    \centering
    \begin{subfigure}[t]{0.26\linewidth}
        \centering
        \includegraphics[width=1.0\linewidth]{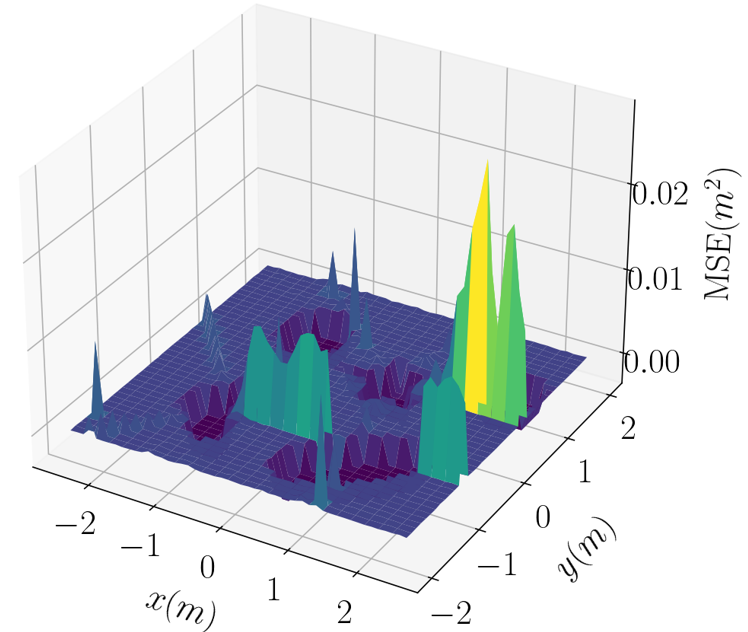}
        \caption{Isometric view}
        \label{subfig:a1}
    \end{subfigure}
    \begin{subfigure}[t]{0.24\linewidth}
        \centering
        \includegraphics[width=1.0\linewidth]{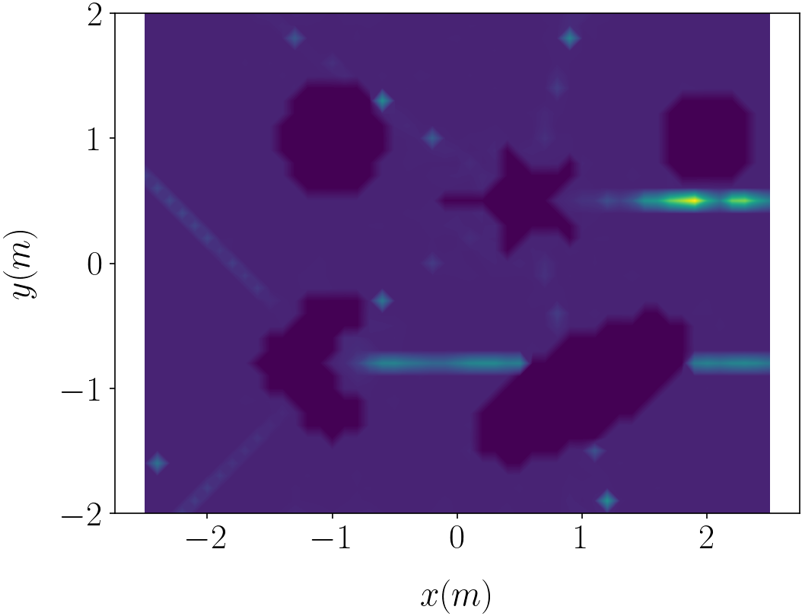}
        \caption{Top view}
        \label{subfig:a2}
    \end{subfigure}  
    \begin{subfigure}[t]{0.24\linewidth}
        \centering
        \includegraphics[width=1.0\linewidth]{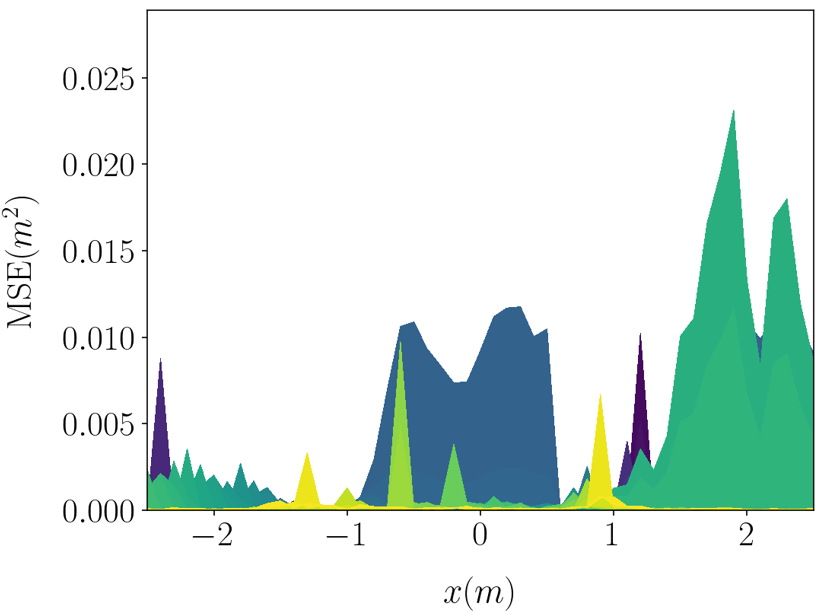}
        \caption{$x\text{-MSE}$ plane view}
        \label{subfig:a3}
    \end{subfigure}
    \begin{subfigure}[t]{0.24\linewidth}
        \centering
        \includegraphics[width=1.0\linewidth]{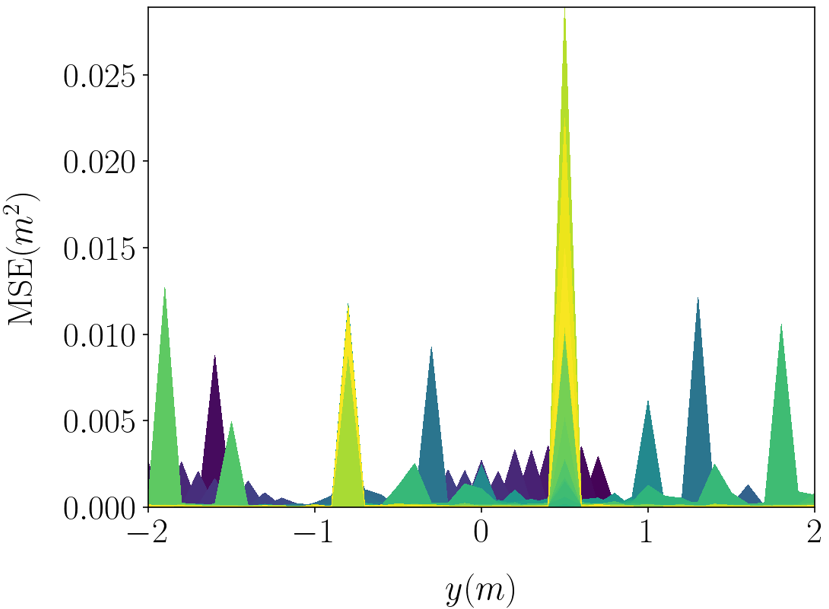}
        \caption{$y\text{-MSE}$ plane view}
        \label{subfig:a4}
    \end{subfigure}
    \caption{3D visualization of the MSE of the testing data. $x$ and $y$ represent the positions of the testing points, while MSE represents the prediction error corresponding to the point $(x,y).$ The maximum MSE is 0.029 at $(1.9, 0.5).$}
    \label{fig: MSE five obstacles}
\end{figure*}
\begin{figure*}[t]
    \centering
    \begin{subfigure}[t]{0.245\linewidth}
        \centering
        \includegraphics[width=1.0\linewidth]{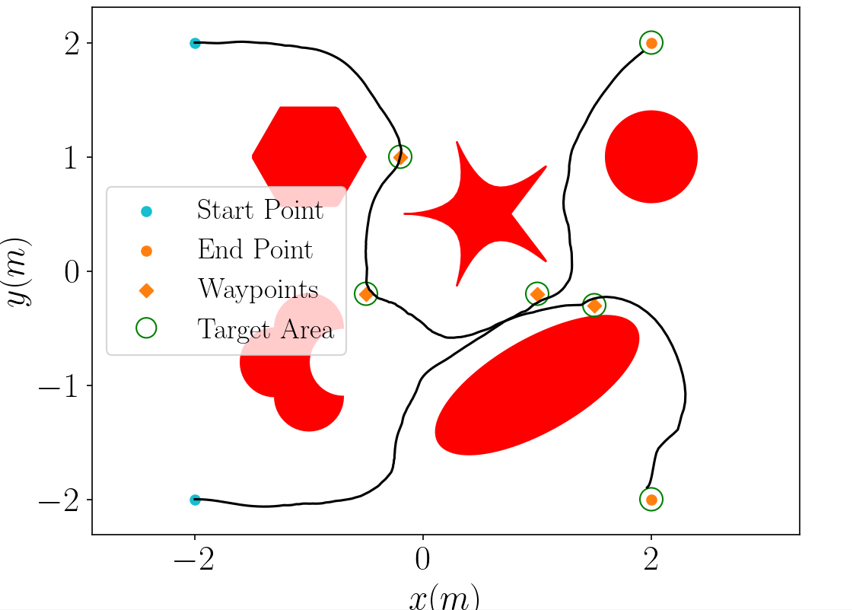}
        \label{subfig:11}
    \end{subfigure}
    \begin{subfigure}[t]{0.245\linewidth}
        \centering
        \includegraphics[width=1.0\linewidth]{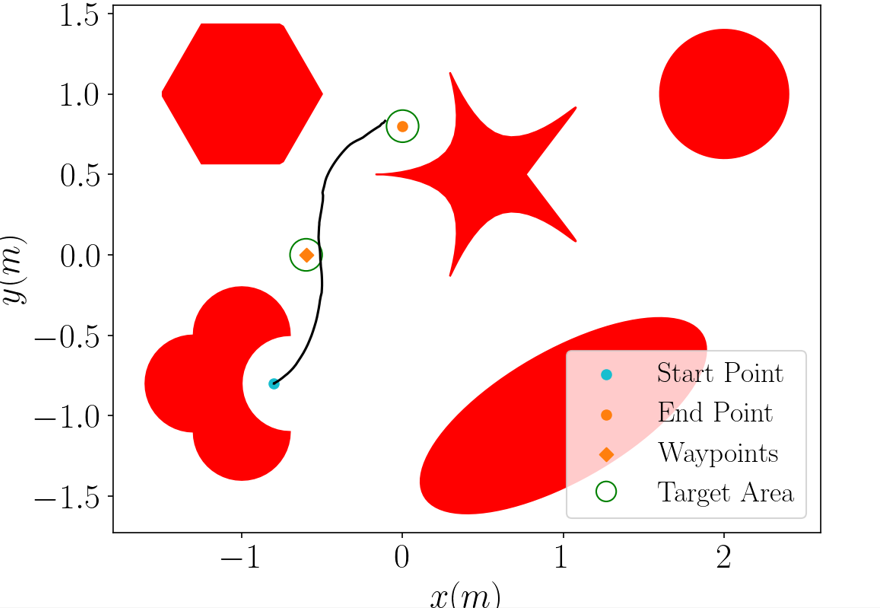}
        \label{subfig:22}
    \end{subfigure}  
    \begin{subfigure}[t]{0.245\linewidth}
        \centering
        \includegraphics[width=1.0\linewidth]{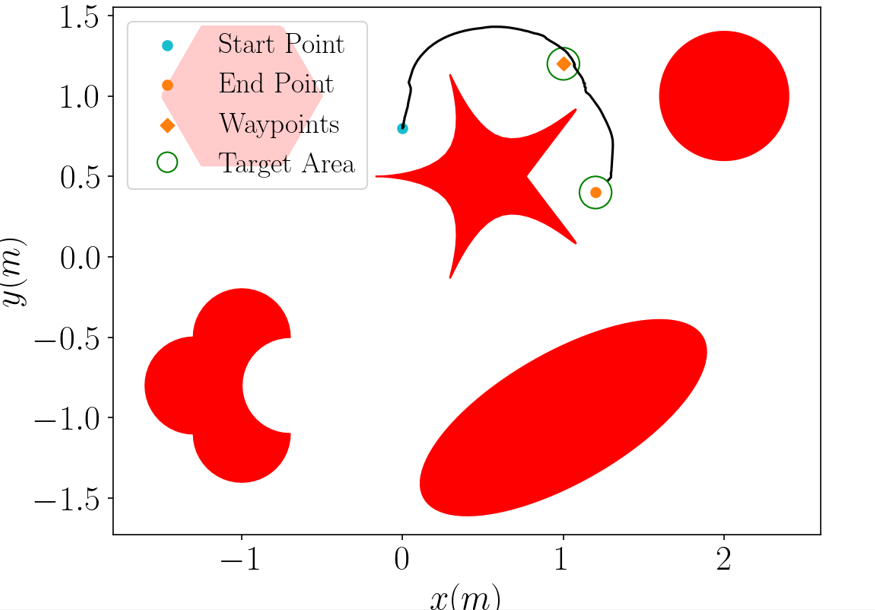}
        \label{subfig:33}
    \end{subfigure}
    \begin{subfigure}[t]{0.245\linewidth}
        \centering
        \includegraphics[width=1.0\linewidth]{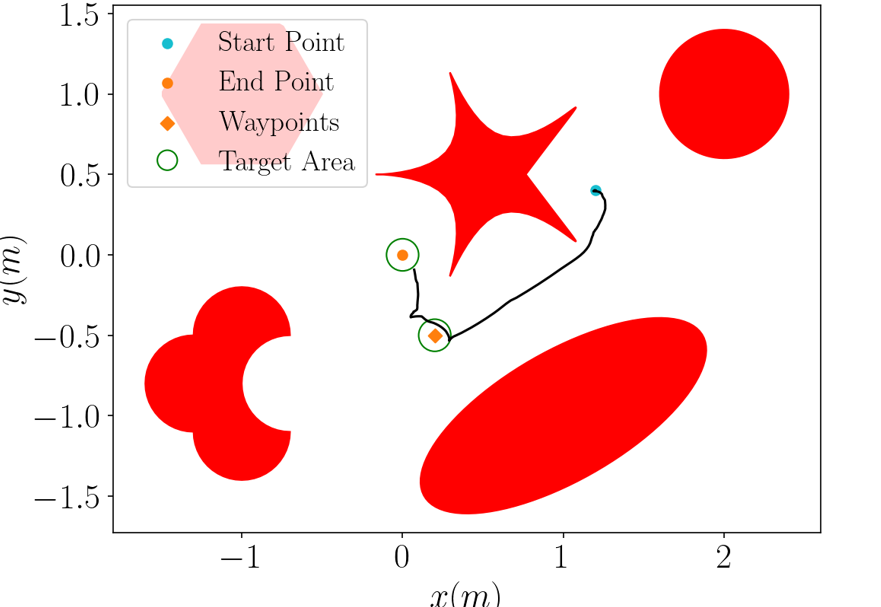}
        \label{subfig:44}
    \end{subfigure}
    \caption{Five closed-loop trajectories (black) with different start and end points controlled by iMPC-DHOCBF with $m_{\text{cbf}}=1, N=24, \gamma_{1}=0.1$. The obstacles are shown in red. The waypoints used in each subfigure are: $(-0.2, 1.0)$, $(-0.5, -0.2)$, $(1.0, -0.2)$, $(1.5, -0.3)$; $(-0.6, 0.0)$; $(1.0, 1.2)$; $(0.2, -0.5)$, respectively. The animation video can be found at \url{https://youtu.be/G9S7y90LRig}.}
    \label{fig:five obstacles navigation}
\end{figure*}

\begin{table*}[]
\vspace*{-0.4cm}
\centering
\caption{Statistical benchmark for computation time and feasibility for iMPC-DHOCBF with SBD.
The randomized states and end points are shared among four approaches and the other irrelevant hyperparameters are fixed for all scenarios. }
\label{tab:impc with SBD}

\resizebox{0.99\textwidth}{!}{
\begin{tabular}{|cc|cccccc|}
\hline
\multicolumn{2}{|c|}{Approaches} & $N = 4$ & $N = 8$ & $N = 12$ & $N = 16$ & $N = 20$ & $N = 24$ \\ \hline
\multicolumn{1}{|c|}{\multirow{2}{*}{\begin{tabular}[c]{@{}c@{}}iMPC-DHOCBF\\ ($m_{\text{cbf}} = 2,\gamma_1 = \gamma_2 = 0.1$)\end{tabular}}} & mean / std (s) & $0.029\pm0.020$ & $0.065\pm0.042$ & $0.126\pm0.069$ & $0.202\pm0.090$ & $0.340\pm0.126$ & $0.446\pm0.138$ \\
\multicolumn{1}{|c|}{} & infeas. rate & 4.5\% & 5.4\% & 5.0\%  &  6.0\% & 5.9\% & 6.4\% \\ \hline
\multicolumn{1}{|c|}{\multirow{2}{*}{\begin{tabular}[c]{@{}c@{}}iMPC-DHOCBF\\ ($m_{\text{cbf}} = 1,\gamma_1 = 0.1$)\end{tabular}}} & mean / std (s) & $0.028\pm0.020$ & $0.053\pm0.030$ & $0.108\pm0.055$ & $0.167\pm0.068$ & $0.235\pm0.082$ & $0.374\pm0.106$ \\
\multicolumn{1}{|c|}{} & infeas. rate & 4.9\% & 5.4\% & 4.9\% & 5.4\% & 5.0\% & 5.3\% \\ \hline
\multicolumn{1}{|c|}{\multirow{2}{*}{\begin{tabular}[c]{@{}c@{}}iMPC-DHOCBF\\ ($m_{\text{cbf}} = 2,\gamma_1 = \gamma_2 = 0.2$)\end{tabular}}} & mean / std (s) & $0.031\pm0.020$ & $0.073\pm0.046$ & $0.141\pm0.070$ & $0.242\pm0.091$ & $0.416\pm0.113$ & $0.545\pm0.144$ \\
\multicolumn{1}{|c|}{} & infeas. rate & 4.6\% & 4.9\% & 6.3\% & 7.5\% & 9.3\% & 10.9\% \\ \hline
\multicolumn{1}{|c|}{\multirow{2}{*}{\begin{tabular}[c]{@{}c@{}}iMPC-DHOCBF\\ ($m_{\text{cbf}} = 1,\gamma_1 = 0.2$)\end{tabular}}} & mean / std (s) & $0.032\pm0.024$ & $0.071\pm0.040$ & $0.139\pm0.055$ & $0.217\pm0.063$ & $0.296\pm0.063$ & $0.435\pm0.063$ \\
\multicolumn{1}{|c|}{} & infeas. rate & 4.5\% & 4.6\% & 4.3\% & 4.9\% & 4.5\% & 5.0\% \\ \hline
\end{tabular}
}
\end{table*}

\subsubsection{SBD Prediction Accuracy:}\hspace{0.1em} The prediction accuracy of SBD is demonstrated in Figs. \ref{fig:neural-network2} and \ref{fig: MSE five obstacles}. In Fig. \ref{fig:neural-network2}, we selected eight different locations from testing data on the map, each represented by a solid dot in a corresponding color. For these eight locations, there are five reference nearest points on the obstacle boundaries (diamonds) and the nearest points predicted by the SBD (plus symbols). When the robot is near the convex boundary of an obstacle, the predicted nearest points closely match the references. However, near concave boundaries, there is some deviation, though not substantial. Note that the predicted nearest point of the black dot is closer than the ground truth, demonstrating that the SBD can interpolate from sparse pixels to predict locally optimal nearest points, as discussed in Rem.~\ref{rem: motivation of DNN}. Figure \ref{fig: MSE five obstacles} shows the relationship between the SBD's prediction error (MSE) and the robot's positions in 3D and projected views. Since testing data are from outside the obstacles, MSE within obstacles is not shown (darkest areas). The MSE is relatively small near convex boundaries (darker colors) and larger near concave boundaries (brighter colors). Higher MSE values are observed near concave boundaries and extend along certain radial lines due to the presence of multiple equidistant nearest points on the obstacle boundaries relative to these lines. During data collection, one of the equidistant points is randomly chosen as a reference. This randomness leads the trained SBD to predict a nearest point as a blend of equidistant candidates, causing noticeable deviation from the reference point and a high MSE.

\subsubsection{Safe Trajectory Generation:}\hspace{0.1em}
Five safe closed-loop trajectories start from corresponding start points and end in corresponding target areas, shown in Fig. \ref{fig:five obstacles navigation}. 
We observe that even with multiple obstacles, including those with complex shapes like the sharp protrusions of a pentagram, the robot navigates safely in free space, regardless of proximity to obstacles or path length. Under the guidance of waypoints, the generated trajectories are smooth. They can be further improved by increasing the number of waypoints, using adaptive warm starts, and refining convergence criteria.

\subsubsection{Computation Time and Feasibility:}\hspace{0.1em}
In order to compare computational time and feasibility of our proposed iMPC-DHOCBF under different hyperparameters, 1000 independent randomized safe states are generated in state constraint $\mathcal{X}$ in \eqref{eq:state-input-constraint2}. To make a fair comparison, the randomized states and end points are shared among four sets of hyperparameters and the other irrelevant hyperparameters are fixed for all scenarios. We can observe in Tab. \ref{tab:impc with SBD} that as $N$ increases, the computation time per step significantly increases. Similarly, when $m_{\text{cbf}}$ increases or $\gamma_{1},\gamma_{2}$ increase, the computation time per step also rises. 
The computation time is primarily affected by 
$N$ and $m_{\text{cbf}}$ because each call to the SBD consumes a certain amount of time. Increases in $N$ and $m_{\text{cbf}}$ linearly raise the number of SBD calls, thus significantly increasing the computation time per step. However, these hyperparameters are shown not to  affect the infeasibility rate of the iMPC-DHOCBF method proportionally. Compared to the iMPC-DHOCBF in Tab. \ref{tab:compuation-time1}, iMPC-DHOCBF triggering SBD still maintains a fast computation speed per step and has a lower infeasibility rate, even in a more complex map.

\subsection{CASE STUDY \Romannum{3}: CIRCULAR ROBOT NAVIGATION IN A NARROW TRACK WITH UNKNOWN BOUNDARIES}
Similar to Sec. \textnormal{V-}\ref{subsec: case study 2}, we will train a DNN for the SBD to predict the nearest points. 
\subsubsection{System Dynamics:}\hspace{0.1em}
The robot is circular with a radius of 0.4, and its geometric center possesses the lateral vehicle model (see Eqs. (2.52), (2.53) in \cite{rajamani2011vehicle}) expressed by 
\begin{small}
\begin{equation}
\label{eq:vehicle}
\begin{bmatrix}
x_{t+1} - x_t \\
y_{t+1} - y_t \\
\psi_{t+1} - \psi_t \\
v_{t+1} - v_t \\
\beta_{t+1} - \beta_t \\
r_{t+1} - r_t
\end{bmatrix}
=
\begin{bmatrix}
v_t \cos(\psi_t + \beta_t)\Delta t \\
v_t \sin(\psi_t + \beta_t)\Delta t \\
r_t\Delta t \\
0 \\
E_t\Delta t \\
F_t\Delta t
\end{bmatrix}+
\begin{bmatrix}
0 & 0 \\
0 & 0 \\
0 & 0 \\
\Delta t & 0 \\
0 & \frac{C_f}{m (v_t+\epsilon)}\Delta t \\
0 & \frac{C_fl_f}{I_z}  \Delta t
\end{bmatrix}\begin{bmatrix} u_{1,t} \\ u_{2,t} \end{bmatrix},
\end{equation}
\end{small}
where 
\begin{small}
\begin{equation}
\begin{split}
E_t=\frac{C_r l_r - C_f l_f}{m (v_t+\epsilon)^2} r_t - \frac{C_f + C_r}{m (v_t+\epsilon)} \beta_t + \frac{g\sin(\phi)}{v_t+\epsilon} - r_t , \\
F_t= \frac{1}{I_z} \left((-C_f l_f + C_r l_r) \beta_t - \frac{C_f l_f^2 + C_r l_r^2}{v_t+\epsilon} r_t \right).
\end{split}
\end{equation}
\end{small}
$\mathbf{x}_{t}=[x_{t},y_{t},\psi_{t},v_{t},\beta_{t},r_{t}]^{T}$ captures the 2-D location, heading angle, linear speed, slip angle and yaw rate; $\mathbf{u}_{t}=[u_{1,t},u_{2,t}]^{T}$ represents linear acceleration ($u_{1}$) and  the front wheel steering angle ($u_{2}$), respectively.
The system is discretized with $\Delta t = 0.05.$ Other parameters are defined as $\phi=0, m=1500, g=9.8, l_{f}=1.2,l_{r}=1.6, I_{z}=2500, C_{f}=C_{r}=10000, \epsilon=11.$ Let \( \mathcal{X} \subset \mathbb{R}^6 \) and \( \mathcal{U} \subset \mathbb{R}^2 \) denote the state and input constraint sets, respectively. We define \( \mathbf{x}_t \in \mathcal{X} \) and \( \mathbf{u}_t \in \mathcal{U} \) if and only if \( x_t, y_t, \psi_t, v_t, r_t, u_{1,t}, u_{2,t} \in [-10, 10] \) and \( \beta_t \in [-1, 1] \).
\subsubsection{System Configuration:}\hspace{0.1em}
We plan to generate five connected closed-loop trajectories, where the start point of each trajectory is the end point of the previous one. The initial location 
 and four waypoints marked as solid dots are $(4, -4.5), (2.5, -2), (2, 2), (-4, 2), (-7.5, -3.3)$ and the end point is $(-2.5, -4.25)$ in Fig. \ref{fig:maze navigation}. The robot can terminate the current trajectory when its center reaches inside circles centered at each waypoint with a radius of 0.1 and then begin a second trajectory. The initial and reference linear speeds, slip angles and yaw rates are 0 and the initial and reference heading angles are calculated by $\psi_{0}=\psi_{r}=atan2(\frac{y_{r}-y_{t}}{x_{r}-x_{t}}),$ where $(x_{t}, y_{t})$ denotes the current location and $(x_{r}, y_{r})$ denotes each waypoint. The robot's operating area is a narrow track enclosed by two bar-shaped obstacles with point contact connections. The width of the track varies within $[\sqrt{2}, 2],$ and the track shape consists of Z-shaped bends and S-shaped bends. The other reference vectors are $\mathbf{u}_{r}=[0,0]^{T}$ and $\omega_{r}=\mathcal{I}_{1\times4}$. 
\subsubsection{Map Processing:}\hspace{0.1em} The scope of the map is $x\in[-10, 6]$ and $y\in[-6, 6].$ We manually set the coordinates of pixels on the two obstacle boundaries in contact with the track, limiting each boundary to a maximum of 400 pixels. This is equivalent to enclosing the free track space with a closed shape formed by two boundary lines from the two obstacles. 

\subsubsection{SBD Training and Evaluation:}\hspace{0.1em} Our neural network's structure is completely identical to that in Sec. \textnormal{V-B-}\ref{subsubsec:SBD structure}. Within the range of the map, we extracted 714875 data entries (random locations across the entire map) for training and another 48761 (random locations inside the track) for testing. We use the same MSE \eqref{eq:MSE} to evaluate the accuracy of the SBD's predictions of the nearest points on obstacle boundaries relative to the robot.
\subsubsection{DHOCBF:}\hspace{0.1em}
The design of DHOCBFs is exactly the same as that in Sec. \textnormal{V-B-}\ref{subsubsec: DHOCBF}. Note that our robot is circular, therefore, equation \eqref{eq:linearized safe set}, which imposes constraints on the robot's center position, should be shifted a distance equal to the radius toward the robot's center.

\subsubsection{MPC Design:}\hspace{0.1em}
The cost function of the CFTOC problem \eqref{eq:impc-dcbf} consists of stage cost
$q(\mathbf{x}_{t,k}^j,\mathbf{u}_{t,k}^{j},\omega_{t,k,i}^{j})= \sum_{k=0}^{N-1} (||\mathbf{x}_{t,k}^{j}-\mathbf{x}_{r}||_Q^2 + ||\mathbf{u}_{t,k}^{j}-\mathbf{u}_{r}||_R^2 +||\omega_{t,k,i}^{j}-\omega_{r}||_S^2)$
and terminal cost $p(\mathbf{x}_{t,N}^{j})=||\mathbf{x}_{t,N}^{j}-\mathbf{x}_{r}||_P^2$, where $Q=P=10\cdot \mathcal{I}_{6}, R= \mathcal{I}_{2}$ and $S=1000\cdot \mathcal{I}_{4}$. In this case study, the slack variable $\omega_{t,k,i}^{j}$ is subject to the constraint in \eqref {eq:convex-hocbf-constraint-2} for iMPC-DHOCBF to enhance feasibility while guaranteeing safety, as discussed in Thm. \ref{thm:safety-feasibility}. 

\subsubsection{Convergence Criteria:}\hspace{0.1em}
same as Sec. \textnormal{V-B-}\ref{subsubsec: convergence criteria}.
\begin{figure*}[!t]
    \vspace*{-0.4cm}
    \centering
    \begin{subfigure}[t]{0.26\linewidth}
        \centering
        \includegraphics[width=1.0\linewidth]{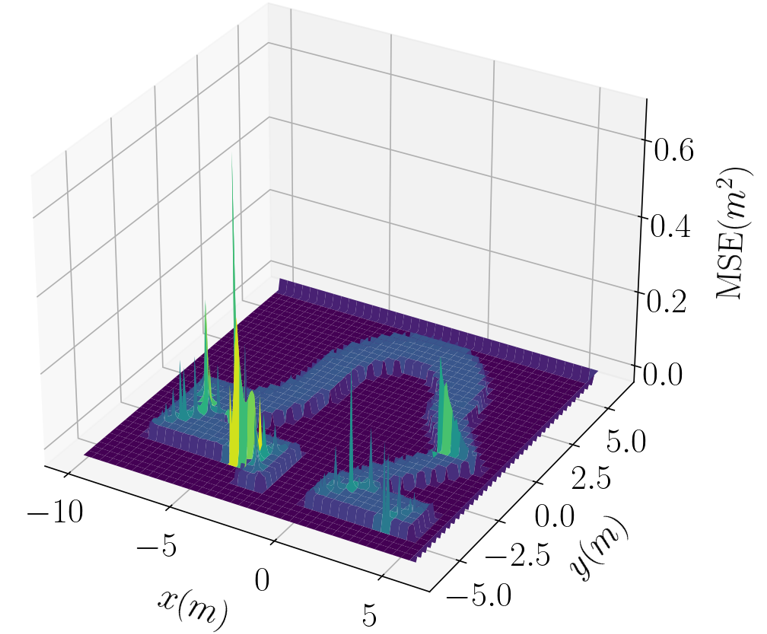}
        \caption{Isometric view}
        \label{subfig:1}
    \end{subfigure}
    \begin{subfigure}[t]{0.24\linewidth}
        \centering
        \includegraphics[width=1.0\linewidth]{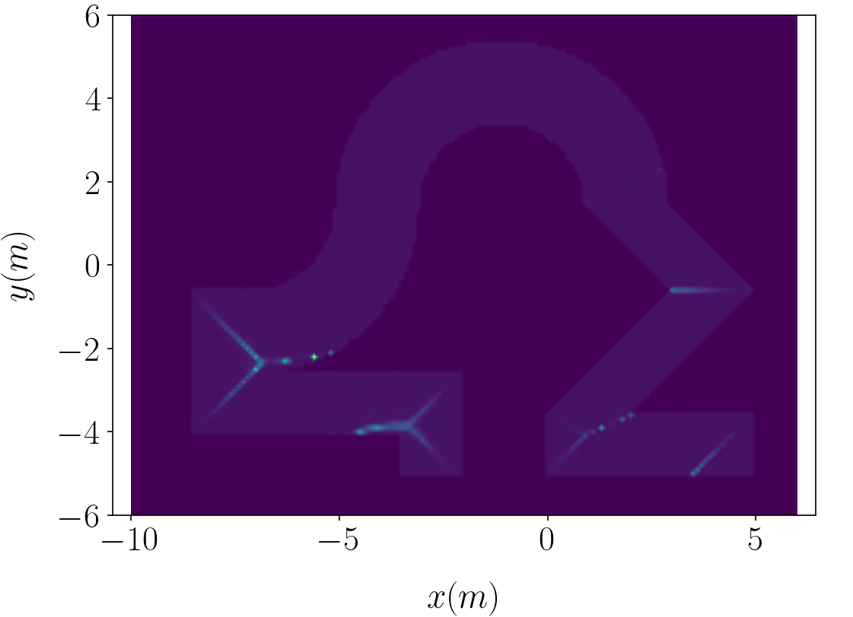}
        \caption{Top view}
        \label{subfig:2}
    \end{subfigure}  
    \begin{subfigure}[t]{0.24\linewidth}
        \centering
        \includegraphics[width=1.0\linewidth]{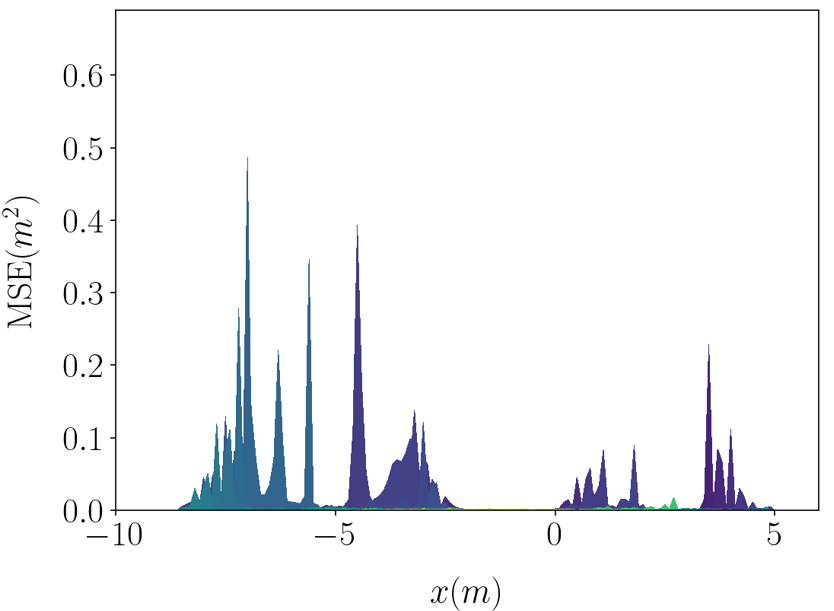}
        \caption{$x\text{-MSE}$ plane view}
        \label{subfig:3}
    \end{subfigure}
    \begin{subfigure}[t]{0.24\linewidth}
        \centering
        \includegraphics[width=1.0\linewidth]{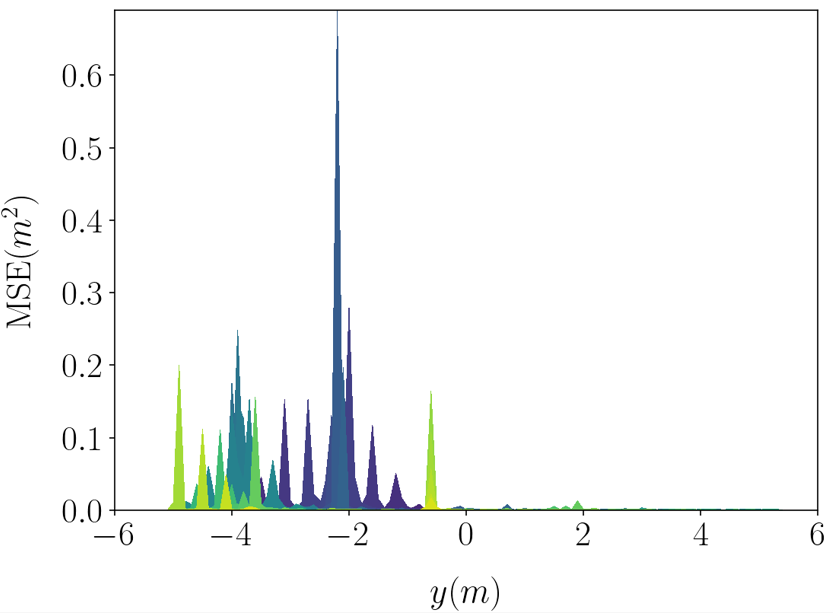}
        \caption{$y\text{-MSE}$ plane view}
        \label{subfig:4}
    \end{subfigure}
    \caption{3D visualization of the MSE of the testing data. $x$ and $y$ represent the positions of the testing points, while MSE represents the prediction error corresponding to the point $(x,y).$ The maximum MSE is 0.690 at $(-5.5, -2.1).$}
    \label{fig: MSE maze}
\end{figure*}
\subsubsection{Solver and CPU Specs:}\hspace{0.1em}
same as Sec. \textnormal{V-B-}\ref{subsubsec: specs}.

\subsubsection{SBD Prediction Accuracy:}\hspace{0.1em} The prediction accuracy of SBD is demonstrated in Fig. \ref{fig: MSE maze}, where we show the relationship between the SBD's prediction error (MSE) and the robot's various positions.
Since the testing data was selected from the robot track, the MSE outside the track on the chart is devoid of values and the darkest. We can observe that when the robot's position is near the convex boundary of the track, the MSE of the predicted nearest point is relatively small (indicated by darker colors). Conversely, when the robot's position is near the concave boundary of the track, the MSE of the predicted nearest point is relatively large (indicated by brighter colors), which is similar to the observations made in Fig. \ref{fig: MSE five obstacles}. Unlike Fig. \ref{fig: MSE five obstacles}, not all concave boundary areas exhibit large prediction errors, e.g., the concave boundary sections of Z-shaped bends display a linear pattern of increasing MSE while concave boundary sections of the S-shaped bends do not show large MSE. This shows that the concavity of a boundary relative to a point outside the boundary also affects the MSE of nearest point prediction. The greater the concavity, the larger the prediction error. Additionally, we noted that the largest MSE is 0.690, which is much larger than the maximum MSE in Fig. \ref{fig: MSE five obstacles}. The peak MSEs in Fig. \ref{fig: MSE maze} are also significantly larger than that in Fig. \ref{fig: MSE five obstacles}. This is because the size of the map also affects the prediction error. Large maps often require significantly large quantities of training data to reduce prediction errors to very low levels. We can enhance safety using the approach described in Rem. \ref{rem: convexity}, or adjust the slack-variable weight $S$ to reduce the conservativeness introduced by prediction errors.

\begin{figure}
    \centering
    \includegraphics[scale=0.25]{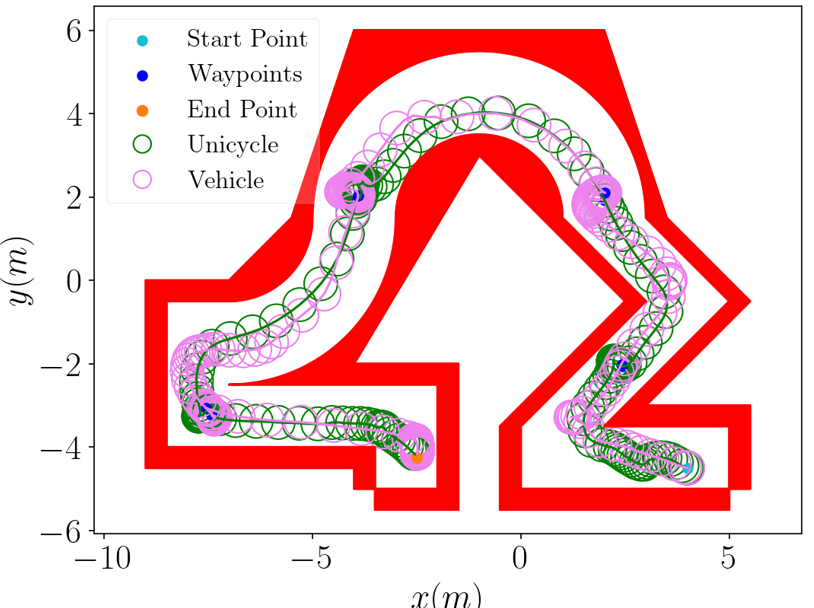}
    \caption{Five connected closed-loop trajectories (pink for the vehicle model and green for the unicycle model) are generated by iMPC-DHOCBF with \( m_{\text{cbf}} = 1 \), \( N = 24 \), and \( \gamma_1 = 0.1 \). Each trajectory starts where the previous ends. The narrow track is enclosed by two red obstacles. The method ensures safe navigation in tight spaces; animation is available at \url{https://youtu.be/G9S7y90LRig}.}
    \label{fig:maze navigation}
\end{figure}
\subsubsection{Safe Trajectory Generation:}
\begin{figure}
    \centering
    \includegraphics[scale=0.23]{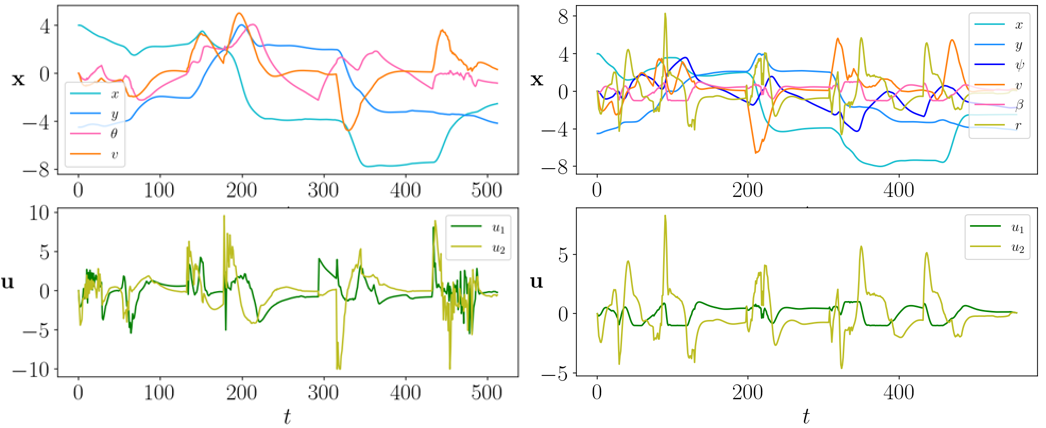}
    \caption{States and control inputs vary over time steps. These system
variables are relative to the motion of the robot governed by the unicycle model (left) and the vehicle model (right) shown in Fig. \ref{fig:maze navigation} and are subject to the corresponding constraints $\mathcal{X}$, $ \mathcal{U}$.}
    \label{fig:states-inputs-all}
\end{figure}

\hspace{0.1em} We evaluated the robot using two models: the vehicle model \eqref{eq:vehicle} and the unicycle model \eqref{eq:unicycle-model}. The parameters and design of control strategy of the unicycle model are identical to those described in Sec. \textnormal{V-}\ref{subsec: case study 2}.
Five closed-loop trajectories—depicted in pink for the vehicle model and green for the unicycle model—connecting the waypoints are shown in Fig. \ref{fig:maze navigation}. 
We can see that even within a narrow track, sometimes requiring sharp turns, the robot governed by both models is still able to navigate safely, demonstrating the scalability of our method across different dynamic constraints. In Fig. \ref{fig:states-inputs-all}, we can observe the trends in the robot’s
states and inputs as they change with each time step during
its movement. Since the reference values for $u_1, u_2$ in the
cost function are all zero, they fluctuate within a small range
around zero as a baseline. All system variables are subject to
the corresponding constraints $\mathcal{X}$ and $ \mathcal{U}$, which highlights the safety and planning
features under constraints of our implementation. In terms of computational complexity, the unicycle model takes an average of 0.6 seconds per time step, while the vehicle model takes 0.7 seconds. The slightly higher computation time of the vehicle model is mainly due to its more nonlinear dynamics (e.g., slip angle $\beta$ and yaw rate $r$), with additional factors detailed in Sec.\textnormal{IV-}E-\ref{subsubsec:state input scale}.

\section{CONCLUSION \& FUTURE WORK}
We proposed an iterative convex optimization procedure for safety-critical model predictive control design. Central to our approach is a learning-based safety boundary detector (SBD) to predict linearized safety boundaries for unsafe sets with arbitrary shapes, the transformation of these safety boundaries into discrete-time high-order control barrier functions (DHOCBFs), and relaxations for the system dynamics and for DHOCBF in the form of linearized constraints. We validate iMPC-DHOCBF on a unicycle robot in three environments (circular obstacles, irregular obstacles, and a narrow track) and extend it to a robot with vehicle dynamics in the narrow-track scenario. Our method outperforms the baseline in computation time and feasibility rate, and both models safely follow pre-designed waypoints through complex environments.
There are still some limitations of iMPC-DHOCBF with SBD that could be ameliorated. One limitation of the proposed method is for locations near the concave boundary of an unsafe set, the SBD's prediction error for the nearest point remains relatively large.
Another limitation is that our DNN-based SBD does not provide a strict a priori bound on its prediction error. Moreover, the feasibility of the optimization and system safety are not always guaranteed at the same time in the whole state space. We will address these limitations in future work.

\vspace*{5pt}

\begin{IEEEbiography}[{\includegraphics[width=1in,height=1.25in,clip,keepaspectratio]{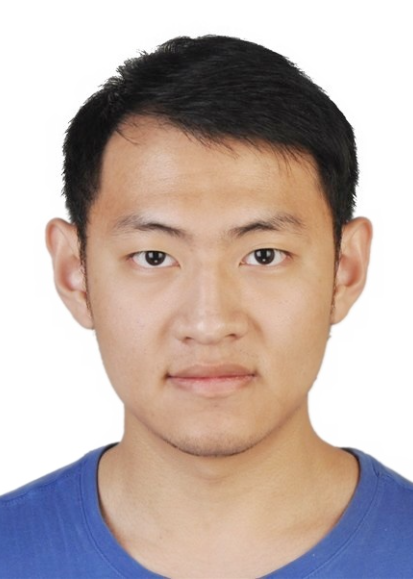}}]{SHUO LIU}{\space}(Student Membe, IEEE) received his M.S. degree in Mechanical Engineering from Columbia University, New York, NY, USA, in 2020 and his B.Eng. degree in Mechanical Engineering from Chongqing University, Chongqing, China, in 2018. He is currently a Ph.D. candidate in Mechanical Engineering at Boston University, Boston, USA. His research interests include optimization, nonlinear control, deep learning, and robotics.
\end{IEEEbiography}

\begin{IEEEbiography}[{\includegraphics[width=1in,height=1.25in,clip,keepaspectratio]{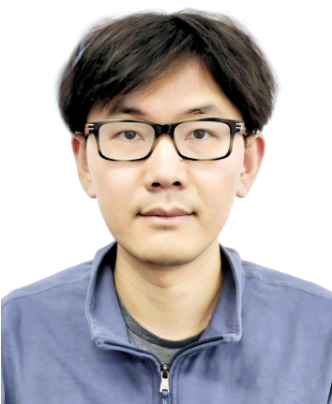}}]{ZHE HUANG}{\space}received his M.S. degree in Robotics from Boston University, Boston, USA, in 2023 and an M.S. degree in Computer Science from the University of Dayton, Dayton, USA, in 2017. He earned his B.S. degree in Electro-Optics from Jianghan University, Wuhan, China, in 2014. His research interests include deep learning, computer vision, robot perception, and robot control.
\end{IEEEbiography}

\begin{IEEEbiography}[{\includegraphics[width=1in,height=1.25in,clip,keepaspectratio]{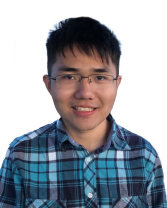}}]{JUN ZENG}{\space}(Member, IEEE) received his Ph.D. in Control and Robotics from the Department of Mechanical Engineering at the University of California, Berkeley, USA, in 2022, his Dipl. Ing. degree from École Polytechnique, France, in 2017, and his B.S.E. degree from Shanghai Jiao Tong University, China, in 2016. His research interests lie at the intersection of optimization, control, planning, and learning, with applications to various robotics platforms.
\end{IEEEbiography}

\begin{IEEEbiography}[{\includegraphics[width=1in,height=1.25in,clip,keepaspectratio]{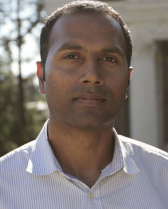}}]{KOUSHIL SREENATH}
{\space}(Senior Member, IEEE) received the M.S. degree in Applied Mathematics and the Ph.D. degree in Electrical Engineering and Systems from the University of Michigan, Ann Arbor, MI, USA, in 2011. He is an Assistant Professor of Mechanical Engineering with the University of California, Berkeley, Berkeley, CA, USA. He was previously an Assistant Professor with Carnegie Mellon University, Pittsburgh, PA, USA, from 2013 to 2017. His research interests include the intersection of highly dynamic robotics and applied nonlinear control.
\end{IEEEbiography}

\begin{IEEEbiography}[{\includegraphics[width=1in,height=1.25in,clip,keepaspectratio]{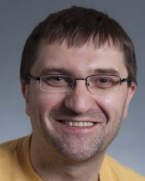}}]{CALIN A. BELTA}{\space}(Fellow, IEEE) received his Ph.D. in Mechanical Engineering from the University of Pennsylvania, Philadelphia, PA, USA, in 2003. He is currently the Brendan Iribe Endowed Professor of Electrical and Computer Engineering and Computer Science at the University of Maryland, College Park, MD, USA. His research focuses on control theory and formal methods, with particular emphasis on hybrid and cyber-physical systems, synthesis and verification, and applications in robotics and biology.
\end{IEEEbiography}

\clearpage
\begin{thebibliography}{99}
\setcounter{enumiv}{0}
\bibitem{ames2014control}A. D. Ames, J. W. Grizzle and P. Tabuada, ``Control barrier function based quadratic programs with application to adaptive cruise control,'' in {\em Proc. 53rd IEEE Conf. Decis. Control}, Los Angeles, CA, USA, 2014, pp. 6271--6278.


\setcounter{enumiv}{1}
\bibitem{ames2016control}A. D. Ames, X. Xu, J. W. Grizzle and P. Tabuada, ``Control barrier function based quadratic programs for safety critical systems,'' {\em  IEEE Trans. Autom. Control.}, vol. 62, no. 8, pp. 3861--3876, Aug. 2017, DOI: 10.1109/TAC.2016.2638961.


\setcounter{enumiv}{2}
\bibitem{zeng2021safety}J. Zeng, B. Zhang and K. Sreenath, ``Safety-critical model predictive control with discrete-time control barrier function,'' in {\em Proc. 2021 Am. Control Conf.}, New Orleans, LA, USA, 2021, pp. 3882--3889.

\setcounter{enumiv}{3}
\bibitem{diehl2005real}M. Diehl, H. G. Bock, and J. P. Schlöder, ``A real-time iteration scheme for nonlinear optimization in optimal feedback control,'' {\em SIAM J. Control Optim.}, vol. 43, no. 5, pp. 1714--1736, Jan. 2005, DOI: 10.1137/S0363012902400713.

\setcounter{enumiv}{4}
\bibitem{paolo2017mpc}G. Paolo, I. Ferrara, and L. Magni, ``MPC for robot manipulators with integral sliding mode generation,'' {\em IEEE/ASME Trans. Mechatronics}, vol. 22, no. 3, pp. 1299--1307, June. 2017, DOI: 10.1109/TMECH.2017.2674701.

\setcounter{enumiv}{5}
\bibitem{scianca2020mpc}N. Scianca, D. De Simone, L. Lanari, and G. Oriolo, ``MPC for humanoid gait generation: Stability and feasibility,'' {\em IEEE Trans. Robot.}, vol. 36, no. 4, pp. 1171--1188, Aug. 2020, DOI: 10.1109/TRO.2019.2958483.

\setcounter{enumiv}{6}
\bibitem{grandia2020nonlinear}R. Grandia, A. J. Taylor, A. Singletary, M. Hutter, and A. D. Ames, ``Nonlinear model predictive control of robotic systems with control Lyapunov functions,'' in {\em Proc. Robot.: Sci. Syst.}, New York City, NY, USA, 2022.

\setcounter{enumiv}{7}
\bibitem{son2019safety}T. D. Son and Q. Nguyen, ``Safety-critical control for non-affine nonlinear systems with application on autonomous vehicle,'' in {\em Proc. 58th IEEE Conf. Decis. Control}, Nice, France, 2019, pp. 7623--7628.

\setcounter{enumiv}{8}
\bibitem{eklund2011switched}J. M. Eklund, J. Sprinkle, and S. S. Sastry, ``Switched and symmetric pursuit/evasion games using online model predictive control with application to autonomous aircraft,'' {\em IEEE Trans. Control Syst. Technol.}, vol. 20, no. 3, pp. 604--620, May. 2012, DOI: 10.1109/TCST.2011.2136435.

\setcounter{enumiv}{9}
\bibitem{frasch2013auto}J. V. Frasch, A. Gray, M. Zanon, H. J. Ferreau, S. Sager, F. Borrelli, and M. Diehl, ``An auto-generated nonlinear MPC algorithm for real-time obstacle avoidance of ground vehicles,'' in {\em Proc. Eur. Control Conf.}, Zürich, Switzerland, 2013, pp. 4136--4141.

\setcounter{enumiv}{10}
\bibitem{liniger2015optimization}A. Liniger, A. Domahidi, and M. Morari, ``Optimization-based autonomous racing of 1:43 scale RC cars,'' {\em Optim. Control Appl. Methods}, vol. 36, no. 5, pp. 628--647, July. 2015.

\setcounter{enumiv}{11}
\bibitem{zhang2020optimization}X. Zhang, A. Liniger, and F. Borrelli, ``Optimization-based collision avoidance,'' {\em IEEE Trans. Control Syst. Technol.}, vol. 29, no. 3, pp. 972--983, May. 2020, DOI: 10.1109/TCST.2019.2949540.

\setcounter{enumiv}{12}
\bibitem{galloway2015torque}K. Galloway, K. Sreenath, A. D. Ames, and J. W. Grizzle, ``Torque saturation in bipedal robotic walking through control Lyapunov function-based quadratic programs,'' {\em IEEE Access}, vol. 3, pp. 323--332, Apr. 2015, DOI: 10.1109/ACCESS.2015.2419630.

\setcounter{enumiv}{13}
\bibitem{ames2019control}A. D. Ames, S. Coogan, M. Egerstedt, G. Notomista, K. Sreenath, and P. Tabuada, ``Control barrier functions: Theory and applications,'' in {\em Proc. 18th Eur. Control Conf.}, Naples, Italy, 2019, pp. 3420--3431.

\setcounter{enumiv}{14}
\bibitem{nguyen20163d}Q. Nguyen, A. Hereid, J. W. Grizzle, A. D. Ames, and K. Sreenath, ``3D dynamic walking on stepping stones with control barrier functions,'' in {\em Proc. 55th IEEE Conf. Decis. Control}, Las Vegas, NV, USA, 2016, pp. 827--834.

\setcounter{enumiv}{15}
\bibitem{nguyen2016exponential}Q. Nguyen and K. Sreenath, ``Exponential control barrier functions for enforcing high relative-degree safety-critical constraints,'' in {\em Proc. 2016 Am. Control Conf.}, Boston, MA, USA, 2016, pp. 322--328.

\setcounter{enumiv}{16}
\bibitem{xiao2021high}W. Xiao and C. Belta, ``High-order control barrier functions,'' {\em IEEE Trans. Autom. Control}, vol. 67, no. 7, pp. 3655--3662, Jul. 2022, DOI: 10.1109/TAC.2021.3105491.

\setcounter{enumiv}{17}
\bibitem{tan2021high}X. Tan, W. S. Cortez, and D. V. Dimarogonas, ``High-order barrier functions: Robustness, safety, and performance-critical control,'' {\em IEEE Trans. Autom. Control}, vol. 67, no. 6, pp. 3021--3028, Jun. 2021, DOI: 10.1109/TAC.2021.3089639.

\setcounter{enumiv}{18}
\bibitem{usevitch2022adversarial}J. Usevitch and D. Panagou, ``Adversarial resilience for sampled-data systems under high-relative-degree safety constraints,'' {\em IEEE Trans. Autom. Control},  vol. 68, no. 3, pp. 1537--1552, Mar. 2023, DOI: 10.1109/TAC.2022.3157791.

\setcounter{enumiv}{19}
\bibitem{xiao2021adaptive}W. Xiao, C. Belta, and C. G. Cassandras, ``Adaptive control barrier functions,'' {\em IEEE Trans. Autom. Control}, vol. 67, no. 5, pp. 2267--2281, May. 2022, DOI: 10.1109/TAC.2021.3074895.

\setcounter{enumiv}{20}
\bibitem{liu2023auxiliary}S. Liu, W. Xiao, and C. A. Belta, ``Auxiliary-variable adaptive control barrier functions for safety critical systems,'' in {\em Proc. 62nd IEEE Conf. Decis. Control}, Singapore, 2023, pp. 8602--8607.

\setcounter{enumiv}{21}
\bibitem{Nguyen2015_RobustCLF}Q. Nguyen and K. Sreenath, ``Optimal robust control for bipedal robots through control Lyapunov function based quadratic programs,'' in {\em Proc. Robot.: Sci. Syst.}, Rome, Italy, 2015.

\setcounter{enumiv}{22}
\bibitem{jankovic2018robust}M. Jankovic, ``Robust control barrier functions for constrained stabilization of nonlinear systems,'' {\em Automatica}, vol. 96, pp. 359--367, Oct. 2018.

\setcounter{enumiv}{23}
\bibitem{clark2021control}A. Clark, ``Control barrier functions for stochastic systems,'' {\em Automatica}, vol. 130, p. 109--688, Aug. 2021.

\setcounter{enumiv}{24}
\bibitem{liu2023feasibility}S. Liu, W. Xiao, and C. A. Belta, ``Feasibility-guaranteed safety-critical control with applications to heterogeneous platoons,'' in {\em Proc. 63rd IEEE Conf. Decis. Control}, Milan, Italy, 2024, pp. 8066--8073.

\setcounter{enumiv}{25}
\bibitem{agrawal2017discrete}A. Agrawal and K. Sreenath, ``Discrete control barrier functions for safety-critical control of discrete systems with application to bipedal robot navigation,'' in {\em Proc. Robot.: Sci. Syst.}, Cambridge, MA, USA, 2017.

\setcounter{enumiv}{26}
\bibitem{ma2021feasibility}H. Ma, X. Zhang, S. E. Li, Z. Lin, Y. Lyu, and S. Zheng, ``Feasibility enhancement of constrained receding horizon control using generalized control barrier function,'' in {\em Proc. IEEE Int. Conf. Ind. Cyber-Phys. Syst.}, 2021, pp. 551--557.

\setcounter{enumiv}{27}
\bibitem{xiong2022discrete}Y. Xiong, D.-H. Zhai, M. Tavakoli, and Y. Xia, ``Discrete-time control barrier function: High-order case and adaptive case,'' {\em IEEE Trans. Cybern.}, vol. 53, no. 5, pp. 3231--3239, May. 2023, DOI: 10.1109/TCYB.2022.3170607.

\setcounter{enumiv}{28}
\bibitem{he2022autonomous}S. He, J. Zeng, and K. Sreenath, ``Autonomous racing with multiple vehicles using a parallelized optimization with safety guarantee using control barrier functions,'' in {\em Proc. IEEE Int. Conf. Robot. Autom.}, 2022, pp. 3444--3451.

\setcounter{enumiv}{29}
\bibitem{li2022bridging}Z. Li, J. Zeng, A. Thirugnanam, and K. Sreenath, ``Bridging model-based safety and model-free reinforcement learning through system identification of low dimensional linear models,'' in {\em Proc. Robot.: Sci. Syst.}, New York City, NY, 2022.

\setcounter{enumiv}{30}
\bibitem{zeng2021decay}J. Zeng, B. Zhang, Z. Li, and K. Sreenath, ``Safety-critical control using optimal-decay control barrier function with guaranteed point-wise feasibility,'' in {\em Proc. Am. Control Conf.}, 2021, pp. 3856--3863.

\setcounter{enumiv}{31}
\bibitem{zeng2021enhancing}J. Zeng, Z. Li, and K. Sreenath, ``Enhancing feasibility and safety of nonlinear model predictive control with discrete-time control barrier functions,'' in {\em Proc. 60th IEEE Conf. Decis. Control}, 2021, pp. 6137--6144.

\setcounter{enumiv}{32}
\bibitem{srinivasan2020synthesis}M. Srinivasan, A. Dabholkar, S. Coogan, and P. A. Vela, ``Synthesis of control barrier functions using a supervised machine learning approach,'' in {\em Proc. IEEE/RSJ Int. Conf. Intell. Robots Syst.}, 2020, pp. 7139--7145.

\setcounter{enumiv}{33}
\bibitem{saveriano2019learning}M. Saveriano and D. Lee, ``Learning barrier functions for constrained motion planning with dynamical systems,'' in {\em Proc. IEEE/RSJ Int. Conf. Intell. Robots Syst.}, 2019, pp. 112--119.

\setcounter{enumiv}{34}
\bibitem{peng2023safe}C. Peng, O. Donca, G. Castillo, and A. Hereid, ``Safe bipedal path planning via control barrier functions for polynomial shape obstacles estimated using logistic regression,'' in {\em Proc. IEEE Int. Conf. Robot. Autom.}, 2023, pp. 3649--3655.

\setcounter{enumiv}{35}
\bibitem{robey2020learning}A. Robey, H. Hu, L. Lindemann, H. Zhang, D. V. Dimarogonas, S. Tu, and N. Matni, ``Learning control barrier functions from expert demonstrations,'' in {\em Proc. 59th IEEE Conf. Decis. Control}, 2020, pp. 3717--3724.

\setcounter{enumiv}{36}
\bibitem{lavanakul2024safety}W. Lavanakul, J. Choi, K. Sreenath, and C. Tomlin, ``Safety filters for black-box dynamical systems by learning discriminating hyperplanes,'' in {\em Proc. Learn. Dyn. Control Conf.}, 2024, pp. 1278--1291.

\setcounter{enumiv}{37}
\bibitem{liu2023iterative}S. Liu, J. Zeng, K. Sreenath, and C. A. Belta, ``Iterative convex optimization for model predictive control with discrete-time high-order control barrier functions,'' in {\em Proc. Am. Control Conf.}, 2023, pp. 3368--3375.

\setcounter{enumiv}{38}
\bibitem{sun2003initial}M. Sun and D. Wang, ``Initial shift issues on discrete-time iterative learning control with system relative degree,'' {\em IEEE Trans. Autom. Control}, vol. 48, no. 1, pp. 144--148, Jan. 2003, DOI: 10.1109/TAC.2002.806668.

\setcounter{enumiv}{39}
\bibitem{thirugnanam2022duality}A. Thirugnanam, J. Zeng, and K. Sreenath, ``Duality-based convex optimization for real-time obstacle avoidance between polytopes with control barrier functions,'' in {\em Proc. Am. Control Conf.}, 2022, pp. 2239--2246.

\setcounter{enumiv}{40}
\bibitem{kramer1992autoassociative}M. A. Kramer, ``Autoassociative neural networks,'' {\em Comput. Chem. Eng.}, vol. 16, no. 4, pp. 313--328, Apr. 1992.

\setcounter{enumiv}{41}
\bibitem{oleynikova2017voxblox}H. Oleynikova, Z. Taylor, M. Fehr, R. Siegwart, and J. Nieto, ``Voxblox: Incremental 3D Euclidean signed distance fields for on-board MAV planning,'' in {\em Proc. IEEE/RSJ Int. Conf. Intell. Robots Syst.}, 2017, pp. 1366--1373.

\setcounter{enumiv}{42}
\bibitem{muja2009fast}M. Muja and D. G. Lowe, ``Fast approximate nearest neighbors with automatic algorithm configuration,'' in {\em Proc. Int. Conf. Comput. Vis. Theory Appl.}, 2009, pp. 331--340.

\setcounter{enumiv}{43}
\bibitem{stellato2020osqp}B. Stellato, G. Banjac, P. Goulart, A. Bemporad, and S. Boyd, ``OSQP: An operator splitting solver for quadratic programs,'' {\em Math. Program. Comput.}, vol. 12, no. 4, pp. 637--672, Feb. 2020.

\setcounter{enumiv}{44}
\bibitem{biegler2009large}L. T. Biegler and V. M. Zavala, ``Large-scale nonlinear programming using IPOPT: An integrating framework for enterprise-wide dynamic optimization,'' {\em Comput. Chem. Eng.}, vol. 33, no. 3, pp. 575--582, Mar. 2009.

\setcounter{enumiv}{45}
\bibitem{lofberg2004yalmip}J. Lofberg, ``YALMIP: A toolbox for modeling and optimization in MATLAB,'' in {\em Proc. IEEE Int. Conf. Robot. Autom.}, 2004, pp. 284--289.

\setcounter{enumiv}{46}
\bibitem{fisac2019tac}
J. F. Fisac, A. K. Akametalu, M. N. Zeilinger, S. Kaynama, J. Gillula, and C. J. Tomlin,
``A general safety framework for learning-based control in uncertain robotic systems,''
{\em IEEE Trans. Autom. Control}, vol. 64, no. 7, pp. 2737--2752, Jul. 2019,
DOI: 10.1109/TAC.2018.2876389.

\setcounter{enumiv}{47}
\bibitem{wang2024risk}
H. Wang and G. Zhao,
``Risk-aware safe feedback motion planning in Gaussian splatting world,''
in \emph{Proc. IEEE Int. Conf. Unmanned Syst. (ICUS)}, Nanjing, China, 2024, pp. 447--452.

\setcounter{enumiv}{48}
\bibitem{wabersich2021tac}
K. P. Wabersich, L. Hewing, A. Carron, and M. N. Zeilinger,
``Probabilistic model predictive safety certification for learning-based control,''
{\em IEEE Trans. Autom. Control}, vol. 67, no. 1, pp. 176--188, Jan. 2021, DOI: 10.1109/TAC.2021.3049335.

\setcounter{enumiv}{49}
\bibitem{rosolia2018learning}
U. Rosolia and F. Borrelli, ``Learning Model Predictive Control for Iterative Tasks: A Data-Driven Control Framework,'' {\em IEEE Trans. Autom. Control}, vol. 63, no. 7, pp. 1883--1896, Jul. 2018, DOI: 10.1109/TAC.2017.2753460.

\setcounter{enumiv}{50}
\bibitem{rajamani2011vehicle}
R. Rajamani, \emph{Vehicle Dynamics and Control}, Springer Science \& Business Media, 2011.

\setcounter{enumiv}{51}
\bibitem{borrelli2017predictive}
F. Borrelli, A. Bemporad, and M. Morari, \emph{Predictive Control for Linear and Hybrid Systems}, Cambridge University Press, 2017.

\setcounter{enumiv}{52}
\bibitem{achiam2017constrained}
J. Achiam, D. Held, A. Tamar, and P. Abbeel, 
``Constrained policy optimization,'' 
in \emph{Proc. Int. Conf. Machine Learning (ICML)}, 2017, pp. 22--31.

\setcounter{enumiv}{53}
\bibitem{belta2017formal}
C. Belta, B. Yordanov, and E. A. Gol, 
\emph{Formal Methods for Discrete-Time Dynamical Systems}, 
vol. 89. Springer, 2017.

\setcounter{enumiv}{54}
\bibitem{qinlearning}
Z. Qin, K. Zhang, Y. Chen, J. Chen, and C. Fan, 
``Learning safe multi-agent control with decentralized neural barrier certificates,'' 
in \emph{Proc. Int. Conf. Learning Representations (ICLR)}, 2021.

\setcounter{enumiv}{55}
\bibitem{tsuzuku2018lipschitz}
Y. Tsuzuku, I. Sato, and M. Sugiyama, 
``Lipschitz-margin training: Scalable certification of perturbation invariance for deep neural networks,'' 
in \emph{Advances in Neural Information Processing Systems (NeurIPS)}, vol. 31, 2018.

\setcounter{enumiv}{56}
\bibitem{katz2017reluplex}
G. Katz, C. Barrett, D. L. Dill, K. Julian, and M. J. Kochenderfer, 
``Reluplex: An efficient SMT solver for verifying deep neural networks,'' 
in \emph{Proc. Int. Conf. Computer Aided Verification (CAV)}, 2017, pp. 97--117.

\setcounter{enumiv}{57}
\bibitem{wong2018provable}
E. Wong and Z. Kolter, 
``Provable defenses against adversarial examples via the convex outer adversarial polytope,'' 
in \emph{Proc. Int. Conf. Machine Learning (ICML)}, 2018, pp. 5286--5295.

\end{thebibliography}
\end{document}